\documentclass[11pt]{article}
\pdfoutput=1
\usepackage{dcolumn}
\usepackage{bm}

\usepackage[square,numbers,compress,sort]{natbib}
\usepackage{graphicx}
\usepackage{amssymb,amsmath}
\usepackage{multirow}
\usepackage{slashed,xfrac}
\usepackage[makeroom]{cancel}
\usepackage{color,url}
\usepackage[colorlinks=true
,urlcolor=blue
,anchorcolor=blue
,citecolor=blue
,filecolor=blue
,linkcolor=blue
,menucolor=blue
,linktocpage=true
,pdfproducer=medialab
,pdfa=true
]{hyperref}

\usepackage{epsfig,psfrag,rotating,soul}
\usepackage{rotfloat}
\usepackage[normalem]{ulem}

\graphicspath{{figs/}}

\evensidemargin \oddsidemargin
\allowdisplaybreaks

\psfrag{lk}{$l_k$}
\psfrag{lm}{$l_m$}
\psfrag{blk}{$\bar {l}_k$}
\psfrag{blm}{$\bar {l}_m$}

\def\thefootnote{\fnsymbol{footnote}}
\def\ne{\boldsymbol{n}_e}
\def\nm{\boldsymbol{n}_\mu}
\def\nt{\boldsymbol{n}_\tau}
\def\mode{|\ne|}
\def\modmu{|\nm|}
\def\modtau{|\nt|}
\def\tme{\theta_{\mu e}}
\def\tte{\theta_{\tau e}}
\def\ttm{\theta_{\tau \mu}}
\def\cme{c_{\mu e}}
\def\cte{c_{\tau e}}
\def\ctm{c_{\tau \mu}} 

\def\modelname{ISS-$\cancel{\rm LFV}\hskip-.1cm_{\mu e}$}
\def\rotationO{\mathcal O}

\definecolor{vdrgreen}{rgb}{0.0, 0.7, 0.0}

\begin{document}
\thispagestyle{empty}

\begin{flushright}
IFT-UAM/CSIC-16-064\\
FTUAM-16-27\\
\end{flushright}

\vspace{0.5cm}

\begin{center}

\begin{Large}
\textbf{\textsc{Lepton flavor violating  $Z$ decays:}} \\
\textbf{\textsc{ A promising window to low scale seesaw neutrinos}}
\end{Large}

\vspace{1cm}

{\sc
V. De Romeri%
\footnote{\tt \href{mailto:valentina.deromeri@uam.es}{valentina.deromeri@uam.es}}%
, M.J. Herrero%
\footnote{\tt \href{mailto:maria.herrero@uam.es}{maria.herrero@uam.es}}%
, X. Marcano%
\footnote{\tt \href{mailto:xabier.marcano@uam.es}{xabier.marcano@uam.es}}%
, F. Scarcella%
\footnote{\tt \href{mailto:francesca.scarcella@studio.unibo.it}{francesca.scarcella@studio.unibo.it}}%
}

\vspace*{.7cm}

{\sl
\vspace*{0.1cm}

Departamento de F\'{\i}sica Te\'orica and Instituto de F\'{\i}sica Te\'orica, IFT-UAM/CSIC,\\
Universidad Aut\'onoma de Madrid, Cantoblanco, 28049 Madrid, Spain

}

\end{center}

\vspace*{0.1cm}

\begin{abstract}
\noindent
In this paper we study the Lepton Flavor Violating $Z$ boson decays $Z \to \tau \mu$ and $Z \to \tau e$ in the context of low scale seesaw models with new heavy Majorana neutrinos whose masses could be reachable at the LHC. Our computations of the decay rates are done in the particular realization given by the Inverse Seesaw Model with six extra heavy neutrinos which are quasi-degenerate in three pseudo-Dirac pairs. In particular, we focus on scenarios that are built {\it ad-hoc} to produce suppressed rates in all the processes involving $\mu$-$e$ transitions, given the fact that these are by far the most strongly constrained by present data. We will fully explore the $Z \to \tau \mu$ and $Z \to \tau e$ rates, together with a set of observables that we find to be the most constraining ones, and we will conclude that sizable rates of up to $2 \times 10^{-7}$, accessible at future colliders, can be reached in this model for Majorana masses in the few TeV range, potentially reachable  at LHC. 
\end{abstract}

\def\thefootnote{\arabic{footnote}}
\setcounter{page}{0}
\setcounter{footnote}{0}

\newpage
\section{Introduction}
\label{intro}

The observation of neutrino oscillations, showing that neutrinos do have masses and that lepton flavor violation (LFV) occurs in the neutrino sector, is at present the most clear experimental evidence that the Standard Model of Particle Physics (SM) is insufficient to explain data and needs to be extended. However, what is the particular new physics responsible for giving mass to the neutrinos and what is the origin of the neutrino flavor oscillations are still open questions that need to be answered.
The minimal {\it ad-hoc} extension would be the addition of right handed (RH) neutrino fields, $\nu_R$, to the SM spectrum, so that neutrinos could obtain a Dirac mass through their Yukawa interaction with the Higgs field, as the rest of the SM fermions. Nevertheless, this requires very small neutrino Yukawa couplings and an additional explanation of why there is not a Majorana mass term for the $\nu_R$ fields, since they are singlets under the full SM gauge group.
One of the most popular extensions that tries to address these questions is the type-I seesaw model \cite{Minkowski:1977sc, GellMann,Yanagida,Mohapatra:1979ia,Schechter:1980gr}, that adds RH neutrinos to the SM spectrum, allows both Dirac and heavy Majorana masses for the neutrinos, and explains the smallness of the experimentally observed light neutrino masses in terms of the small ratio of two very distant mass scales, the Dirac mass and the Majorana mass. This condition demands either a tiny neutrino Yukawa coupling or an extremely heavy Majorana mass scale, of the order of $10^{14-15}$ GeV. Thus, in these high scale seesaw models the differences in their phenomenological predictions with respect to the SM ones, due to the presence of the new very heavy neutrinos, are in general  extremely suppressed 
\cite{Ilakovac:1994kj,Arganda:2004bz, Akhmedov:2014kxa}. 
In contrast, it is well known that this suppression may be alleviated in low scale seesaw models, where the seesaw scale providing the mass to the heavy neutrinos can be successfully set to much lower values, even reachable at present colliders, like the CERN-LHC. These low scale seesaw models are variants of the type-I seesaw model where the smallness of the light neutrinos and the total lepton number (LN) symmetry are controlled instead by some additional small mass parameters. By the use of symmetry arguments that preserve the small size of these new mass parameters the low scale seesaw models then leave the possibility to lower the heavy neutrino mass scale,  below 10 TeV,  while keeping at the same time an interesting phenomenology due to the allowed presence of large neutrino Yukawa couplings.

On the other hand, one of the most interesting aspects of these low scale seesaw models is that the associated extension of the neutral lepton sector may also induce new rare phenomena in the charged lepton sector. In particular,
any observation of lepton flavor violation in the charged lepton sector (cLFV) would automatically imply the presence of new physics beyond the SM and could help throwing light on the question of what is the mechanism that generates  the neutrino masses.
Although cLFV has not been observed yet in Nature, there is an extensive experimental program developing different strategies to look for new physics signals in this charged lepton sector and, indeed, there are already at present very competitive upper bounds on several cLFV processes.
We summarize some of the current upper bounds on cLFV transitions in table \ref{LFVsearch} and the corresponding ones to the  LFV $Z$ gauge boson decays (LFVZD) and LFV $H$ boson decays (LFVHD) in table~\ref{LFVsearchII}.
It is interesting to notice that the LHC is currently improving notably these two latter bounds and that ATLAS is already at the level of LEP results for the LFVZD rates, and even better for $Z\to\mu e$ \cite{Aad:2014bca}.
Furthermore, the sensitivity to LFVZD rates are expected to highly improve at future linear colliders, with an expected sensitivity of $10^{-9}$ \cite{Wilson:I, Wilson:II}, or at a Future Circular  $e^+ e^-$ Collider (such as FCC-ee (TLEP)\cite{Blondel:2014bra}), where it is estimated that up to $10^{13}$ $Z$ bosons would be produced and the sensitivities to LFVZD rates could be improved up to $10^{-13}$.
Therefore, we consider extremely timely to explore the predictions for these LFVZD rates in any new physics scenario that could be related to neutrino physics, as  has been previously done in Beyond the Standard Model  frameworks like those with massive (Majorana and/or Dirac) neutrinos~\cite{Ilakovac:1994kj,Mann:1983dv,Bernabeu:1987gr,Dittmar:1989yg,Korner:1992an,Ilakovac:1999md,Illana:1999ww,Illana:2000ic}, or those using an Effective Field Theory approach~\cite{Perez:2003ad,FloresTlalpa:2001sp,Delepine:2001di,Davidson:2012wn}.

\begin{table}[t!]
\begin{center}
\begin{tabular}{|c|c|c|}
\hline
LFV Observable & Present Bound  $(90\%CL)$ & Future Sensitivity \\
\hline
BR$(\mu\to e\gamma)$ &  $4.2\times10^{-13}$ (MEG 2016)\cite{TheMEG:2016wtm} & $4\times 10^{-14} $ (MEG-II)~\cite{Baldini:2013ke}\\
BR$(\tau\to e\gamma)$ & $3.3\times10^{-8}$ (BABAR 2010)~\cite{Aubert:2009ag} & $ 10^{-9}$ (BELLE-II)~\cite{Aushev:2010bq}\\
BR$(\tau\to \mu\gamma)$ & $4.4\times10^{-8}$ (BABAR 2010)~\cite{Aubert:2009ag} & $ 10^{-9}$ (BELLE-II)~\cite{Aushev:2010bq}\\
BR$(\mu\to eee)$ &   $1.0\times10^{-12}$ (SINDRUM 1988)~\cite{Bellgardt:1987du} & $10^{-16}$ Mu3E (PSI)~\cite{Blondel:2013ia} \\
BR$(\tau\to eee)$ & $2.7\times10^{-8}$ (BELLE 2010)~\cite{Hayasaka:2010np} & $10^{-9,-10}$ (BELLE-II)~\cite{Aushev:2010bq}\\
BR$(\tau\to \mu\mu\mu)$ & $2.1\times10^{-8}$ (BELLE 2010)~\cite{Hayasaka:2010np} & $10^{-9,-10}$ (BELLE-II)~\cite{Aushev:2010bq}\\
BR$(\tau\to \mu\eta)$ & $2.3\times10^{-8}$ (BELLE 2010)~\cite{Hayasaka:2010et} & $10^{-9,-10}$ (BELLE-II)~\cite{Aushev:2010bq}\\
CR$(\mu-e,{\rm Au})$ & $7.0\times10^{-13}$ (SINDRUM II 2006)~\cite{Bertl:2006up}& \\
CR$(\mu-e,{\rm Ti})$ &  $4.3\times10^{-12}$ (SINDRUM II 2004)~\cite{Dohmen:1993mp}&$10^{-18}$ PRISM (J-PARC)~\cite{Alekou:2013eta}\\
CR$(\mu-e,{\rm Al})$ &&$3.1\times10^{-15}$ COMET-I (J-PARC)~\cite{Kuno:2013mha}\\
&&$2.6\times10^{-17}$ COMET-II (J-PARC)~\cite{Kuno:2013mha} \\
&&$2.5\times10^{-17}$ Mu2E (Fermilab)~\cite{Carey:2008zz} \\
\hline
\end{tabular}
\caption{Present upper bounds and future expected sensitivities for cLFV transitions.}\label{LFVsearch}
\vskip .5cm
\begin{tabular}{|c|c|c|c|}
\hline
LFV Observable & Present Bound  $(95\%CL)$ \\
\hline
BR$(Z\to\mu e)$ & $1.7\times10^{-6}$ (LEP 1995)~\cite{Akers:1995gz}, $7.5\times10^{-7}$ (ATLAS 2014)~\cite{Aad:2014bca}\\
BR$(Z\to\tau e)$ & $9.8\times10^{-6}$ (LEP 1995)~\cite{ Akers:1995gz} \\
BR$(Z\to\tau\mu )$ &$1.2\times10^{-5}$ (LEP 1995)~\cite{Abreu:1996mj}, $1.69\times10^{-5}$ (ATLAS 2014)~\cite{Aad:2016blu}\\
BR$(H\to\mu e)$ & $3.6\times10^{-3}$ (CMS 2015)~\cite{CMS:2015udp}\\
BR$(H\to\tau e)$ & $1.04\times10^{-2}$ (ATLAS 2016)~\cite{Aad:2016blu}, $0.7\times10^{-2}$  (CMS 2015)~\cite{CMS:2015udp} \\
BR$(H\to\tau\mu )$ & $1.43\times10^{-2}$ (ATLAS 2016)~\cite{Aad:2016blu}, $1.51\times10^{-2}$ (CMS 2015)~\cite{Khachatryan:2015kon} \\
\hline
\end{tabular}
\caption{Present upper bounds at $95\%$ CL on LFV decays of $Z$ and $H$ bosons.}\label{LFVsearchII}
\end{center}
\end{table}

In this work, we consider the Inverse Seesaw (ISS) \cite{Bernabeu:1987gr,Mohapatra:1986aw, Mohapatra:1986bd, GonzalezGarcia:1991be} as a specific realization of the low scale seesaw models.
In particular, the ISS extends the SM spectrum with three pairs of RH neutrinos with opposite lepton numbers and considers Majorana masses for some of these new fields, which are assumed to be naturally small since they are the only masses that violate LN.
Generically, these small Majorana masses are responsible for explaining the smallness of the light neutrino masses and give the freedom of having simultaneously large Yukawa couplings and moderately heavy RH neutrino masses, reachable at LHC, while keeping the values of the light neutrino masses and mixings in agreement with experimental data.
These features make the ISS an appealing model with a very rich phenomenology that has been studied in various processes like leptonic and semileptonic decays \cite{Abada:2012mc, Abada:2013aba}, electric dipole moments \cite{Abada:2015trh,Abada:2016awd}, lepton magnetic moments~\cite{Abada:2014nwa}, heavy neutrino production at colliders \cite{Chen:2011hc,BhupalDev:2012zg,Das:2012ze, Das:2014jxa,Arganda:2015ija, Das:2015toa, Das:2016hof}, dark matter \cite{Abada:2014zra}, LFV Higgs decays \cite{Arganda:2014dta,Arganda:2015naa} and many other cLFV processes \cite{Abada:2012cq, Abada:2014kba,Abada:2015oba}.
Nevertheless, looking at the present experimental upper bounds in table \ref{LFVsearch}, we see that the constraints on cLFV processes involving $\mu$-$e$ transitions, here called ${\rm LFV}_{\mu e}$ in short, are much stronger that the ones in the other sectors, i.e, cLFV processes involving
$\tau$-$\mu$ and $\tau$-$e$ transitions, named here in short ${\rm LFV}_{\tau \mu}$ and ${\rm LFV}_{\tau e}$, respectively. These very stringent constraints in the $\mu$-$e$ sector motivate the class of models considered here, which incorporate automatically this suppression in their input.
Specifically, we will implement this $\mu$-$e$ suppression requirement within the context of the ISS, by working with the same kind of scenarios that were previously proposed in Refs.~\cite{Arganda:2015ija,Arganda:2014dta,Arganda:2015naa}. 
On the other hand, these particular ISS settings with suppressed ${\rm LFV}_{\mu e}$ rates provide very interesting scenarios for exploring the relevant ISS parameter space directions that may lead to large cLFV rates in the other sectors, $\tau$-$\mu$ and/or $\tau$-$e$.

 Motivated by all the peculiarities exposed above, in this work we perform a dedicated study of the LFVZD rates, in particular BR($Z\to\tau\mu$) and BR($Z\to\tau e$), in the context of these ISS scenarios with an {\it ad-hoc} suppression of ${\rm LFV}_{\mu e}$ rates, which will be called from now on ISS-$\cancel{\rm LFV}\hskip-.1cm_{\mu e}$ in short.
LFVZD processes in the presence of low-scale heavy neutrinos have recently been studied considering the full one-loop contributions~\cite{Abada:2014cca} or computing the relevant Wilson coefficients~\cite{Abada:2015zea}.
In these works, maximum allowed LFVZD rates in the reach of future linear colliders were found when considering a minimal ``3+1'' toy model, with BR($Z\to\tau\mu$) up to $\mathcal O(10^{-8})$ for a neutrino mass in the few TeV range.
For more realistic models, like the (2,3) or (3,3) realizations of the ISS model, and after imposing all the relevant theoretical and experimental  bounds, smaller LFVZD rates  were achieved, BR$(Z\to\tau\mu)\lesssim\mathcal O(10^{-9})$, which would be below the reach of future linear colliders sensitivities and might be accessible only at future circular $e^+e^-$ colliders.
The main difference of our study with the ones previously done relies on the different settings of the ISS parameters, as we will focus on some specific directions that are more difficult to access with a random scan of the ISS parameter space.
In the present paper, we will perform a complementary analysis to the one in Ref.~\cite{Abada:2014cca} and we will show that larger maximum allowed rates for BR($Z\to\tau\mu$) and BR($Z\to\tau e$) can be obtained by considering the particular ISS-$\cancel{\rm LFV}\hskip-.1cm_{\mu e}$ scenarios previously commented, such that for some specific directions of the parameter space they could be reached at future linear colliders.

This work is organized as follows. In Sec.~\ref{model} we briefly review the main features of the ISS model and
describe in detail our geometrical parametrization of the neutrino Yukawa matrix that allows us to find the scenarios with suppressed $\rm{LFV}_{\mu e}$ that we are interested in.
We first analyze in Sec.~\ref{Constraints} the behavior of the most relevant constraining observables to the LFVZD rates in terms of the relevant parameters in these peculiar scenarios. Then we devote Sec.~\ref{Results} to present our numerical results for the LFVZD rates in this ISS-$\cancel{\rm LFV}\hskip-.1cm_{\mu e}$ model and we find out the maximum allowed BR($Z\to\tau\mu$) and BR($Z\to\tau e$) rates being compatible with all the relevant experimental and theoretical constraints.
Finally, we conclude in Sec.~\ref{conclusions}.


\section{The ISS model with suppressed $\mu-e$ transitions}
\label{model}

We consider a realization of the ISS where three pairs of fermionic singlets, ($\nu_R,X$), with opposite LN are added to the SM particle content
and assume, as usual in this model, that  LN is almost conserved, slightly broken only by a small Majorana mass term for the $X$ singlets.
This small scale $\mu_X$ will be, precisely, the responsible for explaining the smallness of the light neutrino masses. 
Apart from this Majorana mass term,  the rest of the ISS Lagrangian terms conserve LN, and these are the
Yukawa interactions between the right- and  left-handed neutrinos, $\nu_R$ and $\nu_L$, and a mass term connecting the two fermionic singlets $\nu_R$ and $X$.
Therefore, we extend the SM Lagrangian with the following terms:
\begin{equation}
 \label{LagrangianISS}
 \mathcal{L}_\mathrm{ISS} = - Y^{ij}_\nu \overline{L_{i}} \widetilde{\Phi} \nu_{Rj} - M_R^{ij} \overline{\nu_{Ri}^C} X_j - \frac{1}{2} \mu_{X}^{ij} \overline{X_{i}^C} X_{j} + h.c.\,,
\end{equation}
where, the indices $i,j$ run from 1 to 3, $L$ is the SM lepton doublet, $ \widetilde{\Phi}=i \sigma_2 \Phi^*$ with $\Phi$ the SM Higgs doublet, $Y_\nu$ is the 3$\times$3 neutrino Yukawa coupling matrix, $M_R$ is a LN conserving complex 3$\times$3 mass matrix, and $\mu_X$ is
a Majorana complex 3$\times$3 symmetric mass matrix that violates LN by two units.
It is worth mentioning that it is possible to add in Eq.~(\ref{LagrangianISS}) another LN violating Majorana mass term for the $\nu_R$ fields, i.e. $\mu_R^{ij} \overline{\nu_{Ri}^C} \nu_{R_j}$ with $\mu_R$ a $3\times3$ symmetric matrix. Assuming this $\mu_R$ Majorana scale to be small, the new term will respect the approximated LN symmetry required by the ISS. 
Nevertheless, $\mu_R$ does not generate light neutrino masses at tree level, whereas the Majorana mass $\mu_X$ does. The effects of this Majorana mass term for $\nu_R$ appear only at one-loop level in the light neutrino masses. 
Furthermore, the effects of the tiny Majorana mass terms for both $\nu_R$ and $X$ fields are negligible for the LFV Z decays studied in this work, which are governed by the Yukawa couplings. 
Therefore, we will set $\mu_R$ to zero for the rest of this work and consider a small $\mu_X$ as the only Lepton Number violating parameter leading to the light neutrino masses. 

After the electroweak symmetry breaking has taken place, the following 9$\times$9 neutrino mass matrix is obtained in the $(\nu_L^c, \nu_R, X)$ basis,
\begin{equation}
\label{ISSmatrix}
 M_{\mathrm{ISS}}=\left(\begin{array}{c c c} 0 & m_D & 0 \\ m_D^T & 0 & M_R \\ 0 & M_R^T & \mu_X \end{array}\right)\,,
\end{equation}
where the Dirac mass matrix is defined as $m_D=v Y_\nu$, with $v=\langle \phi \rangle=174$~GeV.
This 9$\times$9 symmetric mass matrix can be diagonalized by a unitary matrix $U^\nu$, leading to 9 physical Majorana states $n_j\, (j=1,\dots,9)$ with masses given by,
\begin{equation}
\label{ISSmatrixRotation}
U^{\nu^T}M_{\rm ISS}\, U^\nu = {\rm diag}(m_{n_1},\dots,m_{n_9}).
\end{equation}
For completeness, we summarize the relevant neutrino interactions for the observables studied here, {\it i.e}, the ones to the $W$ and $Z$ gauge bosons, to the Higgs boson $H$ and to the Goldstone bosons $G$. In the neutrino mass basis, they are given by the following terms in the Lagrangian:
\begin{align}
\mathcal L_W&=-\dfrac g{\sqrt{2}} \sum_{i=1}^{3}\sum_{j=1}^{9} W^-_\mu \bar\ell_i B_{\ell_i n_j} \gamma^\mu P_L n_j + h.c., \\
\mathcal L_Z &= -\dfrac g{4c_W} \sum_{i,j=1}^{9}Z_\mu\, \bar n_i \gamma^\mu \Big[C_{n_i n_j} P_L - C_{n_in_j}^* P_R\Big]n_j ,\\
\mathcal L_H &=-\dfrac g{2m_W}  \sum_{i,j=1}^{9} H\,\bar n_i C_{n_in_j}\Big[m_{n_i}P_L+m_{n_j} P_R\Big]n_j,\\
\mathcal{L}_{G^{\pm}} &= -\frac{g}{\sqrt{2} m_W}\sum_{i=1}^{3}\sum_{j=1}^{9} G^{-}\bar{\ell_i} B_{\ell_i n_j} \Big[m_{\ell_i} P_L - m_{n_j} P_R \Big]n_j  + h.c\,,\\ 
\mathcal{L}_{G^{0}} & =-\dfrac {ig}{2 m_W} \sum_{i,j=1}^{9}G^0\, \bar n_i  C_{n_in_j} \Big[m_{n_i}  P_L - m_{n_j} P_R\Big]n_j ,  
\end{align}
where, 
\begin{align}
\label{eq:BCmatrices}
B_{\ell_in_j}&=U_{ij}^{\nu*}, \\
C_{n_in_j}&=\sum_{k=1}^{3} U^\nu_{ki} U^{\nu*}_{kj},
\end{align}
and $P_{R,L}=(1\pm \gamma^5)/2$  are the usual chirality projectors. 
The charged leptons are assumed to be in the physical basis in all this work. 

As mentioned above, we assume that $\mu_X$ is much smaller than the other masses $m_D$ and $M_R$. In this situation, six of the physical states, $n_{4,..9}$, are heavy Majorana neutrinos, named here as $N_{1,...6}$, which can be grouped into three pseudo-Dirac pairs with nearly degenerate masses within each pair. 
The other three states, $n_{1,2,3}$, are light Majorana fermions with masses proportional to $\mu_X$, which are therefore identified as the light neutrinos, ${\nu_{\,1,2,3}}$,  measured in the oscillation experiments. The small differences among  the two quasi-degenerate heavy neutrino masses in each pair are also governed by the mass parameter $\mu_X$. 
 The corresponding pairings of these heavy neutrinos  are denoted here as $N_{1/2}$, $N_{3/4}$ and $N_{5/6}$, with 
$m_{N_{1/2}}\leq m_{N_{2/3}}\leq m_{N_{5/6}}$. Generically, the behavior with $\mu_X$, $m_D$ and $M_R$ of the predicted light and heavy neutrino masses in this ISS model with nine Majorana neutrinos follow a similar pattern as in the one generation case, where
the light $\nu$ and the two heavy neutrinos $N_{a/b}$ in the unique pseudo-Dirac pair have masses, for $\mu_X\ll m_D,M_R$, given by: 
\begin{align}
 m_\nu &= \frac{m_{D}^2}{m_{D}^2+M_{R}^2} \mu_X\,\label{mnu},\\
 m_{N_{a/b}}  &= \pm \sqrt{M_{R}^2+m_{D}^2} + \frac{M_{R}^2 \mu_X}{2 (m_{D}^2+M_{R}^2)}~.\label{mN}
\end{align}
Furthermore, assuming the mass hierarchy $\mu_X\ll m_D\ll M_R$, 
the masses of the three heavy pairs are dominated by $M_{R}$, and the 
low energy neutrino data can easily be  accommodated by using the $\mu_X$-parametrization introduced in Ref.~\cite{Arganda:2014dta}:
\begin{equation}
\label{MUXparametrization}
\mu_X=M_R^T m_D^{-1} U_{\rm PMNS}^{*\phantom{\dagger}} \, m_\nu  U_{\rm PMNS}^\dagger\,  m_D^{T^{-1}} M_R,
\end{equation}
where $m_\nu={\rm diag}(m_{\nu_1},m_{\nu_2},m_{\nu_3})$ are the masses of the three lightest neutrinos and $U_{\rm PMNS}$ is the Pontecorvo-Maki-Nakagawa-Sakata (PMNS) unitary matrix \cite{Pontecorvo:1957cp,Maki:1962mu}.  
The main advantage of using this parametrization is that it allows us to consider the heavy neutrino mass matrix $M_R$ and the Yukawa coupling matrix as our input parameters for the ISS model.
For the numerical estimates, unless otherwise specified, we will consider a normal mass ordering for the light neutrinos with the lightest neutrino mass fixed at $m_{\nu_1}=0.01$~eV, and the rest of the masses and mixing angles set to their central values of the global fit \cite{Gonzalez-Garcia:2014bfa}:
\begin{align}
\sin^2\theta_{12}&=0.308^{+0.013}_{-0.012}\,, &
\Delta m^2_{21}&=7.49^{+0.19}_{-0.17}\times10^{-5}{\rm eV}^2\,, \nonumber\\
\sin^2\theta_{23}&=0.574^{+0.026}_{-0.144}\,, &
\Delta m^2_{31}&=2.484^{+0.045}_{-0.048}\times10^{-3}{\rm eV}^2\,, \nonumber\\
\sin^2\theta_{13}&=0.0217^{+0.0013}_{-0.0010} \,.
\end{align}

For the rest of this paper, we work in the basis where $M_R$ is diagonal and assume the simplest case where all its entries are degenerate and real, {\it i.e}, $M_{R_{1,2,3}}\equiv M_R$. 
In order to avoid potential constraints from lepton electric dipole moments (EDM), we also consider only real values for the $Y_\nu$ matrix, as well as for the PMNS matrix. 
In this situation, the one-loop induced cLFV processes are driven by powers of the combination $Y_\nu Y_\nu^T$, instead of $Y_\nu Y_\nu^\dagger$, and it turns out to be useful and instructive to apply the geometrical interpretation discussed in \cite{Arganda:2014dta}, where the 9 entries of the Yukawa matrix are interpreted in terms of the components of three generic neutrino vectors in flavor space ($\ne,\nm,\nt$),
\begin{equation}
Y_\nu=\left(\begin{array}{ccc}
 Y_{\nu_{11}} & Y_{\nu_{12}} & Y_{\nu_{13}}\\
 Y_{\nu_{21}} & Y_{\nu_{22}} & Y_{\nu_{23}}\\
 Y_{\nu_{31}} & Y_{\nu_{32}} & Y_{\nu_{33}}\end{array}\right)\equiv f 
 \left(\begin{array}{c} \ne \\ \nm \\ \nt \end{array}\right),
\end{equation}
which for the  relevant combination in cLFV  processes give:
 \begin{equation}
 \label{Yneusq}
Y_\nu Y_\nu^T= f^2 \left(\begin{array}{ccc}
 \mode^2 & \ne\cdot\nm   & \ne\cdot\nt \\
 \ne\cdot\nm & \modmu^2 &  \nm\cdot\nt\\
 \ne\cdot\nt &   \nm\cdot\nt  & \modtau^2\end{array}\right).
 \end{equation}
This means that the input parameters that determine the $Y_\nu$ matrix can be seen as  the 3 modulus of these three vectors ($\mode,\modmu,\modtau$), the 3 relative {\it flavor} angles between them ($\tme,\tte,\ttm$), with $\theta_{ij}\equiv \widehat{\,\boldsymbol{n}_i \boldsymbol{n}_j}$, and 3 extra angles ($\theta_1,\theta_2,\theta_3$) that parametrize a global rotation $\rotationO$ of these 3 vectors that does not change their relative angles. In addition, we have introduced the parameter $f$ that characterizes the global Yukawa coupling strength.
Since the combination $Y_\nu Y_\nu^T/f^2$ is symmetric, it only depends on 6 parameters that we take to be the 3 modulus ($\mode,\modmu,\modtau$) and the cosine of the three flavor angles ($\cme,\cte,\ctm$), with $c_{ij}\equiv\cos\theta_{ij}$.
The names of the angles are motivated by the fact that the cosine of the angle $\theta_{ij}$ controls the LFV transitions in the $\ell_i$-$\ell_j$ sector, which we write in short as ${\rm LFV}_{\ell_i \ell_j}$. It is interesting to notice that 
the global rotation $\rotationO$ does not enter in the $Y_\nu Y_\nu^T$ combination and, therefore, it will not affect any of the cLFV  processes studied in this work. 

As mentioned in the introduction, experimental searches in table \ref{LFVsearch} indicate that the existence of LFV in the $\mu$-$e$ sector is by far much more constrained that in the other $\tau$-$\mu$ and $\tau$-$e$ sectors. Therefore, and leaving apart the issue of which could be the theoretical origin of these remarkable differences among the transitions of the various sectors, it may suggest indeed a realistic absence of LFV transitions in the $\mu$-$e$ sector in Nature, motivating the study of models that incorporate these peculiarities automatically in their input settings. In particular, the $\mu$-$e$ suppression can be easily realized with our geometrical interpretation by just assuming that $\ne$ and $\nm$ are orthogonal vectors, i.e, $\cme=0$.  
Such a condition defines a family of ISS scenarios that can be parametrized using the following Yukawa matrix:
\begin{equation}\label{YukawaAmatrix}
Y_\nu=A\cdot \rotationO \quad{\rm with}\quad 
A\equiv f \left(\begin{array}{ccc} \mode & 0 & 0 \\ 0 & \modmu & 0 \\ \modtau \cte & \modtau \ctm & \modtau\sqrt{1-\cte^2-\ctm^2}\end{array}\right),
\end{equation}
where $\rotationO$ is the above commented orthogonal rotation matrix, which does not enter in the product $Y_\nu Y_\nu^T$, and we have factorized out again the parameter $f$ that controls the global strength of the Yukawa coupling matrix. 
Notice that the $Y_\nu$ matrix in Eq.~(\ref{YukawaAmatrix}) is the most general one that satisfies the condition $\cme=0$. 

\begin{figure}[t!]
\begin{center}
\includegraphics[width=.49\textwidth]{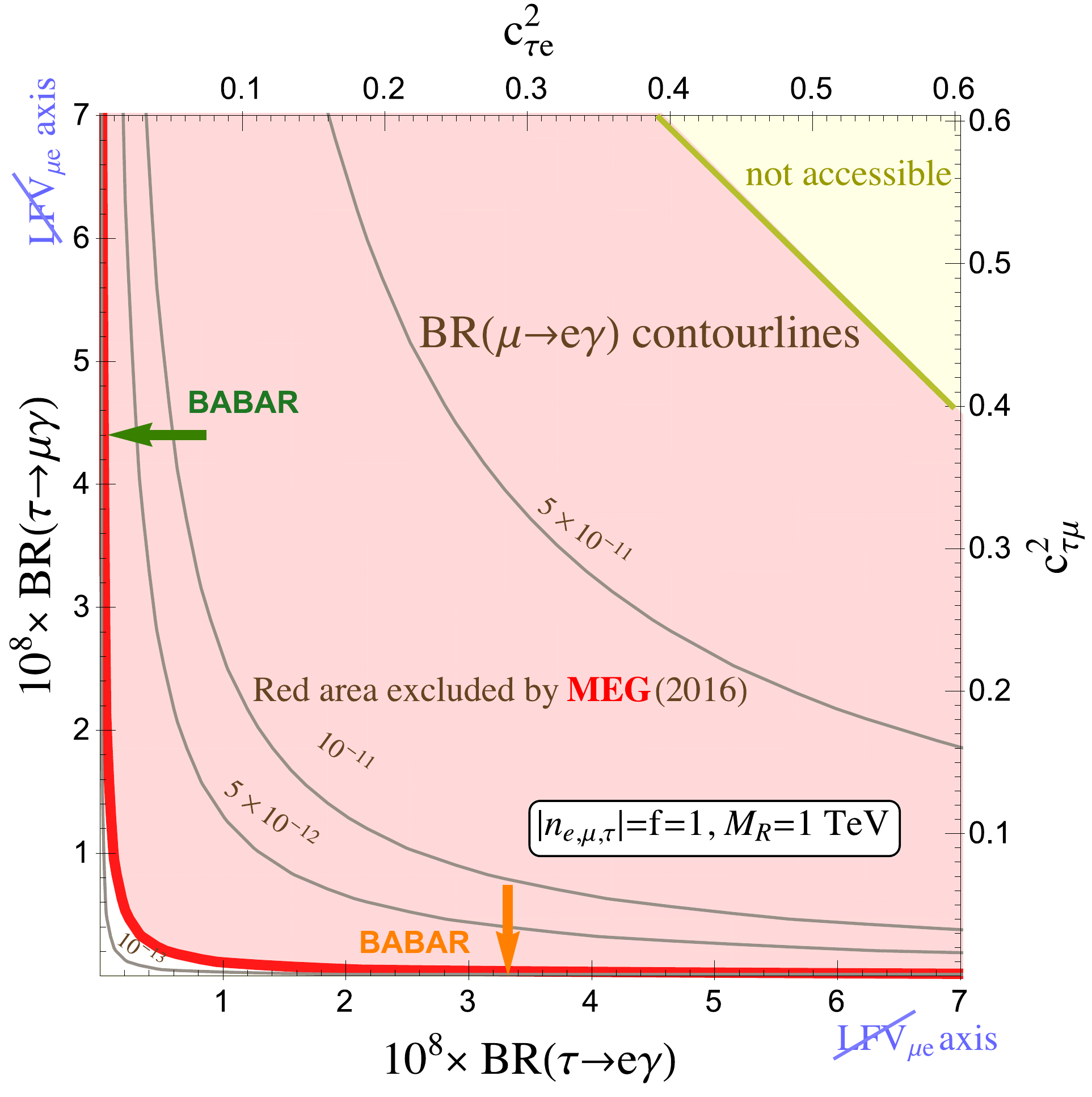}
\includegraphics[width=.5\textwidth]{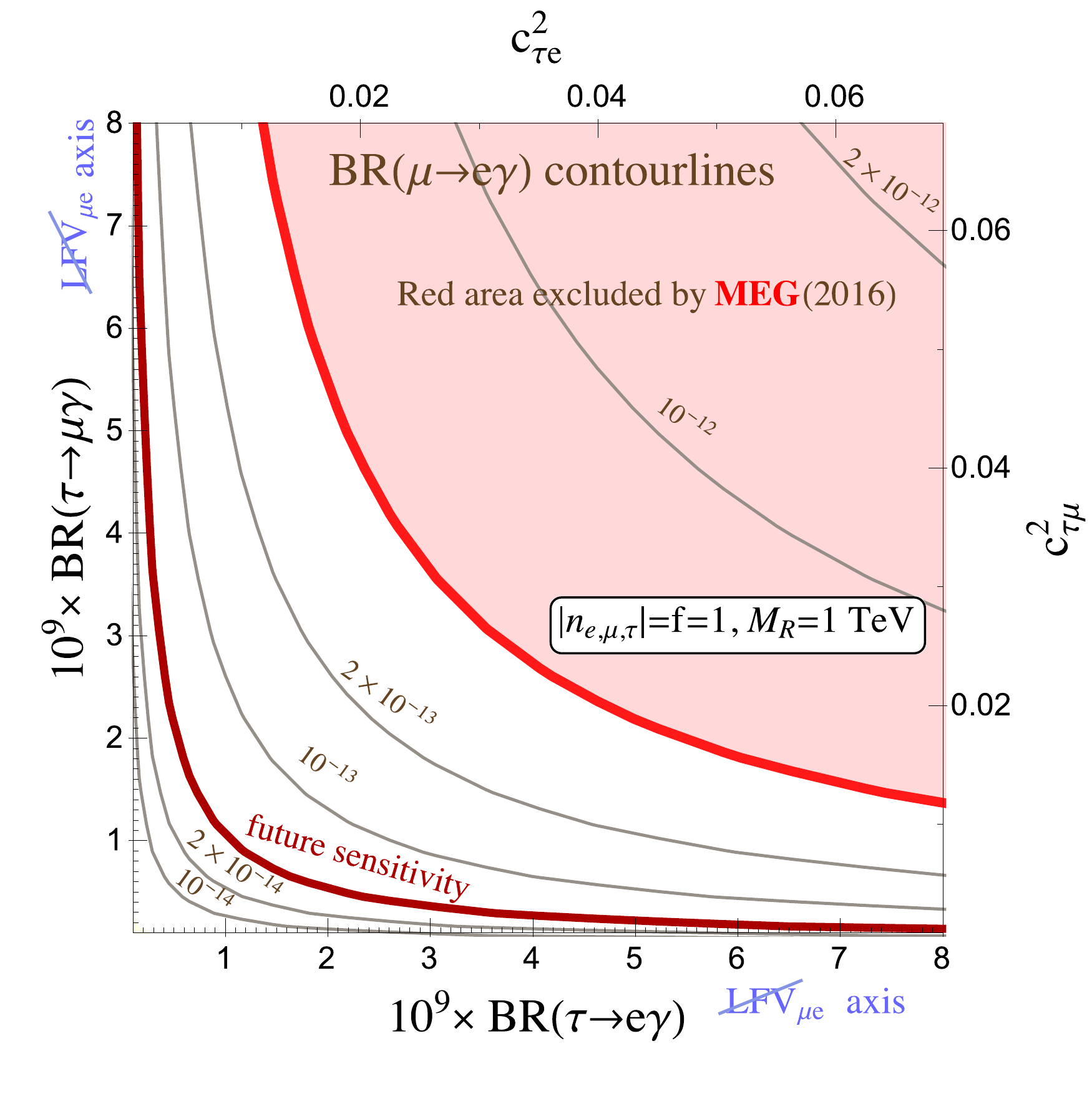}
\caption{
Left panel: Contour lines for BR($\mu\to e\gamma$) as a function of BR($\tau\to e\gamma$) and BR($\tau\to \mu\gamma$) rates, for fixed $M_R=1$~TeV, $|\boldsymbol{n}_{e,\mu,\tau}|=f=1$ values and varying $c_{\tau e}^2$ and $c_{\tau\mu}^2$ from 0 to $0.6$, as shown in the right and top axes. The yellow area represents the region that cannot be accessed with real Yukawa matrices. The red area is excluded by the upper bound on $\mu\to e\gamma$ of BR($4.2\times10^{-13}$) from MEG~\cite{TheMEG:2016wtm}, while the orange (green) arrow marks the present upper bound  BR$(\tau\to e\gamma)<3.3 \times 10^{-8}$ (BR$(\tau\to\mu\gamma)<4.4 \times 10^{-8}$) from Babar~\cite{Aubert:2009ag}. Right panel: Zoom on the lower left corner of the plot in the left panel which allows for a better reading of the region allowed by present experimental data. The extra darker red line represents the future expected sensitivity of $4\times10^{-14}$ by MEG-II~\cite{Baldini:2013ke}.
} \label{radiativedecays}
\end{center}
\end{figure}

In order to better understand the implications on cLFV  phenomenology of these ISS scenarios, we first explore in this section the LFV radiative decays that, as we have said, are one of the most constrained cLFV observables. 
All the numerical estimates and plots in this work are made using the full one-loop formulas of the radiative decays in the neutrino mass basis \cite{Ilakovac:1994kj}, which are provided in Appendix~\ref{app:LFVdecays} for completeness. Nevertheless, for the purpose of the following discussion, it proves convenient to comment first on the dependence of the relevant parameters by using the following approximate expression - which is very simple and has been proven to work quite well in the present ISS context \cite{Arganda:2014dta}, as long as $v Y_\nu \ll M_R$: 
\begin{equation}\label{RadiativeApprox}
{\rm BR}(\ell_m\to \ell_k\gamma)\approx \frac{\alpha_W^3 s_W^2}{1024\pi^2 m_W^4}\frac{m_{\ell_m}^5}{\Gamma_{\ell_m}} \frac{v^4}{M_R^4} \Big|\big(Y_\nu Y_\nu^T\big)_{km} \Big|^2.
\end{equation}
From this equation, we can easily see that the LFV radiative decays of the $\tau$ lepton depend on the most relevant parameters, $f$, $M_R$ and $c_{\tau\ell}$ as follows:
\begin{equation}\label{RadiativeApproxTAU}
{\rm BR}(\tau\to \ell\gamma)\sim \frac{f^4}{M_R^4}~c_{\tau\ell}^2\quad{\rm with }~\ell=e,\mu.
\end{equation}
The case of $\mu\to e\gamma$ is different, since the assumption $\cme=0$ cancels the leading order contribution given by the approximate formula in Eq.~(\ref{RadiativeApprox}), so the first relevant contribution in this observable is of higher order in the expansion series in powers of the Yukawa coupling. Specifically it is of the type ($Y_\nu Y_\nu^T Y_\nu Y_\nu^T$). 
Consequently, it is suppressed with respect to Eq.~(\ref{RadiativeApproxTAU}) and the predicted rates for this relevant contribution turn out to depend on the product of both $\cte$ and $\ctm$:
\begin{equation}\label{RadiativeApproxMUEG}
{\rm BR}(\mu\to e\gamma)\sim \frac{f^8}{M_R^8}~\cte^2 \ctm^2.
\end{equation}
Therefore, to get a non zero value of this BR$(\mu \to e \gamma)$ one needs the other two parameters $\cte$ and $\ctm$, triggering the respective radiative decays $\tau \to \mu \gamma$ and $\tau \to e \gamma$, to be non-vanishing simultaneously.  

This generic suppression of the BR$(\mu \to e \gamma)$ rates is also illustrated numerically in Figure~\ref{radiativedecays}. This plot shows the full one-loop numerical results for BR($\mu\to e\gamma$), computed with the complete formulas in Appendix~\ref{app:LFVdecays},  and displays the above commented correlations with the corresponding full one-loop predictions of BR($\tau \to \mu \gamma$) and BR($\tau \to e \gamma$) 
via the parameters $\ctm$ and $\cte$, respectively. The contour lines for BR($\mu\to e\gamma$) are obtained by varying $\ctm^2$ and $\cte^2$ within the interval $(0,0.6)$, which in turn provide predictions for BR($\tau \to \mu \gamma$) and BR($\tau \to e \gamma$) that are represented in the vertical and horizontal axes respectively. This is for the simple case with $|\boldsymbol{n}_{e,\mu,\tau}|=f=1$, $M_R=1$~TeV and $\rotationO=\mathbb I$ (remember that the result is independent on the choice of $\rotationO$), but similar qualitative conclusions  can be obtained for other choices of these parameters. Notice that since we are assuming a real non-singular Yukawa matrix, Eq.~(\ref{YukawaAmatrix}) imposes the condition $\cte^2+\ctm^2<1$, making the yellow area, where $\cte^2+\ctm^2\geq1$, not accessible in our analysis. We also find that the rates for $\tau\to \mu\gamma$ ($\tau\to e\gamma$) can in general be large, for the values of the parameters selected in this plot, of the order of the present upper bound from BaBar~\cite{Aubert:2009ag}, marked here with a green (orange) arrow, and that they depend just on $\ctm^2$ ($\cte^2$), in agreement with the approximate expression in Eq.~(\ref{RadiativeApproxTAU}).
We also learn that the predictions for BR($\mu\to e\gamma$) are 3-4 orders of magnitude smaller than the $\tau$ radiative decay rates, as expected from Eq.~(\ref{RadiativeApproxMUEG}), but they are still above the upper bound from the MEG experiment for most of the parameter space. 
In fact, the MEG bound excludes everything but the area close to the axes. 
Eventually, the BR($\mu\to e\gamma$) goes asymptotically to zero when approaching the axes.
When lying just on top of these axes, the predictions for BR($\mu\to e\gamma$) completely vanish  (see Eq.~(\ref{RadiativeApproxMUEG})), implying that BR($\tau\to e \gamma$) must be small in order to allow for large BR($\tau\to\mu\gamma$),  and viceversa. 

\begin{table}[t!]
\begin{center}

\begin{tabular}{|c|c|c|c|c|l|}
\hline
 Scenario Name& $\ctm$ & $\mode$ & $\modmu$ & $\modtau$ & \hskip1.5cm   Example 	\\
\hline
TM-1 & $1/\sqrt2$ & 1 & 1 & 1 & $Y_\nu=f \left(\begin{array}{ccc}1 & 0 & 0\\ 0&1&0\\0&\sfrac{1}{\sqrt2} &\sfrac{1}{\sqrt2}\end{array}\right)$\\
TM-2 & $1$ & 1 & 1 & 1 & $Y_\nu\simeq f\left(\begin{array}{ccc}1 & 0 & 0\\ 0&1&0\\0&1 &0\end{array}\right)$\\
TM-3 & $1/\sqrt2$ & 0.1 & 1 & 1 & $Y_\nu=f\left(\begin{array}{ccc}0.1 & 0 & 0\\ 0&1&0\\0&\sfrac{1}{\sqrt2} &\sfrac{1}{\sqrt2}\end{array}\right)$\\
TM-4 & $1$ & 0.1 & 1 & 1 & $Y_\nu\simeq f\left(\begin{array}{ccc}0.1 & 0 & 0\\ 0&1&0\\0&1 &0\end{array}\right)$\\
TM-5 & $1$ & $\sqrt2 $& $1.7$ &$ \sqrt3 $& $Y_\nu=f \left(\begin{array}{ccc} 0&1&-1 \\ 0.9&1&1 \\ 1& 1& 1 \end{array}\right)~(  Y_{\tau\mu}^{(1)}$ in \cite{Arganda:2014dta})\\
TM-6 & $1/3$ & $\sqrt2 $& $\sqrt3$ & $\sqrt3 $& $Y_\nu=f\left(\begin{array}{ccc}0&1&1 \\ 1&1&-1 \\ -1& 1& -1  \end{array}\right)~(  Y_{\tau\mu}^{(2)}$ in \cite{Arganda:2014dta})\\
TM-7 & $0.1$ & $\sqrt2 $& $\sqrt3 $& $1.1 $& $Y_\nu=f\left(\begin{array}{ccc}0&-1&1 \\ -1&1&1 \\ 0.8& 0.5& 0.5  \end{array}\right)~(  Y_{\tau\mu}^{(3)}$ in \cite{Arganda:2014dta})\\
TM-8 & $1$ & $1/2$ & $1/3$ & $1/4$ & $Y_\nu\simeq f\left(\begin{array}{ccc}1 & 0 & 0\\ 0&0.5&0\\0&0.08 &0.32\end{array}\right)$\\
 \hline
\end{tabular}
\caption{Examples of TM scenarios with $\tau-\mu$ transitions that we consider in the numerical estimates. The cases
where $Y_\nu$ is defined with '$\simeq$' instead of '$=$' means that $\ctm=0.99$ instead of 1 in order to have a non-singular  $Y_\nu$ matrix, as required by  Eq.~(\ref{MUXparametrization}).
Three scenarios,  TM-5, TM-6 and TM-7, were previously introduced in Ref.~\cite{Arganda:2014dta} under the names of $Y_{\tau\mu}^{(1)}$, $Y_{\tau\mu}^{(2)}$ and $Y_{\tau\mu}^{(3)}$, respectively.  Equivalent scenarios for the TE class are easily obtained by exchanging $\mu$ and $e$ in these TM ones. }\label{TMscenarios}
\end{center}
\end{table}

Therefore, we focus our analysis on these two directions in the ISS parameter space with suppressed $\mu$-$e$ transitions, identified with the two axes in Figure~\ref{radiativedecays}, that define  our  \modelname ~model.
We then consider two classes of scenarios: the TM scenarios along the ${\rm LFV}_{\tau\mu}$ axis ($\cte=0$) that may give sizable rates for $\tau$-$\mu$ transitions, but always give negligible contributions to ${\rm LFV}_{\mu e}$ and ${\rm LFV}_{\tau e}$; and the TE scenarios along  the ${\rm LFV}_{\tau e}$ axis ($\ctm=0$) that may lead to large rates only for the $\tau$-$e$ transitions.
In table \ref{TMscenarios} we list the specific examples that we will use for the numerical estimates of our selected TM scenarios, where we have also included, for comparison, the three particular scenarios that were previously introduced in Ref.~\cite{Arganda:2014dta}.
Equivalent examples for the TE scenarios are obtained by exchanging $\mu$ and $e$ everywhere in the previous TM scenarios. 

\begin{figure}[b]
\begin{center}
\includegraphics[width=0.925\textwidth]{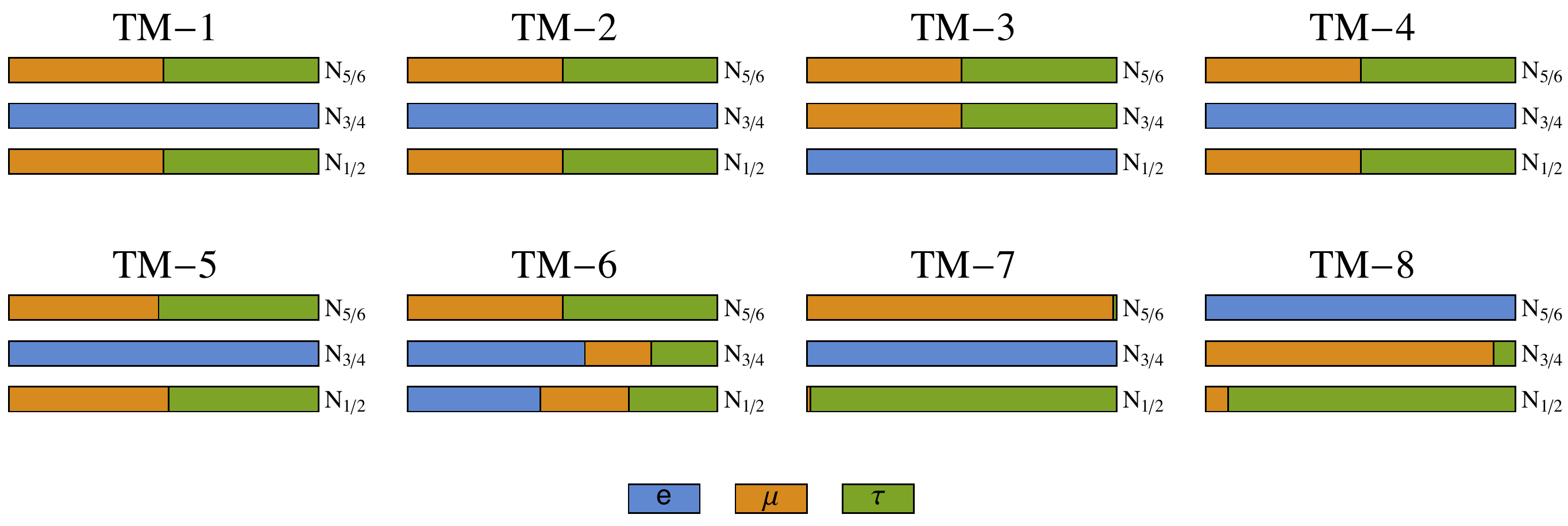}
\end{center}
\caption{Heavy neutrino flavor mixings,  as defined in  Eq.~(\ref{mixeq}), within the ISS scenarios of Table~\ref{TMscenarios}.}\label{mixingbars}
\end{figure}

Finally, to complete the description of this family of scenarios, we remark that they are built mostly for suppressing the loop generated ${\rm LFV}_{\mu e}$ processes, therefore each of them could give very different results for other tree level observables, including those preserving flavor. 
One example of the latter is the heavy neutrino production in association with a charged lepton of a specific flavor, which can be used to define the flavor of the heavy neutrinos~\cite{Arganda:2015ija}.
In Figure~\ref{mixingbars} we show the predicted flavor pattern of the six heavy neutrinos, $N_i$ $(i=1..6)$ grouped in pairs, for our eight selected scenarios TM-1, through TM-8. The length of the colored bars is calculated as 
\begin{equation}
S_{\ell N_i}= \dfrac{\displaystyle{|B_{\ell N_i}|^2}}{\displaystyle{\sum_{\ell=e,\mu,\tau} |B_{\ell N_i}|^2}} \,,
\label{mixeq}
\end{equation}
and, therefore, represents the relative mixing of the heavy neutrino $N_i$ with a given flavor $\ell$.
We learn from Figure~\ref{mixingbars} that, although all these TM scenarios share the property of suppressing the LFV $\mu$-$e$ and $\tau$-$e$ rates while maximizing the $\tau$-$\mu$ ones,  the heavy neutrino flavor mixing pattern is different in each scenario.  A common feature in almost all scenarios but TM-6, is that there is always one heavy pseudo-Dirac neutrino pair that is dominantly $e$-flavored (the bar filled in blue). 

 \begin{figure}[t!]
\begin{center}
\includegraphics[width=.48\textwidth]{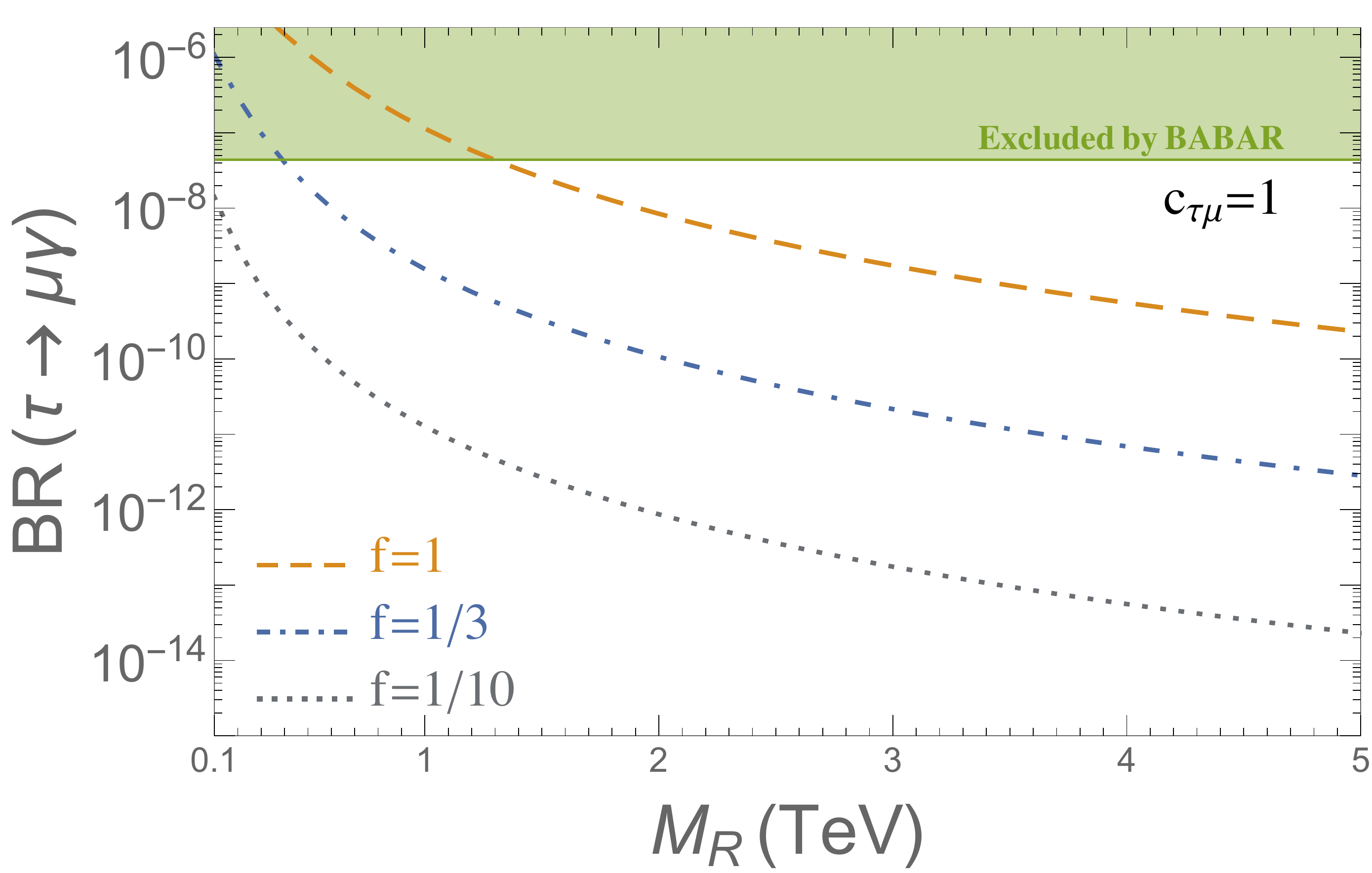}\,\,\,
\includegraphics[width=.48\textwidth]{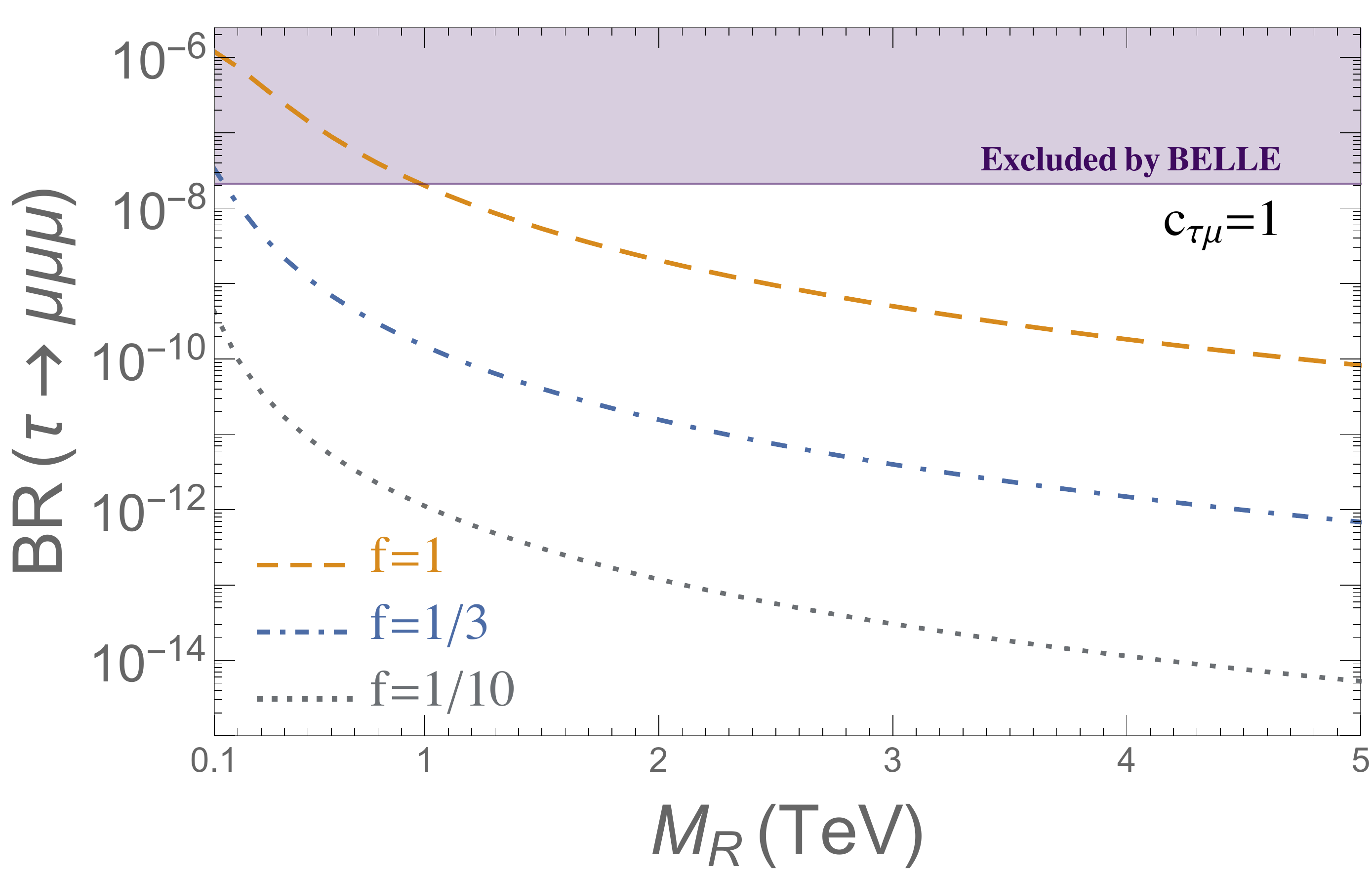}
\includegraphics[width=.48\textwidth]{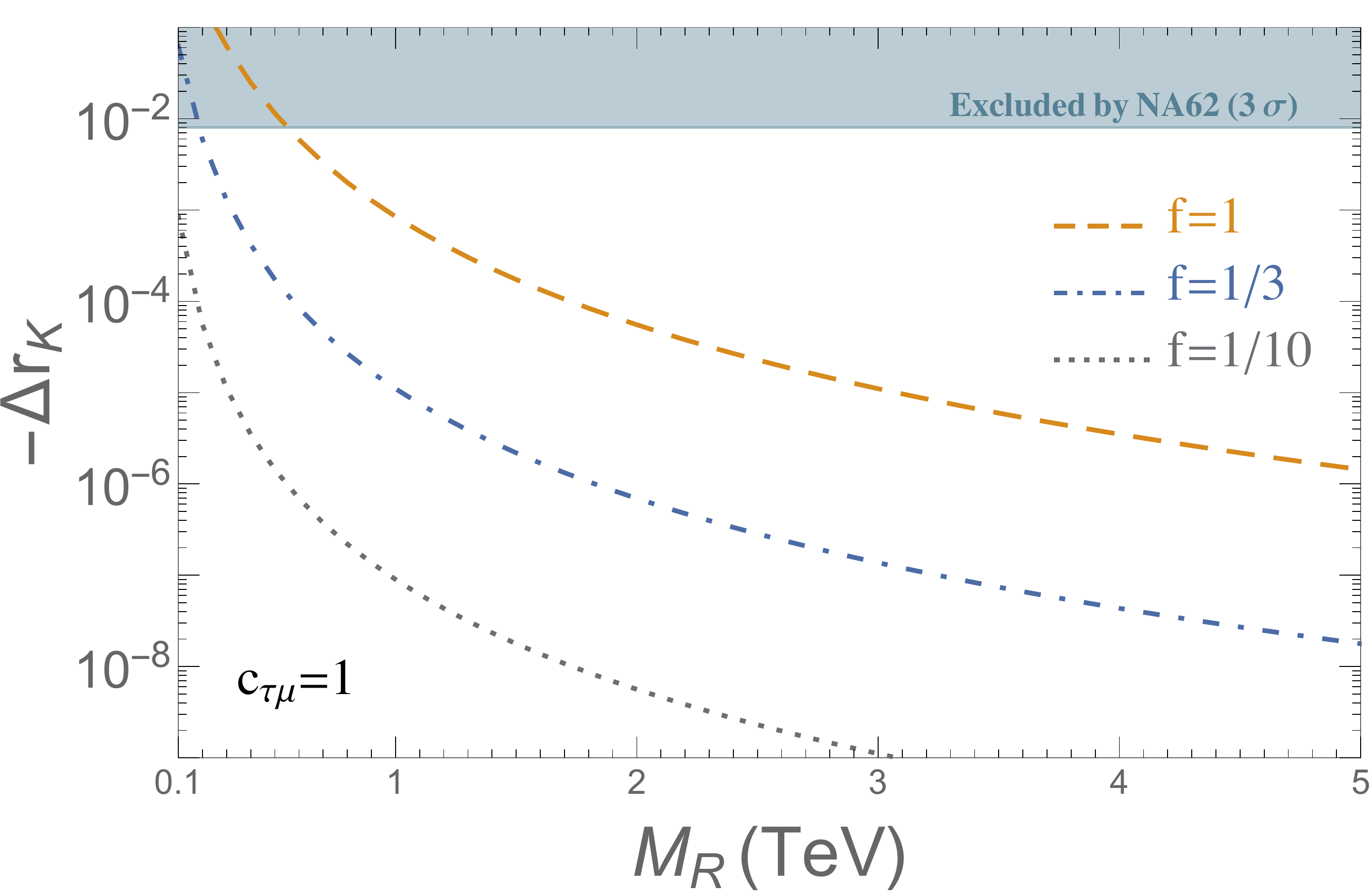}\,\,
\includegraphics[width=.48\textwidth]{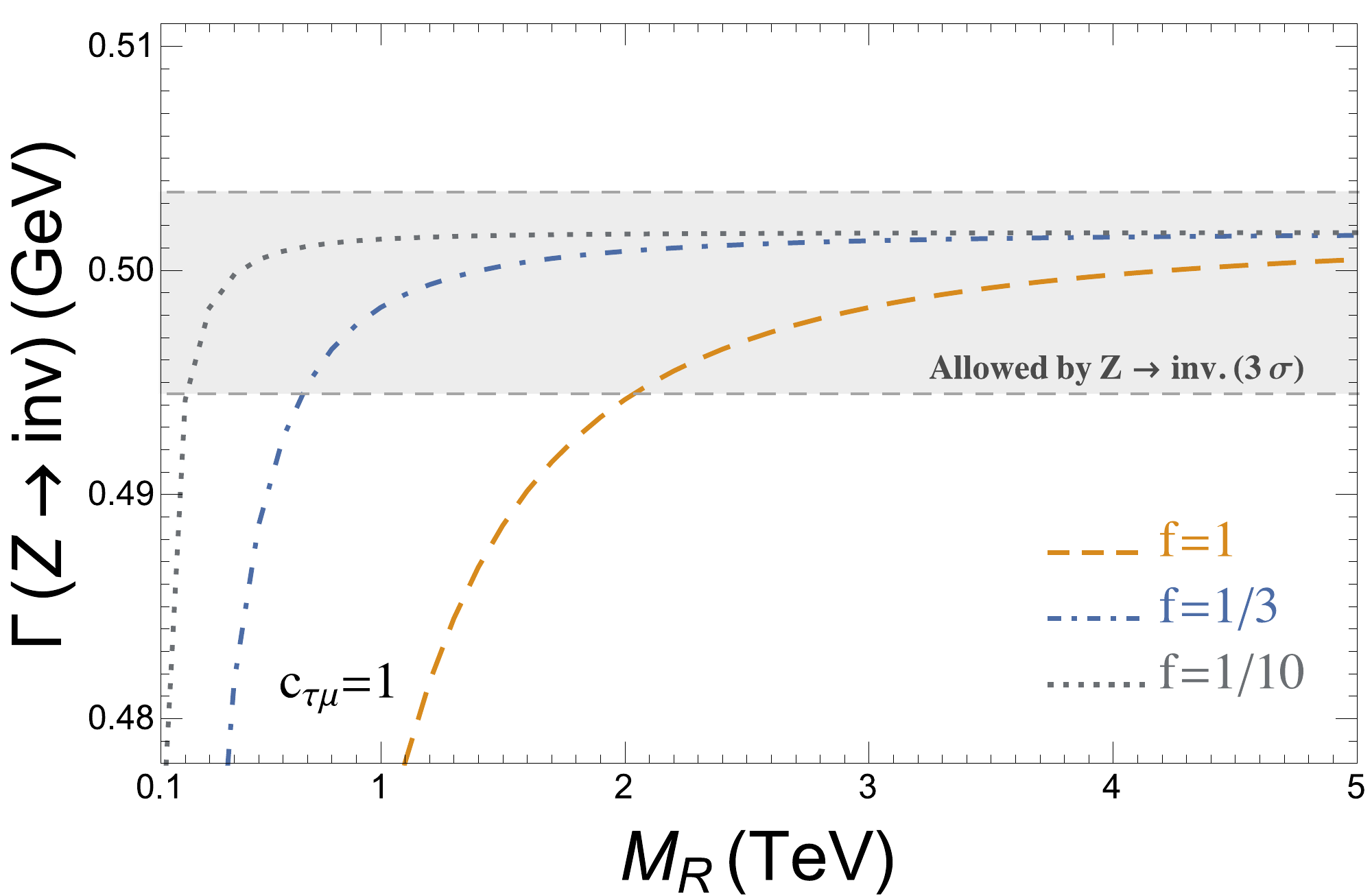}
\includegraphics[width=.48\textwidth]{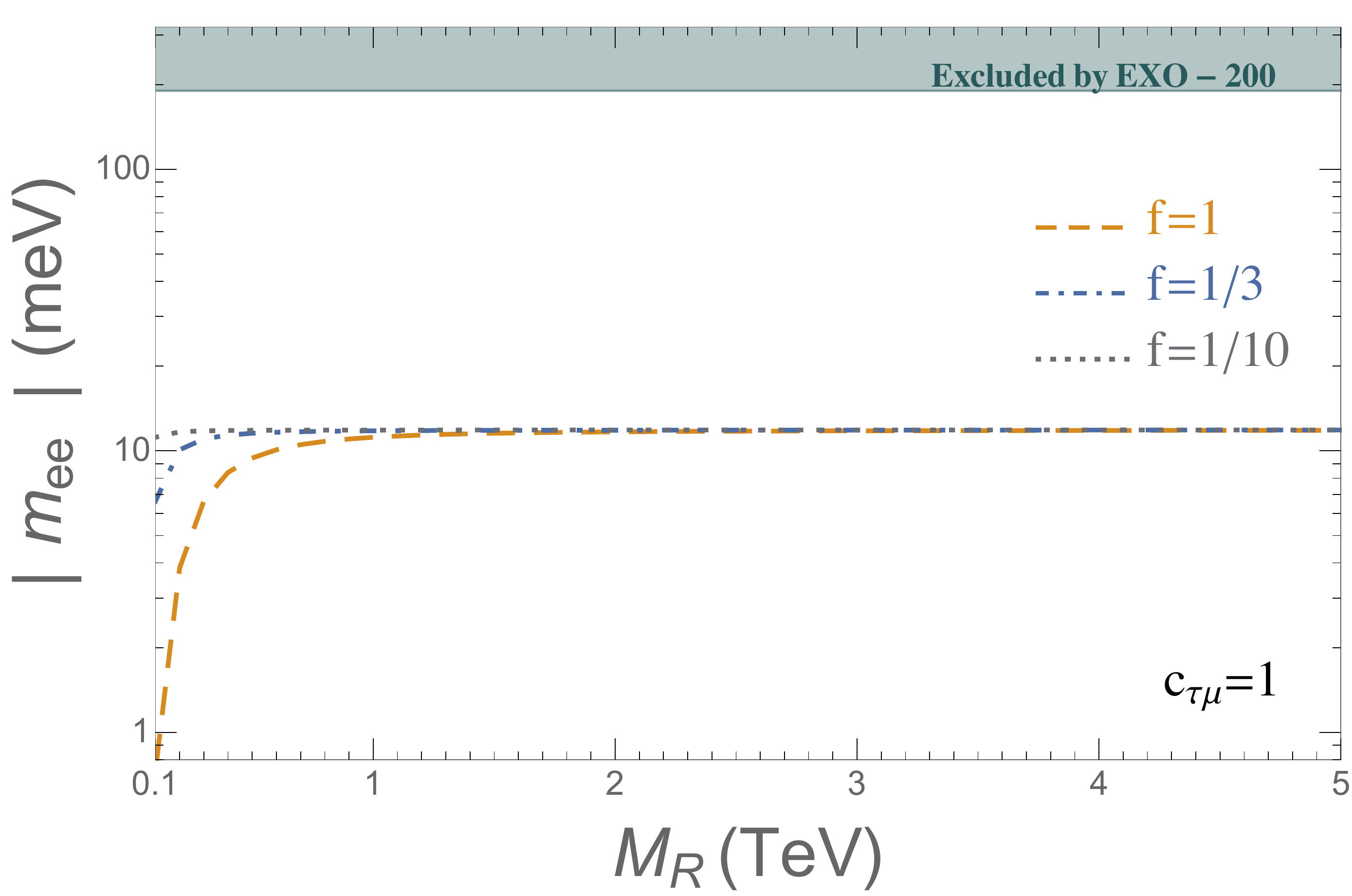}\,\,\,\,\,\,\,\,
\includegraphics[width=.465\textwidth]{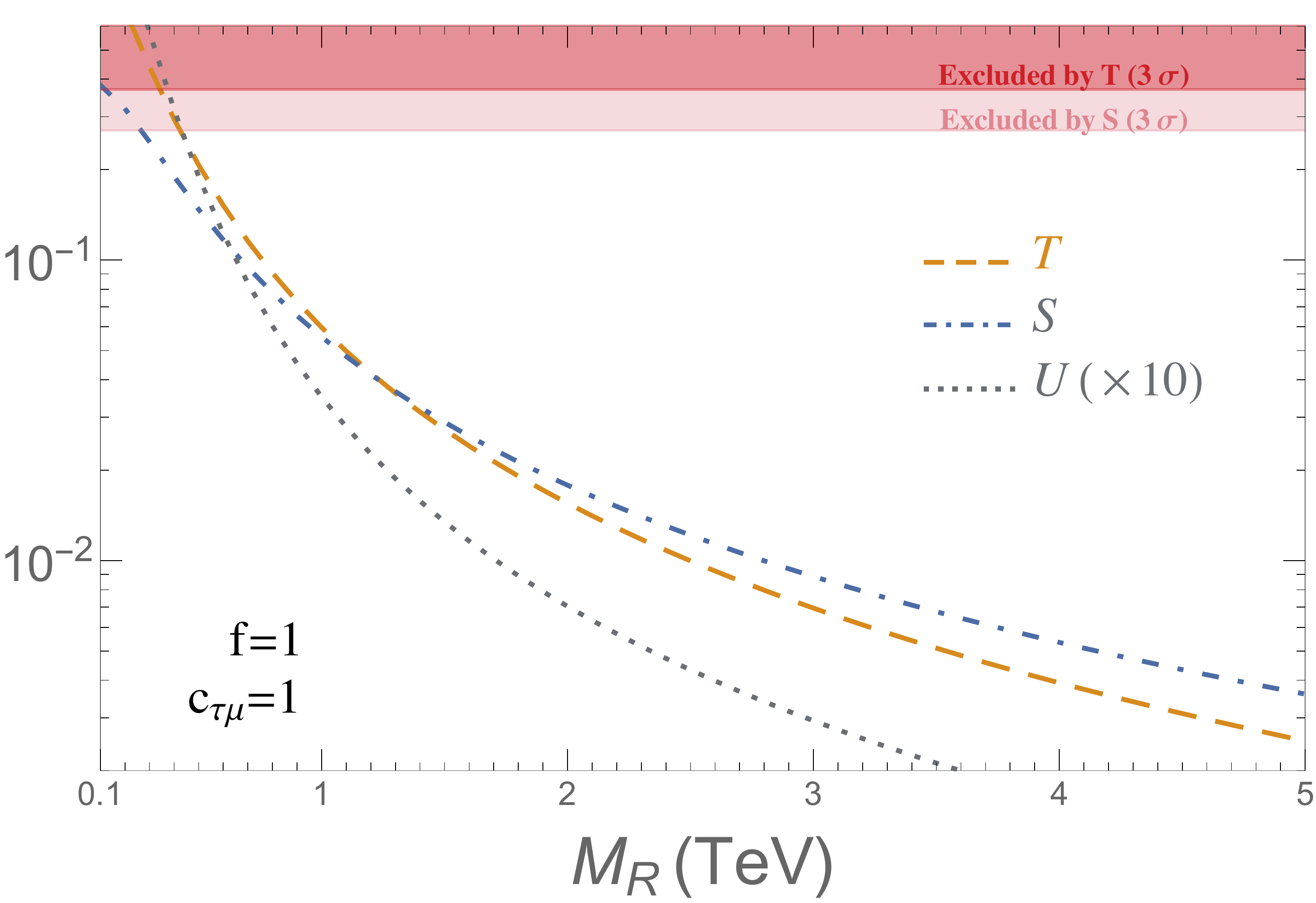}
\caption{
Predictions for the observables constraining the LFVZD rates as functions of $M_R$ in the ISS. Specifically, from top to bottom and from left to right panels, the corresponding plots are for BR($\tau\to \mu\gamma$), BR($\tau\to \mu \mu \mu$), $-\Delta r_k$, $\Gamma(Z \to {\rm inv})$, 
$|m_{ee}|$, and the Electroweak Precision Parameters $S$, $T$, $U$ (the latter enhanced by a factor of 10 to see it more clearly). In all plots we set $|\boldsymbol n_{e,\mu,\tau}|=1$. The upper shadowed bands in all plots, except the $Z$ invisible width, are the excluded regions by present data. For $\Gamma(Z\to {\rm inv.})$ the shadowed band is the experimentally allowed region at $3\sigma$ sigma level.
} 
\label{exconstraints}
\end{center}
\end{figure}

\section{Observables constraining the LFVZD rates} 
\label{Constraints}

As we discussed in the previous section, strong experimental upper bounds on ${\rm LVF}_{\mu e}$ transitions suggest that sizable LFV rates involving a $\tau$ lepton can be achieved just in some particular directions in the ISS  parameter space with $\mu$-$e$ suppression insured, which we have referred to as \modelname $\,$ scenarios. 
Our aim is to study the LFV Z decays (LFVZD) in this kind of scenarios, looking for their maximum rates that are allowed by all the relevant experimental and theoretical constraints. Our searches in these particular directions of the ISS parameter space are in contrast 
to the previous analysis of the LFVZD rates done in Ref.~\cite{Abada:2014cca},  where the ISS parameter space was uniformly scanned, making more difficult to access these peculiar scenarios that give large rates for LFV transitions with $\tau$ leptons allowed by all the constraints, including the more stringent ones coming from $\mu$-$e$ transitions. 
In this section we make our selection for the set of observables constraining the LFVZD rates which we have checked are the most relevant ones for the present study.  We will provide numerical predictions  for all the relevant observables in the very specific parameter space directions of the \modelname $\,$ explored here and compare them directly to their experimental bounds.  Alternative checks of the allowed ISS parameter space make use of global fits \cite{Antusch:2006vwa,FernandezMartinez:2007ms,delAguila:2008pw, Antusch:2008tz,Antusch:2014woa,Fernandez-Martinez:2016lgt}, but we prefer to make the explicit computations of the selected observables here in order to focus our search in the optimal directions of our model.

Generically, the addition of heavy Majorana neutrinos to the particle content of the SM has a phenomenological impact on several observables, including lepton flavor and lepton number violating  processes, via their mixing with the active neutrinos.
Therefore, we want to ensure that our analysis complies with the relevant theoretical and experimental constraints, in all the regimes of RH neutrino masses considered. 
We briefly discuss in the following the constraints that we have found to be the most relevant ones for the present work and which we consequently include in our analysis. For this study we have used our own \textit{Mathematica} code which includes all the relevant formulas for the constraining observables that are taken from the literature and that we include in the appendices for completeness. The main numerical results are summarized in Figures  \ref{exconstraints} and \ref{thconstraints}.
\subsection{LFV lepton decays}
As mentioned above, there are strong experimental upper bounds on cLFV transitions such as the LFV lepton radiative decays and LFV lepton three body decays. 
Since, by construction, the \modelname~scenarios that we are studying suppress the LFV in two of the three $\ell_i$-$\ell_j$ sectors, the most relevant LFV constraints are $\tau\to\mu\gamma$ and $\tau\to\mu\mu\mu$ for the TM scenarios and $\tau\to e\gamma$ and $\tau\to eee$ for the TE scenarios.
We compute the rates for these observables using the full one loop formulas given in Appendix~\ref{app:LFVdecays} that we take from Refs.~\cite{Ilakovac:1994kj,Alonso:2012ji} and compare them with their experimental upper limits from Babar and Belle, respectively, given in Table \ref{LFVsearch}.
In the TM scenarios, and equivalently in the TE ones, we find that the rates for these two observables, BR($\tau\to\mu\gamma$) and BR($\tau\to\mu\mu\mu$), are independent of $\mode$ and $\rotationO$, they decouple as expected with $M_R$ and grow with $f$, $\modmu$, $\modtau$ and $\ctm$. The full radiative decays rates indeed follow the behavior of the approximate formula in Eq.~(\ref{RadiativeApproxTAU}). In Figure \ref{exconstraints}, the predictions for these LFV decays are shown as functions of  the two most relevant parameters, $M_R$ and $f$.  We find that for the TM scenarios with large $\tau-\mu$ transitions given by $\ctm=1$, typically, the maximum  allowed values for $f$  are  
 $\sim {\cal O}(1-0.5)$ for $M_R=$1 TeV and the minimum allowed values for $M_R$  are $\sim  {\cal O}$(1-2) TeV for $f=1$. We find similar conclusions for the TE scenarios, regarding the $\tau \to e \gamma$ and $\tau \to eee$ decays, by simply exchanging $\mu$ and $e$ everywhere in the previous TM results.

\subsection{Lepton Flavor Universality}

It has been shown~\cite{Abada:2012mc, Abada:2013aba} that leptonic and semileptonic decays of pseudoscalar mesons ($K,\ D$, $\ D_s$, $B$) could also put important constraints on the mixing between the active and the sterile neutrinos in the ISS. In particular, the most severe bounds arise from the violation of lepton universality in leptonic kaon decays\footnote{We do not consider other lepton universality tests in view of the fact that they give similar bounds, as in the case of $\Delta r_\pi$, or they are less constraining, like the ones involving $\tau$ leptons \cite{Abada:2013aba}.}.
In the following, we will apply this constraint by considering the contributions of the sterile neutrinos to the $\Delta r_k$ parameter, defined as:
\begin{equation}
\Delta r_k = \dfrac{R_K}{R_K^{\rm SM}}-1\quad{\rm with}\quad R_K=\dfrac{\Gamma(K^+\to e^+\nu)}{\Gamma(K^+\to \mu^+\nu)}.
\label{eq:deltarkandRK}
\end{equation}
The comparison of the theoretical calculation in the SM~\cite{Cirigliano:2007xi,Finkemeier:1995gi}  with the recent measurements from the NA62 collaboration~\cite{Goudzovski:2011tc,Lazzeroni:2012cx} shows that the experimental measurements agree with the SM prediction within $1\sigma$:
\begin{equation}
\Delta r_k=(4\pm4)\times 10^{-3}.
\label{eq:deltarkbound}
\end{equation}
We compute the new physics contributions to $\Delta r_k$  using the formulas listed in Appendix \ref{app:Deltark} that we take from \cite{Abada:2012mc} and we apply the bound in Eq.~(\ref{eq:deltarkbound}) by excluding any solution that falls outside the 3$\sigma$ region. We have found that the deviations from this band become more important when the ratio between $\mode$ and $\modmu$ is different from one. Even if in general the constraints on $f$ and $M_R$ that are obtained  from this observable are weaker than those obtained from the previous LFV lepton decays, $\Delta r_k$ turns out to set relevant constraints in some textures - notably TM-7 and TM-8 - and in most textures at low $M_R$.

\subsection{The invisible decay width of the Z boson}
The presence of sterile neutrinos affects the tree level predictions of the $Z$ invisible width even if they are above the kinematical threshold, since they modify the couplings of the active neutrinos to the $Z$ boson. 
The $Z$ invisible decay width was measured in LEP to be~\cite{Agashe:2014kda}: 
\begin{equation}
\Gamma(Z\to {\rm inv.})_{\rm Exp}= 499\pm1.5~{\rm MeV},
\label{eq:Zinvbound}
\end{equation}
which is about 2$\sigma$ below the SM prediction:
\begin{equation}
\Gamma(Z\to {\rm inv.})_{\rm SM}= \sum_{\nu}\Gamma(Z\to \nu\bar\nu)_{\rm SM}= 501.69\pm0.06~{\rm MeV}.
\end{equation}
We compute the tree level predictions using the formulas provided in Ref.~\cite{Abada:2013aba} and we further include the $\rho$ parameter that accounts for the part of the radiative corrections coming from SM loops, {\it i. e.},
\begin{equation}\label{ZinvISS}
\Gamma(Z\to{\rm inv.})_{\rm ISS} = \sum_{\substack{i,j=1\\ i\leq j}}^{3} \Gamma(Z\to n_i n_j)_{\rm ISS} = \rho \Gamma(Z\to {\rm inv.})_{\rm ISS}^{\rm tree}\, ,
\end{equation}
where $n_i$ runs over all kinematically allowed neutrinos and $\rho$ is evaluated as:
\begin{equation}
 \rho =\frac{\Gamma(Z\to {\rm inv.} )_{\rm SM}}{\Gamma(Z\to {\rm inv.})_{\rm SM}^{\rm tree}}.
\end{equation}
The analytical formula for the tree level partial width of the $Z$ decay into neutrinos within the ISS is given  in Appendix~\ref{app:Zinv}.
We have also estimated the size of the extra loop corrections induced by the new heavy neutrino states using the formulas of Ref.~\cite{Fernandez-Martinez:2015hxa} and found out that they are numerically very small compared with the SM loop corrections, so we will neglect them in the following.  
Moreover, we found that the $Z$ invisible width only depends on $M_R$, $f$ and the modulus $|\boldsymbol{n}_{e,\mu,\tau}|$, while it is not dependent on $\mathcal O$ and on the flavor angles ($c_{\tau\mu}, c_{\tau e}$), as it was expected, since when adding all the possible neutrino final states in Eq.~(\ref{ZinvISS}) the dependence on $\mathcal O$ and on the flavor angles appearing in each channel disappears in the sum.
Regarding the comparison with data we require our predictions to be within the 3$\sigma$ experimental band (Eq.~(\ref{eq:Zinvbound})). As we can see in Figure \ref{exconstraints}, the $Z$ invisible width provides in general quite strong constraints, indeed comparable or even tighter  in some cases than the  previous constraints from 
the LFV lepton decays.  For instance, for scenarios with $\ctm=1$ and $f=1$, this observable also excludes $M_R$ values 
lower than around 1-2 TeV, similar to the constraints from $\tau \to \mu \gamma$. 
\subsection{Neutrinoless double beta decay}
The ISS mechanism calls upon the introduction of singlet neutrinos with Majorana masses thus allowing for LN violating processes such as neutrinoless double beta decay \cite{Benes:2005hn}. 
Within the ISS framework where $6$ sterile fermions are added to the SM particle content, the effective neutrino
mass $m_{ee}$ is given  by~\cite{Abada:2014nwa,Blennow:2010th,Abada:2014vea} 
\begin{equation} \label{eq:22bbdecay}
 m_{ee}\,\simeq \,\sum_{i=1}^{9}  (B_{e n_i})^2 \,p^2
\frac{m_{n_i}}{p^2-m_{n_i}^2} \simeq 
\left(\sum_{i=1}^3 (B_{e n_i})^2 \, m_{n_i}\right)\, 
+ p^2 \, \left(\sum_{i=4}^{9} 
(B_{e n_i})^2  \,\frac{m_{n_i}}{p^2-m_{n_i}^2}\right)\,,
\end{equation}
where $p^2 \simeq - (125 \mbox{ MeV})^2$ is an average estimate over different values - depending on  
the decaying nucleus - of the virtual momentum of the neutrino exchanged in the process.

The neutrinoless double beta decay has not been observed yet by any of the current experiments, actively searching for it. On the other hand, the experiments with highest sensitivity such as GERDA~\cite{Agostini:2013mzu}, 
EXO-200~\cite{Auger:2012ar,Albert:2014awa} and KamLAND-ZEN~\cite{Gando:2012zm} have allowed to set strong bounds on the neutrino effective mass. These bounds on the effective
neutrino Majorana mass - determining the 
amplitude of the neutrinoless double beta decay rate - lie in the range 
\begin{equation}
| m_{ee}| \lesssim 140\text { meV} - 700\text { meV}\,.
\end{equation}
In our analysis, we apply the most recent
constraint from~\cite{Albert:2014awa}, $|m_{ee}| \lesssim 190$ meV. 

As can be seen in Figure \ref{exconstraints}, the maximum value of $|m_{ee}| \sim 10$ meV is reached at large $M_R \gtrsim$~1~TeV and for all studied values of $f$. We have checked that this asymptotic value depends on the mass of the light active neutrinos, i.e.:
\begin{equation}
m_{\nu_1} \sim 0.01 (0.1) \rm eV \rightarrow |m_{ee}| \sim 0.01 (0.1) eV
\end{equation}
The overall conclusion is that this observable is much less constraining than the others in what regards our study of LFV $Z$ decay rates.  
\subsection{Electroweak Precision Observables}
We take into account constraints from electroweak precision data by computing the $S$, $T$ and $U$ parameters \cite{Peskin:1991sw} and comparing our predictions to the experimental results~\cite{Agashe:2014kda}:
\begin{equation}
S=-0.03\pm0.10\,, \qquad T=0.01\pm0.12\,, \qquad U=0.05\pm0.10\,.
\end{equation}
We use the formulas from Ref.~\cite{Akhmedov:2013hec} (which we report in Appendix \ref{app:STU}) and we consider only the parameter space points that give predictions within the 3$\sigma$ bands. As can be seen in Figure \ref{exconstraints}
the constraints on the ISS model from the Electroweak Precision Observables (EWPO), $S$, $T$ and $U$, are in general weaker than from the LFV lepton decays and from the $Z$ invisible width. 
We have found that the most constraining EWPO is the $T$ parameter and next, although quite close, the $S$ parameter. 
For instance, for $f=1$ and $\ctm=1$ we find that $M_R$ below around $300$ GeV are excluded by $T$.


\begin{figure}[t!]
\begin{center}
\includegraphics[width=.48\textwidth]{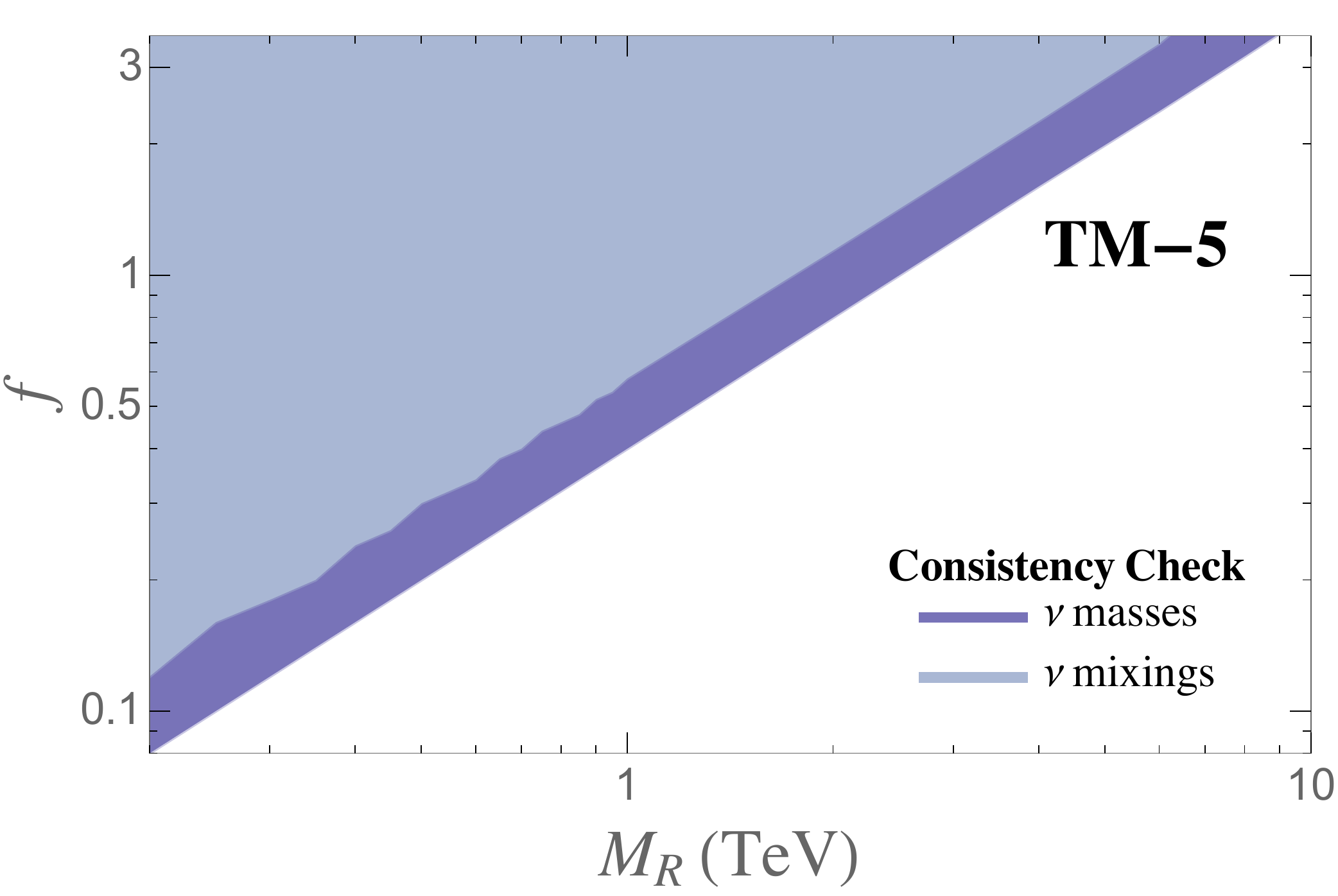}
\includegraphics[width=.49\textwidth]{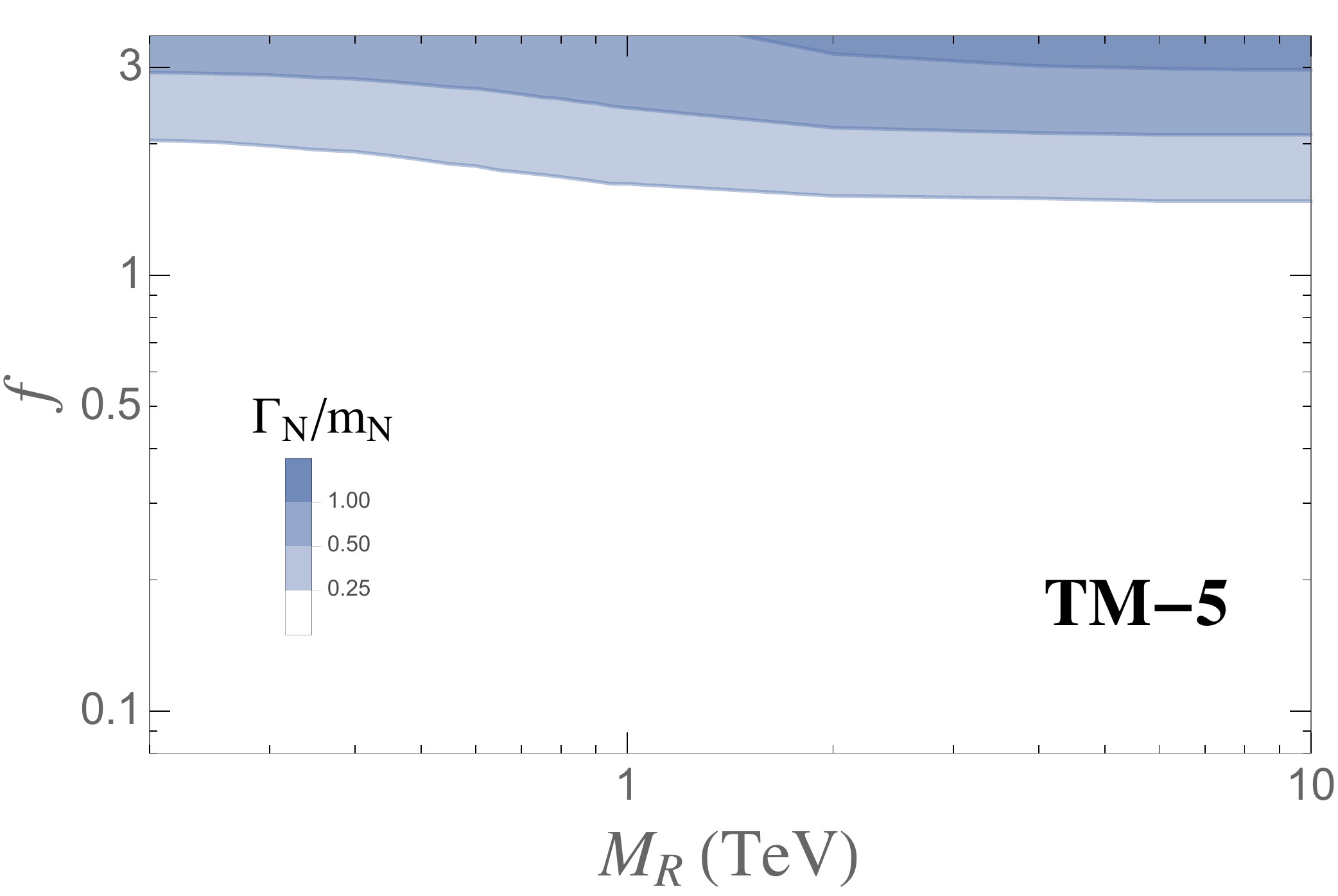}
\includegraphics[width=.48\textwidth]{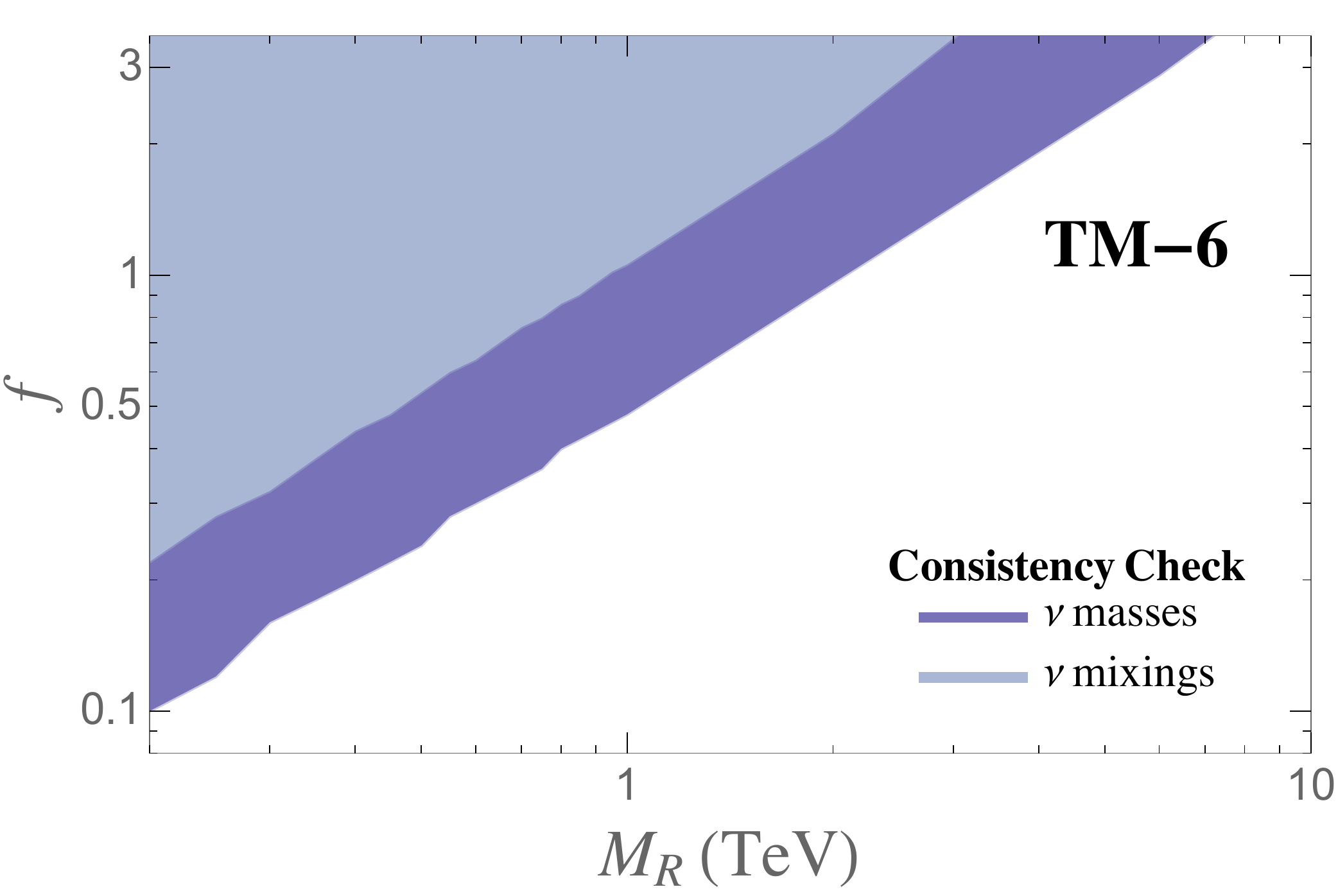} 
\includegraphics[width=.48\textwidth]{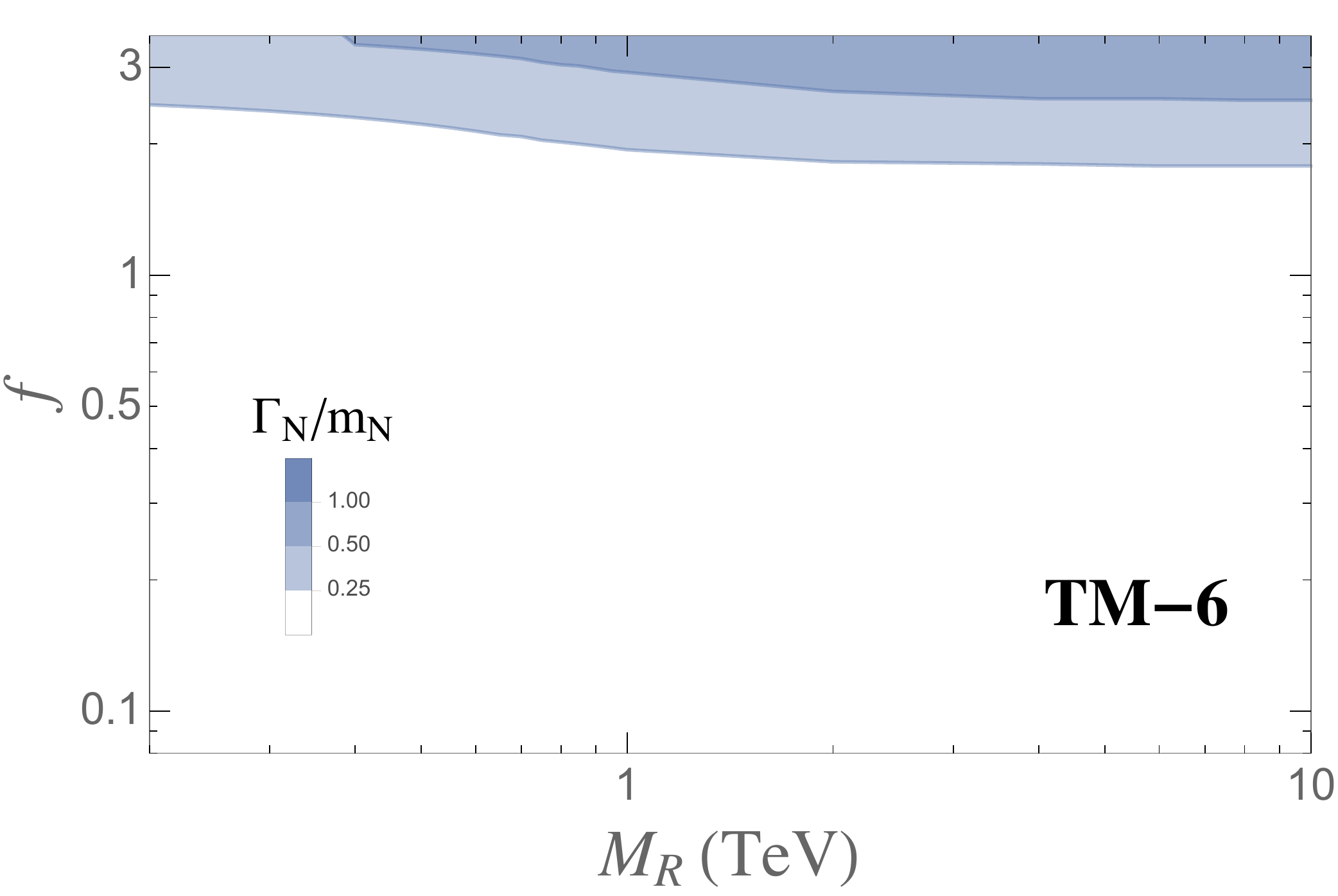}
\includegraphics[width=.48\textwidth]{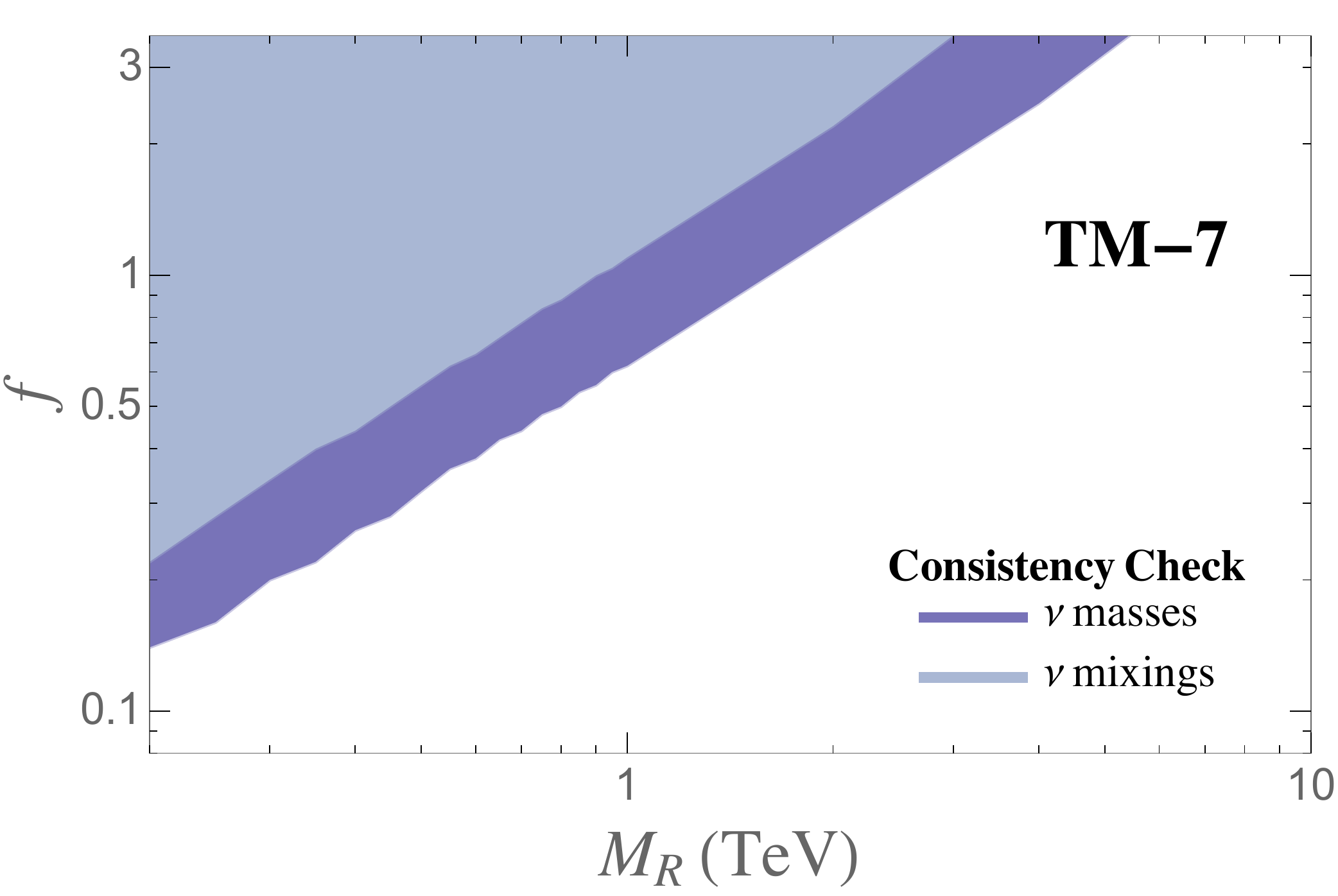} 
\includegraphics[width=.48\textwidth]{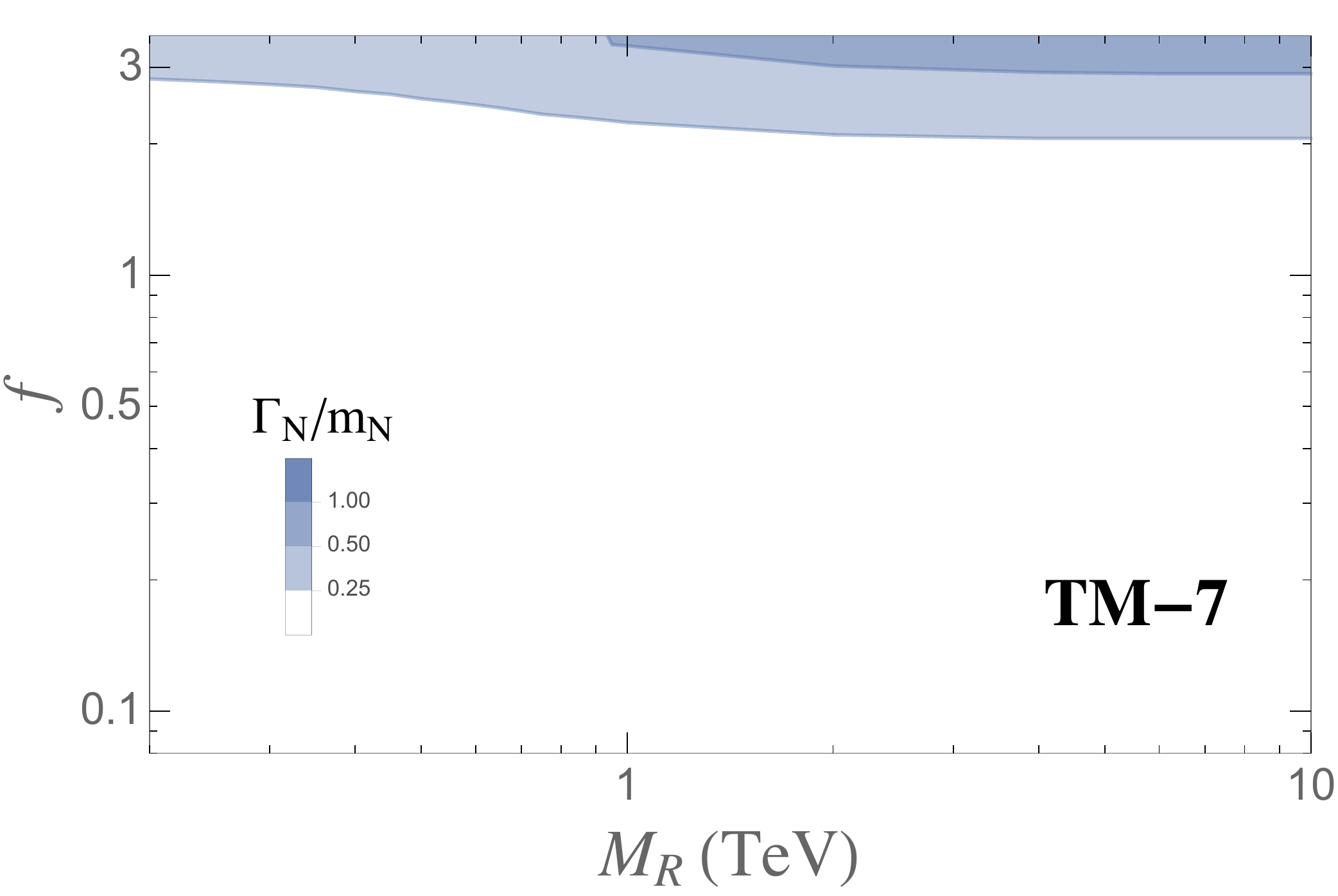}
\caption{Theoretical constraints from the requirement of perturbativity (three plots on the right) and from the consistency of the $\mu_X$ parametrization (three plots on the left), for the scenarios TM-5, TM-6 and TM-7. The regions excluded by the constraints are the shadowed areas.} \label{thconstraints}
\end{center}
\end{figure}

\subsection{Perturbativity constraints}
In this work, we are considering sizable neutrino Yukawa couplings, so we should check that they are still within the perturbative regime. In order to impose perturbativity either one may  choose a direct constraint on the maximum allowed size of the Yukawa matrix entries like, for instance, $|(Y_\nu)_{ij}|^2 /(4 \pi)<1$ or,  alternatively, one may apply a constraint on an observable that grows with this Yukawa coupling, like it is the case of the total width of the heavy neutrinos.  We choose here this second method and ensure we are considering perturbative couplings by requiring that the total decay width of each  heavy neutrino is always less than the corresponding heavy neutrino mass. In particular we have explored in Figure \ref{thconstraints} the following three assumptions to comply with the perturbative unitary
condition:
\begin{equation}
\frac{\Gamma_{N_i}}{m_{N_i}}<1,\frac12, \frac{1}{4}\quad {\rm for}~ i=1,\dots,6.
\label{widthconstraint}
\end{equation}
Notice that this condition is controlled in our ISS scenarios mainly by the global strength parameter $f$. 
Regarding the computation of the total decay width, in the limit $M_R\gg m_D$ that we work with, the possible decay channels are reduced.
For $M_R\gg m_D$ the heavy neutrino masses are close to $M_R$, with small differences of $\mathcal O(m_D^2M_R^{-1})$ between the different pseudo-Dirac pairs and, therefore, assuming they are practically degenerate,  their potential decays into other heavy neutrinos are suppressed. In consequence, the dominant decay channels are simply $N_j\to Z\nu_i, H\nu_i$ and  $W^\pm \ell_i^\mp$, and the total neutrino width can be easily computed by adding the corresponding partial widths of these four decays.
For instance, the decays into $W^+\ell_i^-$ or  $W^-\ell_i^+$ have a partial width given by:
\begin{equation}
\Gamma_{N_j \to W \ell_i}=\frac{\sqrt{\big(m_{N_j}^2-m_{\ell_i}^2-m_W^2\big)^2-4 m_{\ell_i}^2 m_W^2}}{16\pi  m_{N_j}^3}~\big| \overline {F_W} \big|^2,
\end{equation}
with
\begin{equation}
\big| \overline {F_{W}} \big|^2 =\frac{g^2}{4m_W^2} \big|B_{\ell_iN_j}\big|^2 
 \times\Big\{ \big( m_{N_j}^2-m_{\ell_i}^2\big)^2+m_W^2 (m_{N_j}^2+m_{\ell_i}^2)-2m_W^4\Big\}\,.
\end{equation}
When summing over all flavors, $i=1,2,3$, in the final state the four ratios turn out to be approximately equal{~\cite{Atre:2009rg}: 
\begin{equation}
{\rm BR}(N_j \to H \nu)={\rm BR}(N_j \to Z \nu) =
{\rm BR}(N_j \to W^+ \ell^-) ={\rm BR}(N_j \to W^- \ell^+)=25\%\,.
\end{equation} 
In Figure \ref{thconstraints} we show the numerical predictions for the constraints from the perturbativity requirement in three examples of Table \ref{TMscenarios}, concretely TM-5, TM-6 and TM-7, and by trying the three choices in Eq.~(\ref{widthconstraint}). 
We find that this perturbativity requirement is not much sensitive to $M_R$, giving an excluded area in the $(M_R, f)$ plane that is a band nearly horizontal and located at the top, which constrains basically just the size of the global Yukawa coupling $f$, in the most restricted scenarios, to be below order 2-3.  In the following, we will take the  second choice, $1/2$, in Eq.~(\ref{widthconstraint}) as our constraining condition.

\subsection{Constraints from the $\boldsymbol{\mu_X}$ parametrization}
As explained in section \ref{intro}, we are using the $\mu_X$-parametrization of Eq.~(\ref{MUXparametrization}) to accommodate light neutrino data. 
In order to check the validity of this parametrization, 
we require that both the predicted light neutrino mass squared differences and the neutrino mixing angles (more specifically, the corresponding entries of the $U_\nu$ matrix that refer to the light neutrinos  sub-block) that we obtain from the diagonalization of the full neutrino mass matrix (Eq.~(\ref{ISSmatrix})), lie within the 3$\sigma$ experimental  bands~\cite{Gonzalez-Garcia:2014bfa,Tortola:2012te,Fogli:2012ua,Forero:2014bxa,Adhikari:2016bei}.  The predictions for the constraints found in the three  examples, TM-5, TM-6 and TM-7 are shown in Figure \ref{thconstraints}. 
As can be seen in this figure,  the bounds obtained from the constraints on the active neutrinos squared mass differences are in these three scenarios stronger than the ones from the light neutrino mixing matrix entries.   
For other scenarios, like TM-8, we have checked that this can  be reversed, {\it i.e.} the constraints from the neutrino mixings can be stronger than from the neutrino masses.  
In general, we found that the area in the $(M_R,f)$ parameter space that is allowed by all the experimental bounds studied in the previous sections is also allowed by the consistency checks of the $\mu_X$ parametrization, meaning that the parametrization works well for the  parameter space allowed by data.
Finally, we have also compared the validity of this parametrization for two values of the input lightest neutrino mass, $0.1$eV and $0.01$eV (the chosen value for Figure \ref{thconstraints}) and we have concluded that the $\mu_X$-parametrization works better for the case with a smaller value of the light neutrino mass.


\section{Results for the Lepton Flavor Violating Z decays}
\label{Results}
In order to study the LFVZD rates within the \modelname ~ model, we have taken the full one-loop formulas from Ref.~\cite{Illana:1999ww} and we have adapted them to this model, {\it i.e}, we have rewritten them in terms of the proper physical neutrino masses and couplings that we have specified in section~\ref{model}. We include these formulas, for completeness, in Appendix \ref{app:LFVZD} where we have also adapted the loop functions to the usual notation in the literature. The various contributing one-loop diagrams are also displayed in Figure~\ref{Diagrams}, for completeness. We evaluate them numerically with our code and with the help of the {\it LoopTools} \cite{Hahn:1998yk} package for {\it Mathematica}.
We focus our analysis on the particular case of BR($Z\to\tau\mu$) = BR($Z\to\tau\bar\mu$) + BR($Z\to\mu\bar\tau$) within the TM scenarios, which are defined by taking $\cte=0$ in Eq.~(\ref{YukawaAmatrix}).
The relevant parameters in this case are $M_R$, $f$, $\modtau$, $\modmu $ and $\ctm$. 
Nevertheless, all our conclusions for BR($Z\to\tau\mu$) in the TM scenarios can be directly translated to BR($Z\to\tau e$) in the TE scenarios  just by replacing $\modmu $ and $\ctm$ by $\mode$ and $\cte$, respectively. 

\begin{figure}[t]
\begin{center}
\includegraphics[width=.48\textwidth]{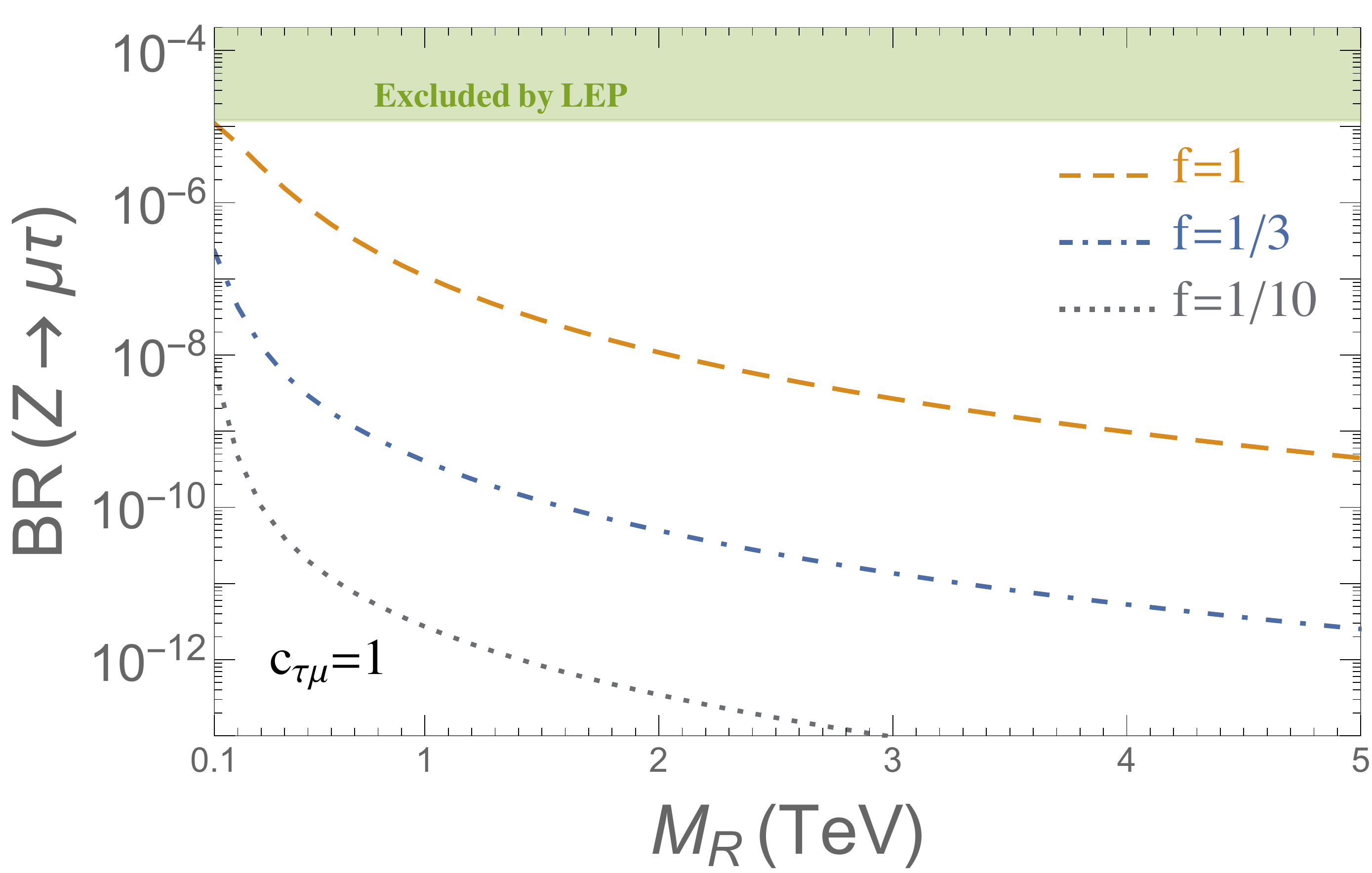}\, \,
\includegraphics[width=.48\textwidth]{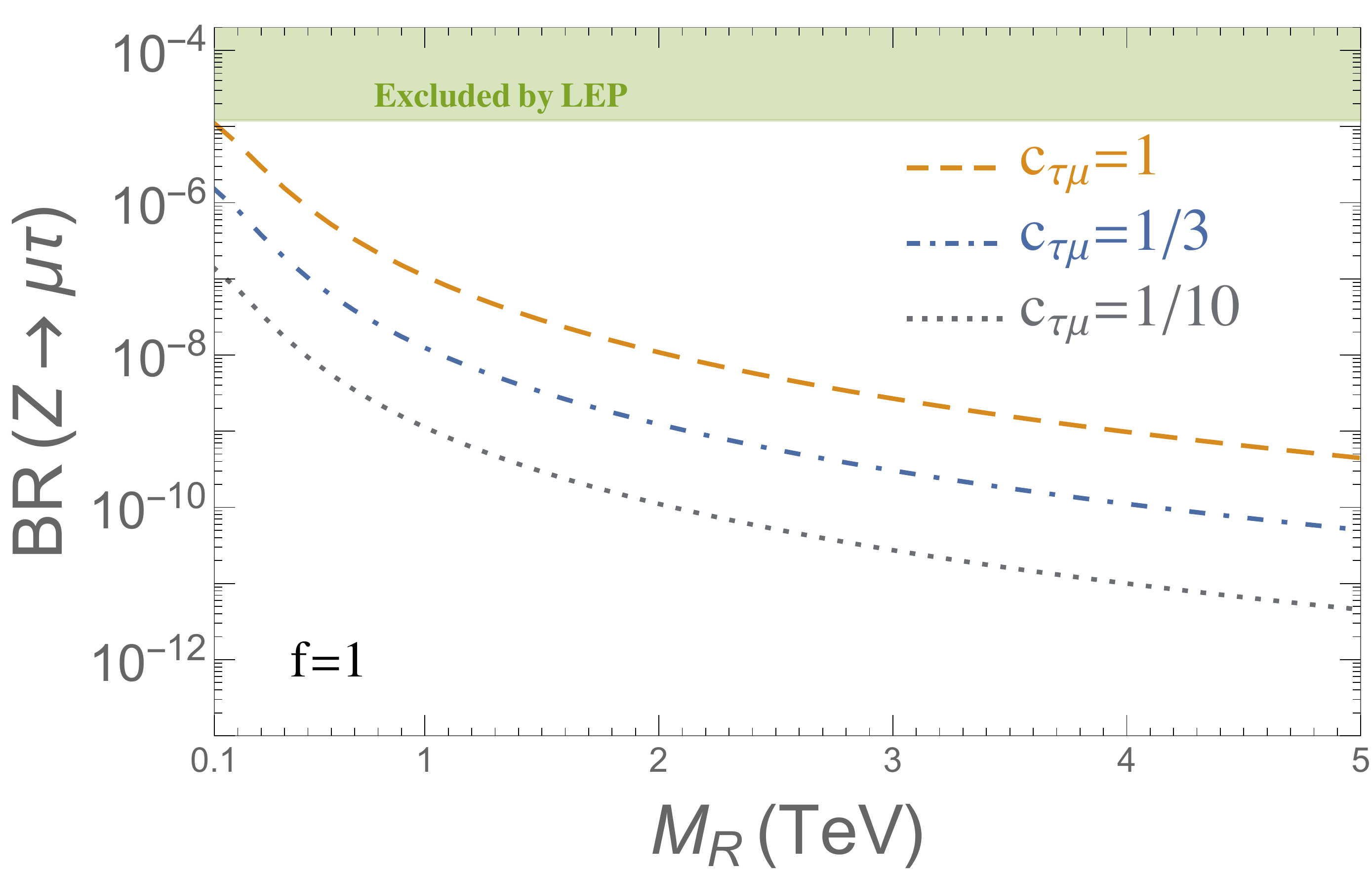}
\includegraphics[width=.48\textwidth]{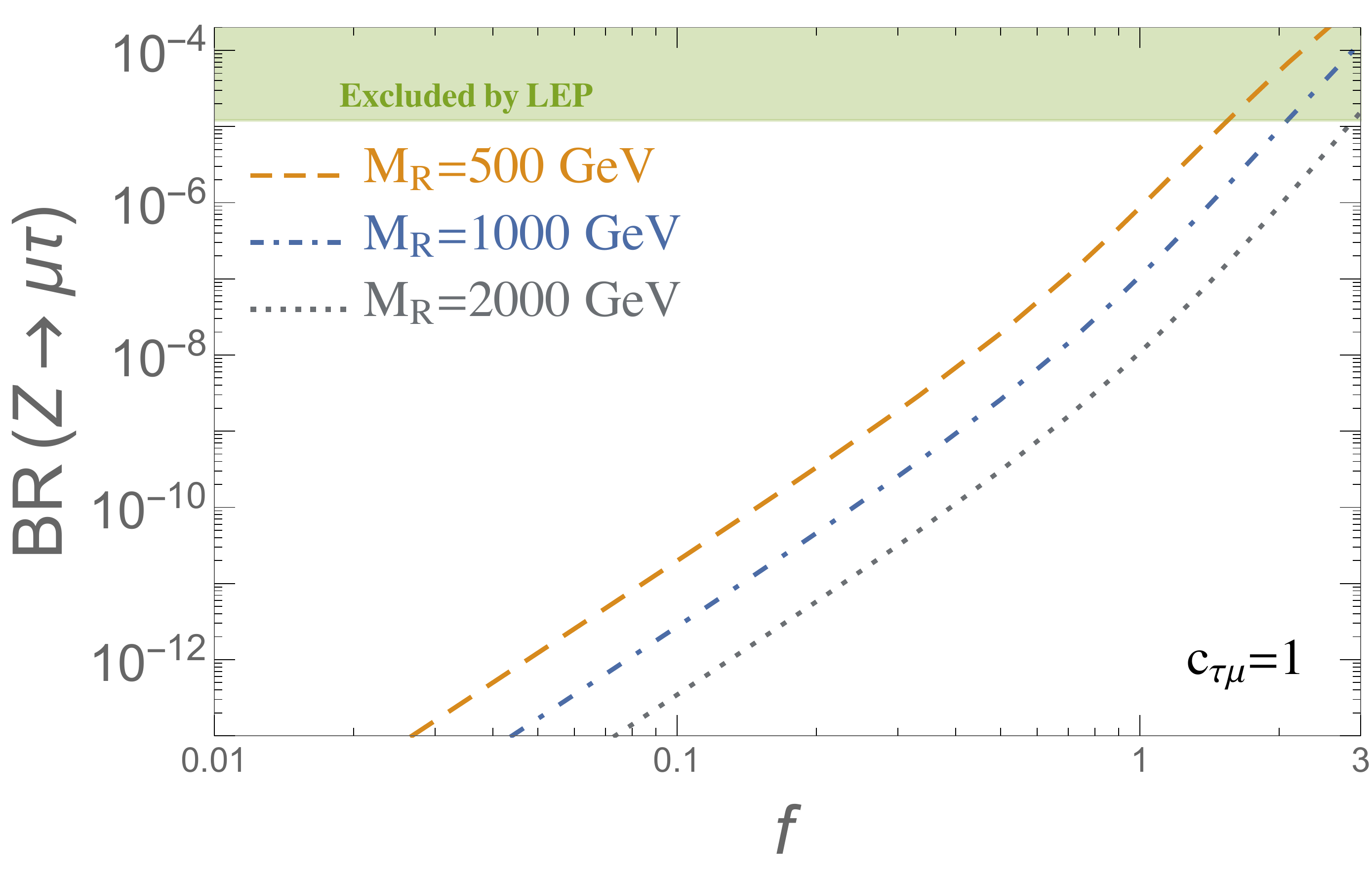}~~
\includegraphics[width=.48\textwidth]{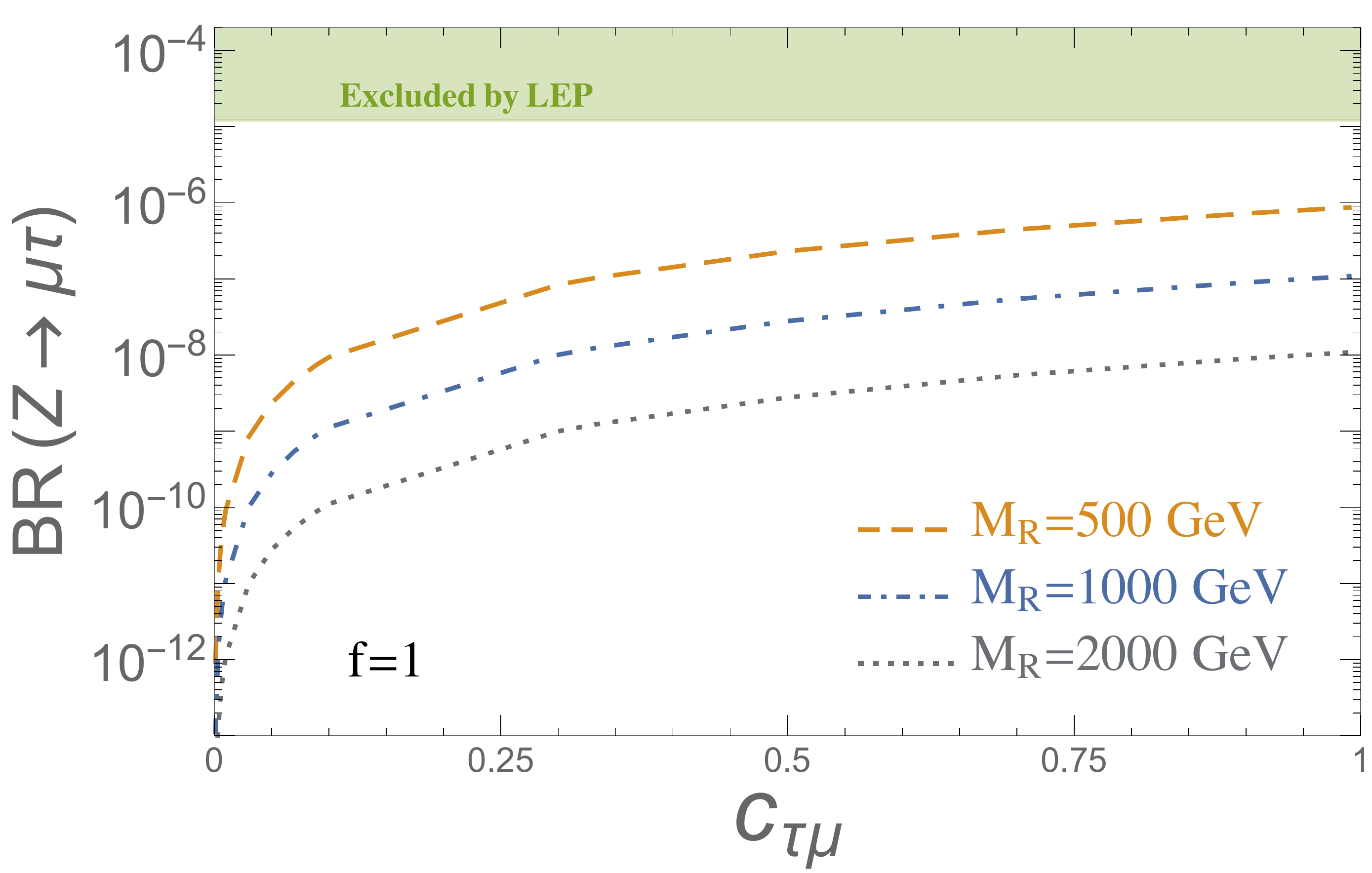}
\caption{Predictions for BR($Z\to\tau\mu$) within the ISS model as a function of the heavy neutrino mass parameter $M_R$ (two upper panels), the neutrino Yukawa coupling strengh $f$ (lower left panel) and $\ctm$ (lower right panel) for various choices of the relevant parameters. In all plots we have fixed, $\cte=0$ and $|\boldsymbol{n}_{e,\mu,\tau}|=1$.  
The upper shadowed areas (in green) are excluded by LEP~\cite{Abreu:1996mj}. Similar results for BR($Z\to\tau e$) by exchanging $\cte$ and $\ctm$.
}\label{Ztaumu}
\end{center}
\end{figure}

We display in Figure~\ref{Ztaumu} the behavior of the BR($Z\to\tau\mu$) rates with the $M_R$, $f$ and $\ctm$ parameters for fixed values of $\mode=\modmu=\modtau=1$, $\cte=0$ and $\rotationO=\mathbb I$.  
As can be seen in this figure, our ISS model gives in general large rates for the LFV $Z$ decay rates, indeed close to the upper bound from LEP (and also close to present LHC sensitivity) in the upper left corner of the two upper plots and in the upper right corner of the two lower plots. 
We also see that the rates decrease with the heavy scale $M_R$ and grow with the Yukawa coupling strength $f$, as expected. We found this growth to be approximately as $f^4$ in the low $f$ region and as $f^8$ in the high $f$ region of the studied interval of this parameter. This suggests that, in contrast to the radiative decays, the two kinds of contributions $Y_\nu Y_\nu^\dagger$ and 
$Y_\nu Y_\nu^\dagger Y_\nu Y_\nu^\dagger$ participate in this observable.

In the lower right panel, we observe that the rates also grow with $\ctm$, albeit the dependence is milder, approximately as $\ctm^2$.
Although not shown here, we have also studied the dependence of the decay rates with the modulus of the vectors, $|{\bf n}_i|$, finding that the predictions for BR($Z\to\tau\mu$) grow with both $\modmu$ and $\modtau$, while they are constant with $\mode$, as expected. 
Finally, we checked that the results do not depend on the global rotation $\rotationO$, as argued when the parametrization for the $Y_\nu$ coupling matrix was motivated. 

Before going to the final analysis of the maximum LFV $Z$ decay rates that are allowed by all the constraints, we find interesting first to compare the predictions of these LFV $Z$ decays with the predictions of the three body LFV lepton decays in our particular ISS scenarios with suppressed $\mu$-$e$ transitions. 
In the left panel of Figure~\ref{tau3mu} we study the behavior of BR($\tau \to \mu \mu \mu$) with respect to $M_R$ for the scenario TM-5 and fixed value of $f=1$, displaying separately the total and the contributions from the $\gamma$ penguin, boxes and $Z$ penguin.
We see that the full contribution is mostly coming from the latter. 
The fact that the BR($\tau\to\mu\mu\mu$) rates are dominated by the $Z$ penguin contributions implies a strong correlation between $\tau\to\mu\mu\mu$ and $Z\to\tau\mu$, as already found in Ref.~\cite{Abada:2014cca}. 
We have also checked in some examples of the ISS parameter space that our numerical predictions of these two observables are in agreement with that reference.  

We study this correlation in the right panel of Figure~\ref{tau3mu}, where we consider three of the scenarios given in table \ref{TMscenarios}, TM-5, TM-6 and TM-7, varying the values of the parameters within the ranges of $f\in(0.1,2)$ and $M_R\in(0.2,10)$~TeV. Both observables grow with $f$ and decrease with $M_R$ in approximately the same way, due to the $Z$ penguin dominance in the three body decays. 
Although the predicted rates in each scenario are obviously different, see for instance the positions of the reference points with $f=1$ and $M_R=3$~TeV, we clearly see that there is a strong correlation between the two observables in our ISS model. 
We can also conclude from this plot that by considering just the constraints from the three body decays, {\it i.e.} the present upper bound on $\tau\to\mu\mu\mu$ from Belle, already suggests a maximum allowed rate of BR($Z\to\tau\mu)\sim 2 \times 10^{-7}$, which is clearly within the reach of future linear colliders ($10^{-9}$ in the most conservative option).
 Interestingly, comparing the future expected sensitivities for both observables, we find some parameter space points where the LFVZD rates are in the reach of future linear colliders
while the cLFV three body decay rates would not be accessible in other facilities, like BELLE-II.
This fact suggests that experiments looking for LFVZD would be able to provide additional information about the model that complements the results of other searches, like the ones in Table \ref{LFVsearch}.
We found a similar correlation between BR$(\tau\to eee)$ and BR($Z\to\tau e$) in the TE scenarios. 

\begin{figure}[t!]
\begin{center}
\includegraphics[width=0.49\textwidth]{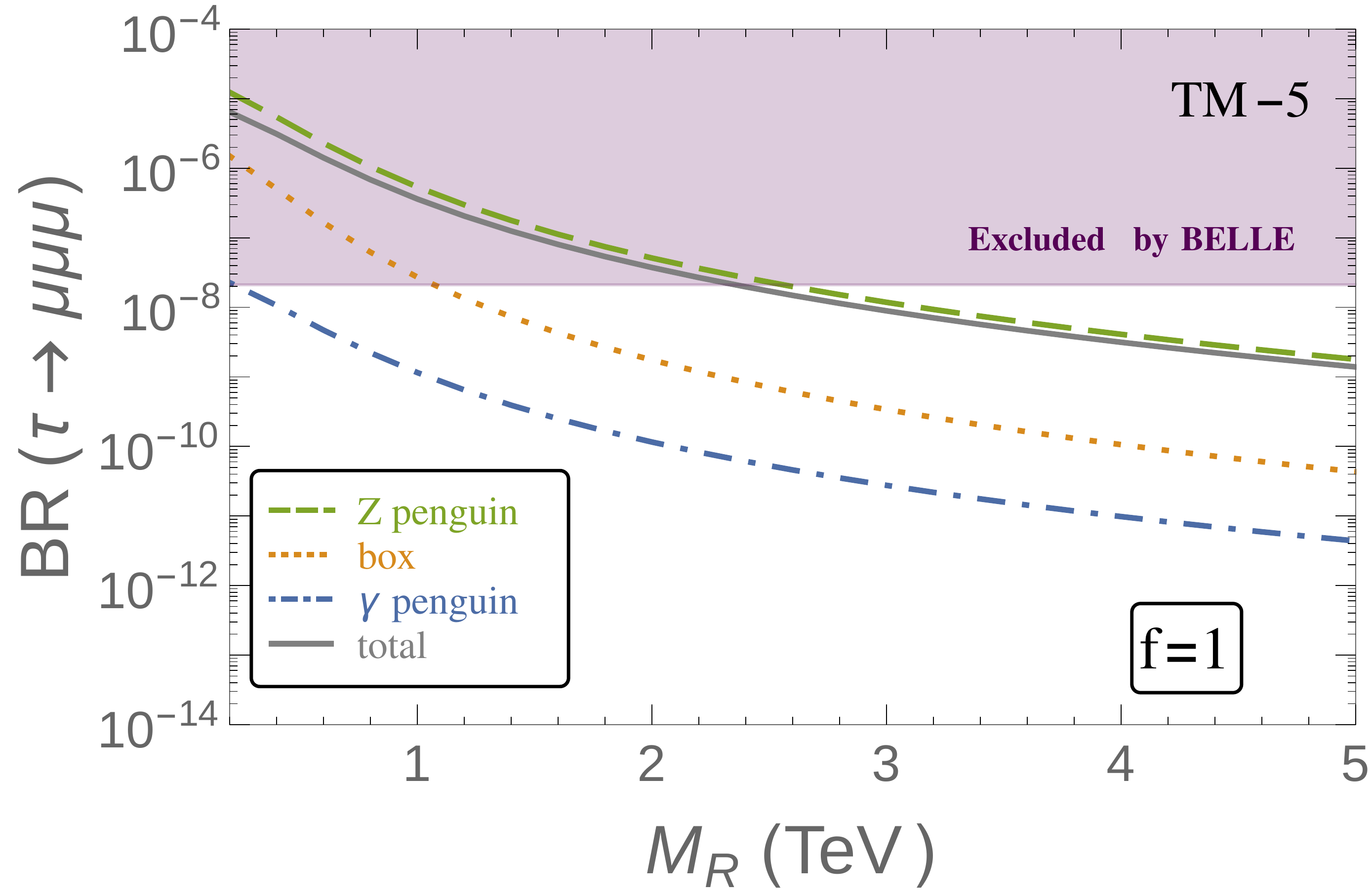}
\includegraphics[width=.49\textwidth]{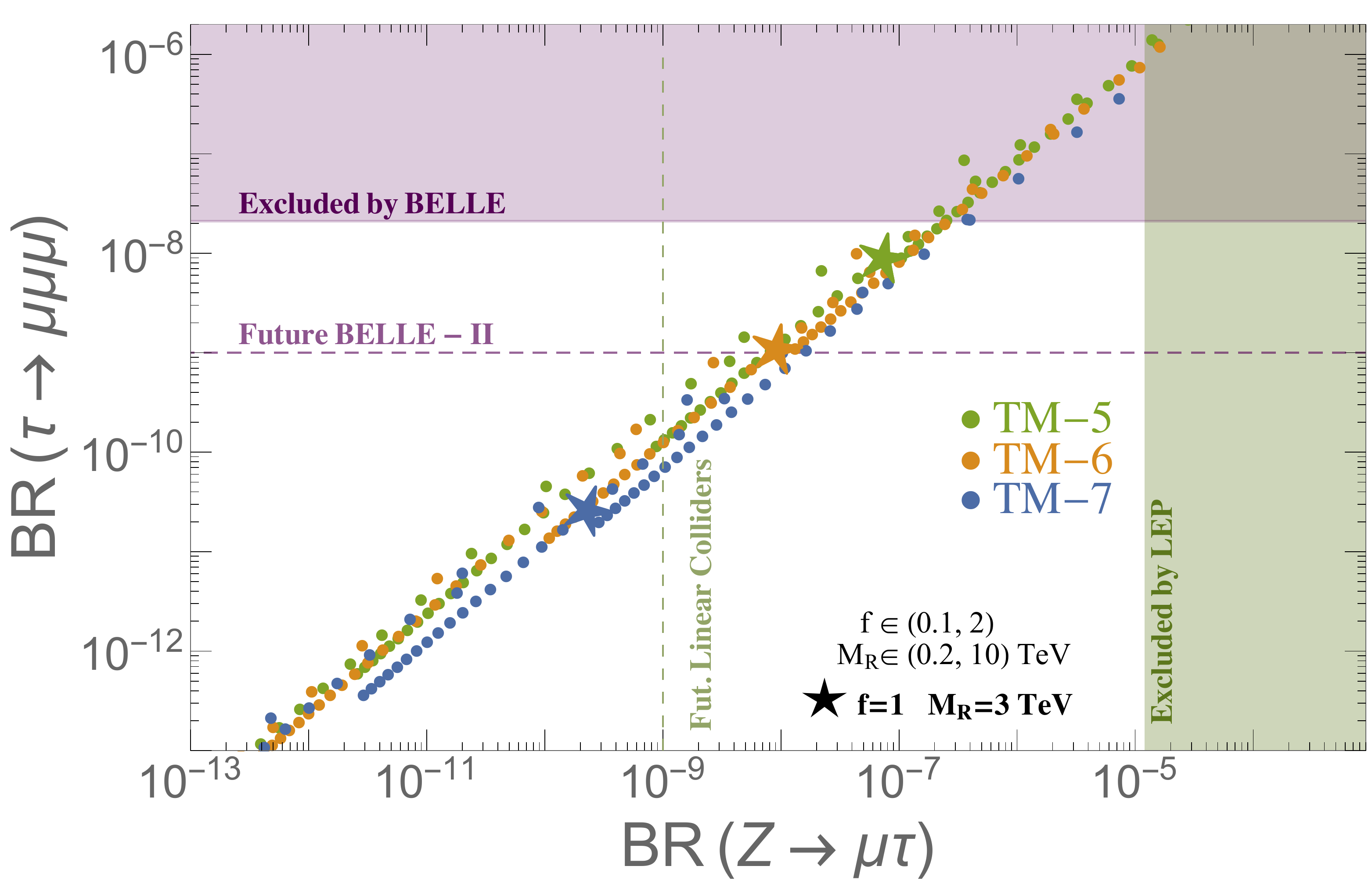}
\caption{Left panel: BR($\tau\to\mu\mu\mu$) as a function of $M_R$ for $f=1$ in the TM-5 scenario. The full prediction (gray solid line) is decomposed in its contributions from  $\gamma$ penguin (blue dot-dashed), boxes (yellow dotted) and $Z$ penguin (green dashed), the dominant one.
Right panel: correlation plot for BR($Z\to\mu\tau$) and BR($\tau\to\mu\mu\mu$) for scenarios TM-5 (green), TM-6 (yellow) and TM-7 (blue) defined in table \ref{TMscenarios}. The dots are obtained by varying $f\in(0.1,2)$ and $M_R\in(0.2,10)$~TeV and the stars are for the reference point $f=1$ and $M_R=3$~TeV.
Purple (green) shadowed area is excluded by Belle~\cite{Hayasaka:2010np} (LEP~\cite{Abreu:1996mj}), while the dashed line denotes expected future sensitivity from Belle-II (future linear colliders).
}\label{tau3mu}
\end{center}
\end{figure}

In the following we present our full analysis of the LFVZD rates in the \modelname, including all the most relevant constraints.
For this analysis we have explored the $(M_R, f)$ plane for the eight TM scenarios given in table \ref{TMscenarios} and provide numerical predictions for the BR($Z \to l_i l_j $) rates together with the predictions of the most constraining observables and their present bounds.  

We show in Figure~\ref{ZtaumufMRplane} the results for BR($Z \to \tau \mu$) together with the constraints from: $\tau \to \mu\mu\mu$, $\tau\to\mu\gamma$, $Z \to {\rm inv.}$, 
$\Delta r_K$ and the EWPO ($S$,$T$ and $U$). 
As in the previous section, we show our results only for the ${\rm LFV}_{\tau\mu}$ sector in the TM scenarios, although the conclusions are very similar for ${\rm LFV}_{\tau e}$ in the TE scenarios.  
We use different colors in the shadowed areas to represent the exclusion regions from each of the constraints listed above.
Specifically, the purple area is excluded by the upper bound on BR($\tau\to\mu\mu\mu$), the green area by BR($\tau\to\mu\gamma$), the yellow area by the $Z$ invisible width, the cyan area by $\Delta r_k$ and the area above the pink solid line is excluded by the $S$, $T$, $U$ parameters. 
Although we are not explicitly showing them here, we have also checked that the total parameter space allowed by all these constraints are also permitted by our requirements on perturbativity and on the validity of the $\mu_X$ parametrization. 
Notice that in some scenarios some of the colored areas are hidden below the excluded regions corresponding to the more constraining observables.
\begin{figure}[H]
\begin{center}
\includegraphics[width=.49\textwidth]{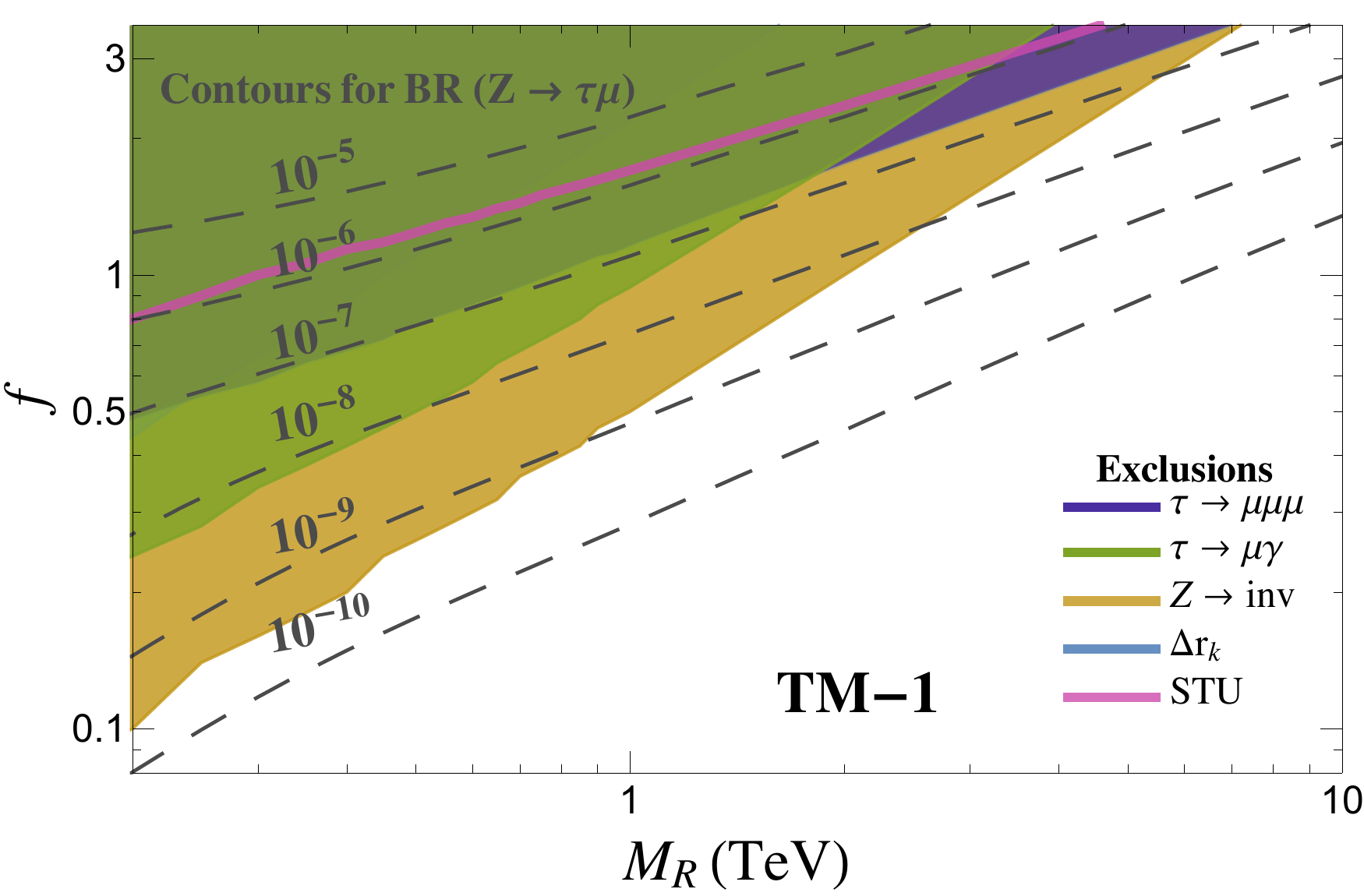}
\includegraphics[width=.49\textwidth]{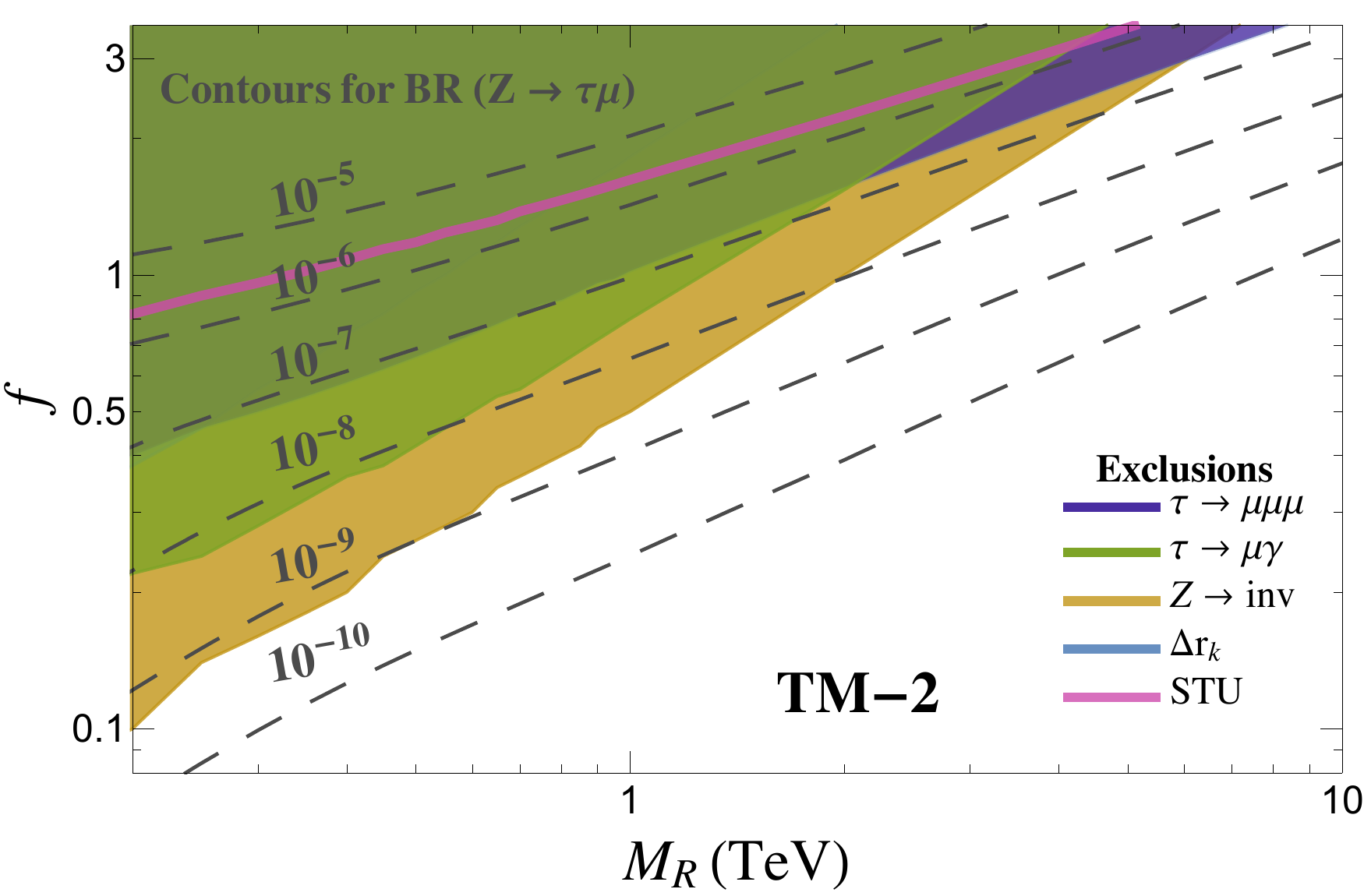}
\includegraphics[width=.49\textwidth]{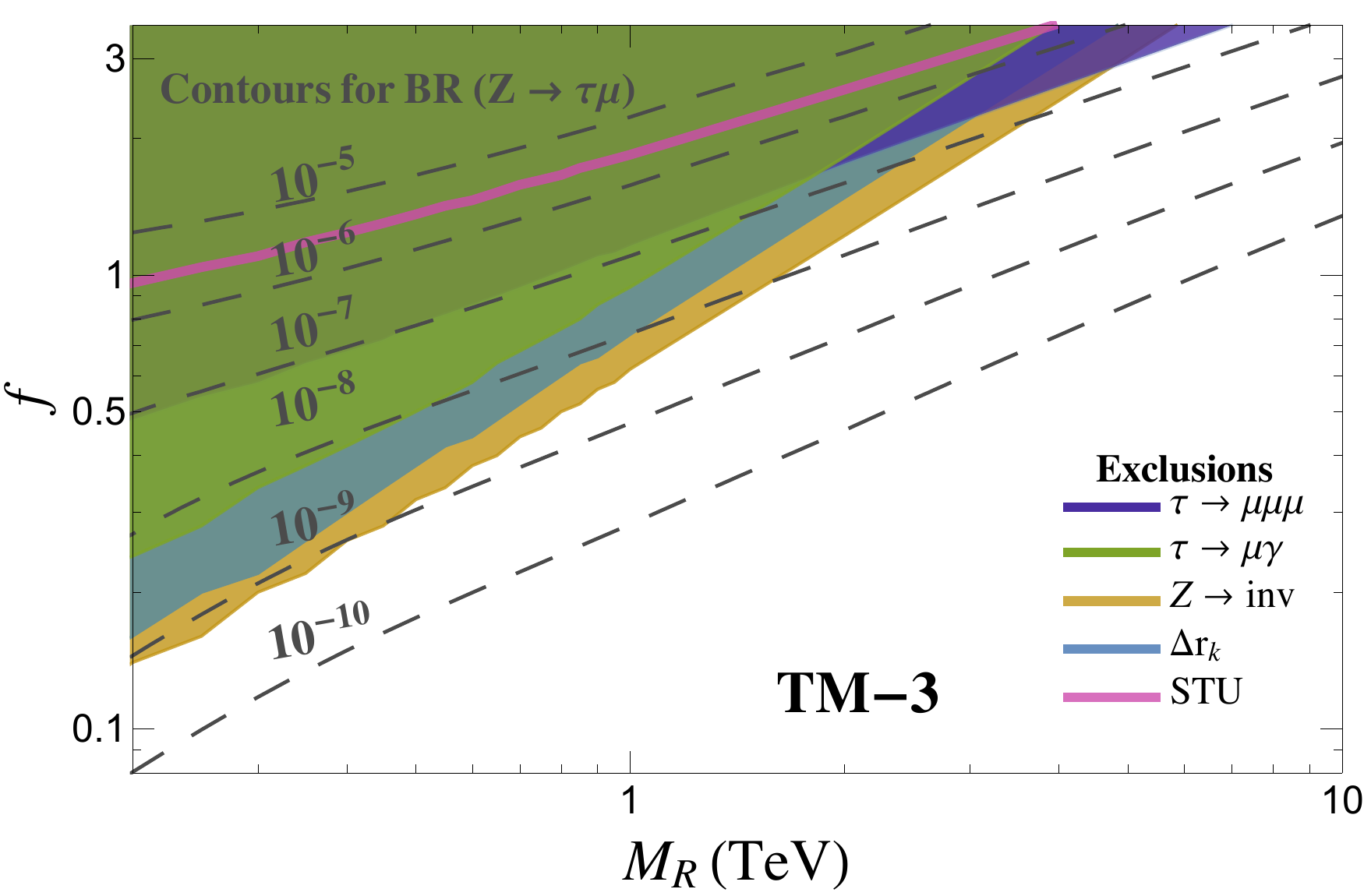}
\includegraphics[width=.49\textwidth]{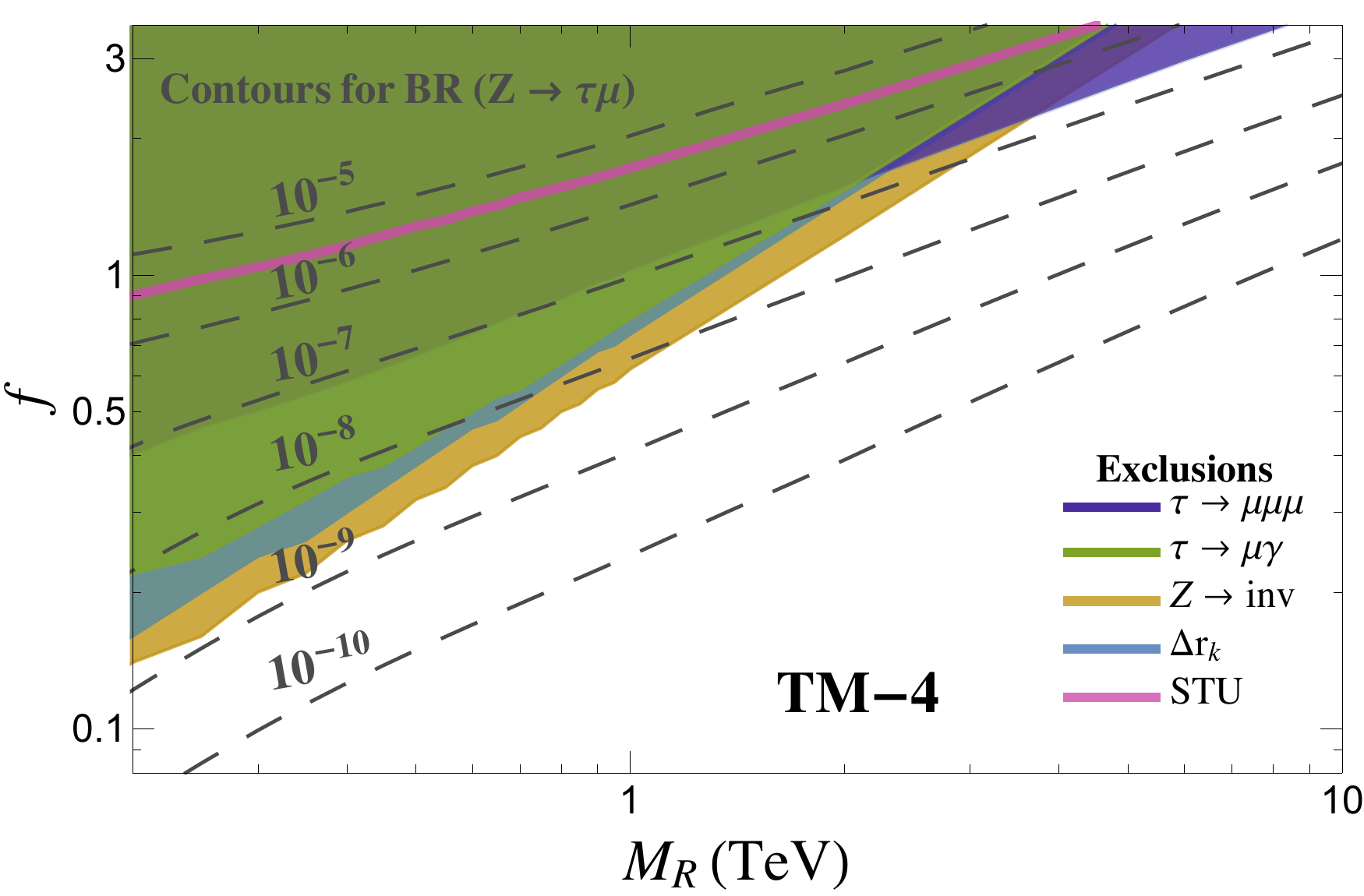}
\includegraphics[width=.49\textwidth]{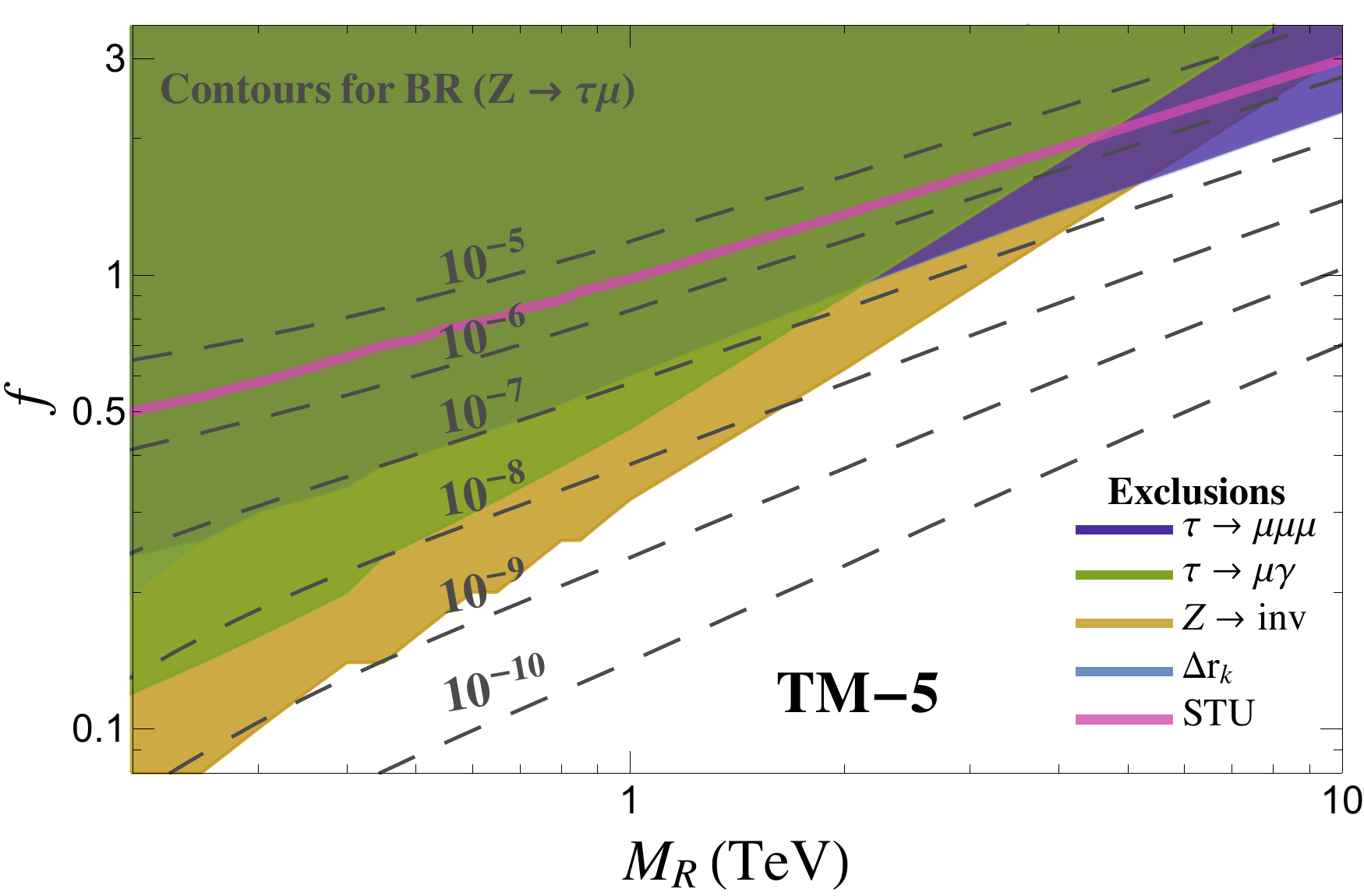}
\includegraphics[width=.49\textwidth]{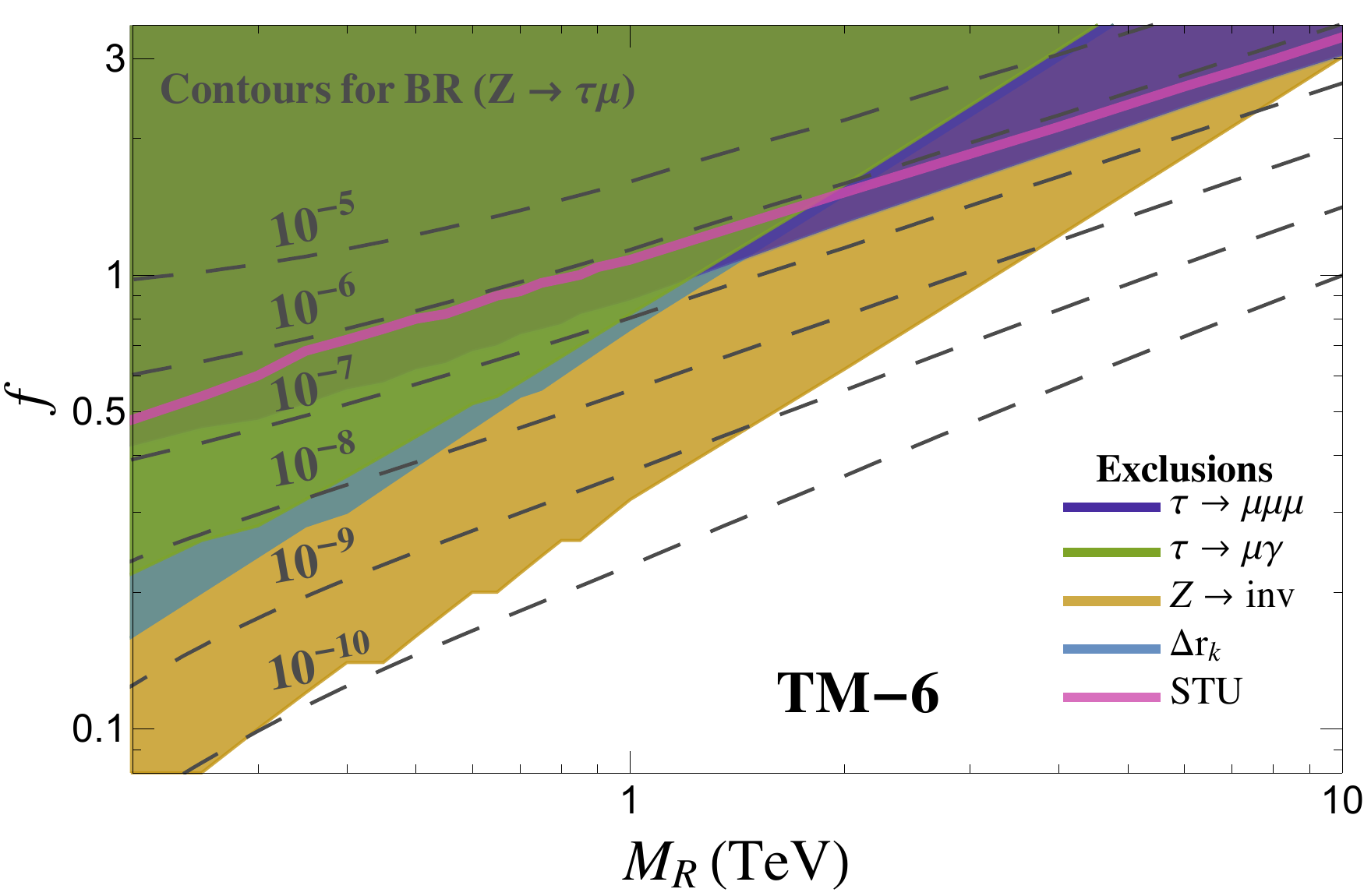}
\includegraphics[width=.49\textwidth]{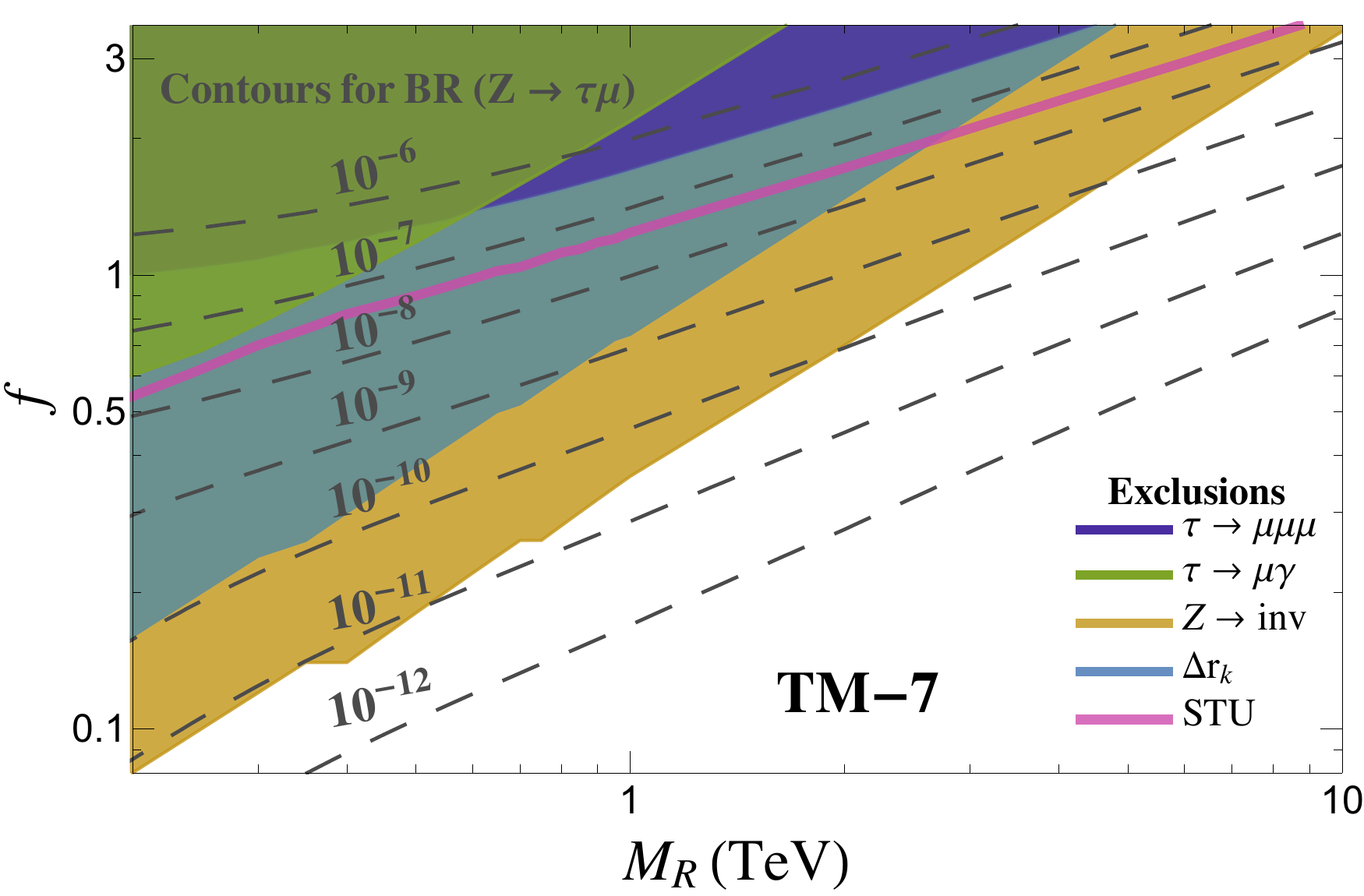}
\includegraphics[width=.49\textwidth]{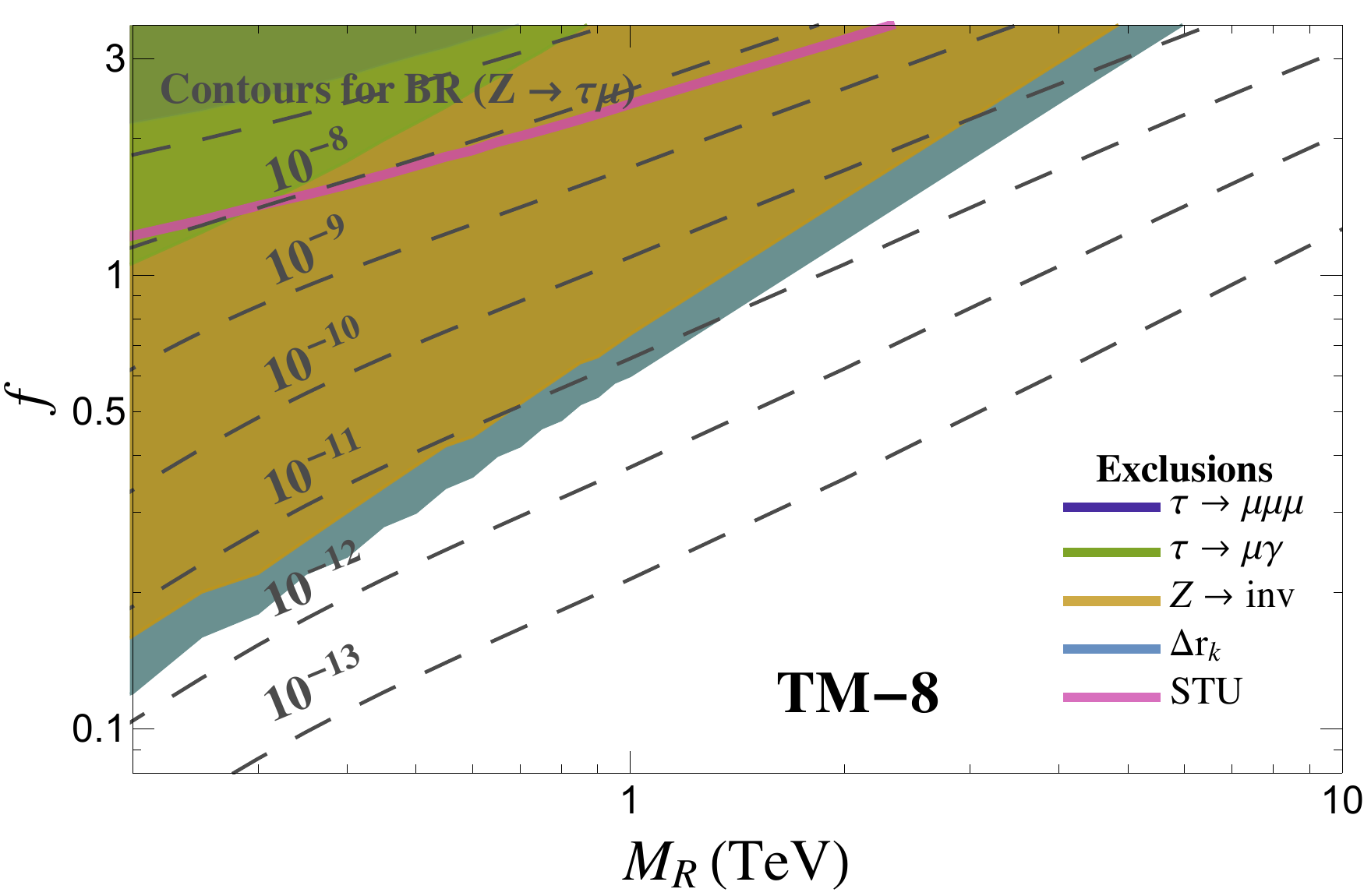}
\caption{Contour lines for BR($Z\to\tau\mu$) (dashed lines) in the ($M_R,f$) plane of the ISS model for the eight TM scenarios in table \ref{TMscenarios}. Shadowed areas are the excluded regions by $\tau\to\mu\mu\mu$ (purple), $\tau\to\mu\gamma$ (green), Z invisible width (yellow) and  $\Delta r_k$ (cyan). The region above the pink solid line is excluded by the $S$, $T$, $U$ parameters. We obtain similar results for BR($Z\to\tau e$) in the TE scenarios by exchanging $\mu$ and $e$ in these plots of the TM scenarios.}\label{ZtaumufMRplane}
\end{center}
\end{figure}
\noindent
On top of all the bounds, we display in Figure~\ref{ZtaumufMRplane} the predicted contour lines for BR($Z\to\tau\mu$) as dashed lines.  
As expected from the correlation studied in Figure~\ref{tau3mu}, we see that these contour lines have approximately the same slope as the border of the exclusion region from BR($\tau\to\mu\mu\mu$), and in particular the line corresponding to BR$(Z\to\tau\mu)=2 \times 10^{-7}$, is very close to the upper bound line of the three body decay in all the TM scenarios ({\it i.e.}, the border of the purple line). 
Furthermore, in the large $M_R$ - large $f$ region of these plots we see that for several TM scenarios, concretely TM-2, TM-3, TM-4 and TM-5, the BR($\tau\to\mu\mu\mu$) is the most constraining observable. 

In contrast, in the low $M_R$ and $f$ region, the most constraining cLFV observable is the radiative decay $\tau\to\mu\gamma$.
On the other hand, regarding the flavor preserving observables, it is clear that the EWPO do not play a relevant role here, but both $\Delta r_K$ and the invisible $Z$ width put relevant constraints in some scenarios. In particular, $\Delta r_K$ is the most constraining observable in the case of TM-8, and the $Z$ invisible width is so in the scenarios TM-1, TM-6 and TM-7.   
We also learn that, typically, the $Z$ invisible width is  the most constraining observable in the region of low $M_R$ values, whereas BR($\tau\to\mu\mu\mu$) is  the most constraining observable in the region of high $M_R$ values. Thus, generically, it is the crossing of these two excluded areas in the ($M_R, f$) plane that gives the focus area of the maximum allowed LFV $Z$ decay rates, BR$(Z\to\tau\mu)\sim 2 \times 10^{-7}$, and this crossing occurs at different values of $M_R$ and $f$ in each scenario.  
For example, in the TM-4 and TM-5 scenarios it happens at $M_R\sim 2-4$~TeV and for $f\sim {\cal O}(1)$, while in the TM-6 $M_R$ is around 10~TeV and $f\sim {\cal O}(2)$. 
On the other hand, if we focus our attention on the mass range of interest for present direct neutrino production searches at LHC, say masses around 1~TeV and below, we observe that the allowed BR$(Z\to\tau\mu)$ rates are smaller than this maximum value $2\times10^{-7}$; nevertheless they are still in the reach of future linear colliders ($10^{-9}$) for some scenarios, like TM-4 or TM-5.

Finally, we would like to end this section by comparing our results in Fig.~\ref{ZtaumufMRplane} with previous results in the literature. In particular, we focus on Ref.~\cite{Abada:2014cca}, which to our knowledge is the only reference that provides predictions for the maximum allowed LFV Z decay rates in the same model, namely, the ISS with  three RH neutrinos and three additional sterile states, which they refer to as the (3,3)-ISS realization. Their numerical results are scatter plots generated by random scans over the model parameter space, and for accommodating neutrino data they use the Casas-Ibarra parametrization, that provides the neutrino Yukawa couplings as output. 
The setup of the ISS model in their analysis is therefore different from ours, since we use the $\mu_X$ parametrization and $Y_\nu$ is an input.  They provide the results in terms of their parameters $\tilde \eta$ and $<m_{4-9}>$, whereas we present our results as contour lines in terms of $f$ and $M_R$ for the eight selected TM-i scenarios. 
The comparison can be done as follows: 1) their $<m_{4-9}>$ can be compared roughly with our $M_R$; 2) their parameter $\tilde \eta$ is defined as: $\tilde \eta= 1-|{\rm Det}(\tilde U_{\rm PMNS})|$ where 
$\tilde U_{\rm PMNS}=(1-\eta)U_{\rm PMNS}$, and the matrix $\eta$ is frequently used in the literature to encode the deviation of $\tilde U_{\rm PMNS}$ from unitarity (see, for instance, Refs.~\cite{Antusch:2006vwa,FernandezMartinez:2007ms, Antusch:2008tz,Antusch:2014woa,Fernandez-Martinez:2016lgt}). The $\tilde U_{\rm PMNS}$ can be compared with the complex conjugate of the $3 \times 3$ sub-block of our unitary $9 \times 9$ matrix $U^\nu$ in Eq.~(3) (or, similarly, with $B_{ln}$ in Eq.~(9)); 3) Our predicted maximum allowed BR's of the LFV Z decays are close to the contour line $10^{-7}$ and correspond to the region in the upper right corner of our plots for TM1-5 in Fig.~\ref{ZtaumufMRplane}.
In order to compare with the results in Ref.~\cite{Abada:2014cca} we have evaluated with our code the corresponding $\eta$ and $\tilde \eta$ for the mentioned region and we have compared them with the upper bounds found in the more recent literature about global constraints on heavy neutrino mixing. Concretely, we have used the recent results in Ref.~\cite{Fernandez-Martinez:2016lgt}  and we have applied them to the three sigma level in order to be in agreement with our choice for the rest of the constraints in this work\footnote{We warmly acknowledge Josu Hern\'andez-Garc\'ia for his valuable help in extracting the $3\sigma$ constraints from Ref.~\cite{Fernandez-Martinez:2016lgt}.}. This translates into an upper bound of $\tilde \eta^{\rm max} \simeq 6 \times 10^{-3}$. We have then checked that our predictions of LFV Z decays BR~$\sim10^{-7}$ in the upper-right part of our plots (with $f=1-3$ and $M_R=6-10$ TeV) enter into the area allowed by this $\tilde \eta^{\rm max}$ constraint. 
More concretely, the crossing of the contourline for  BR$(Z \to \tau \mu )\sim 10^{-7}$ with the contourline for $\tilde \eta^{\rm max} \simeq 6 \times 10^{-3}$ happens at $M_R \simeq $ 10, 8, 7, 6, and 7 TeV for TM-1, TM-2, TM-3, TM-4 and TM-5 respectively. 
In contrast, the results in Ref.~\cite{Abada:2014cca} show that their maximum allowed LFV Z BR's are about $10^{-9}$ and these are placed approximately at $<m_{4-9}> \sim 10-100$ TeV
and $\tilde \eta \sim 10^{-4}$. We also see that their predictions for the allowed LFV Z BR's are placed at $\tilde\eta$ below a maximum value which is in agreement with the above commented $\tilde \eta^{\rm max} \simeq 6 \times 10^{-3}$. 
Moreover, as previously noticed, we also agree with Ref.~\cite{Abada:2014cca} in the correlation found between BR($Z\to\tau\mu$) and BR($\tau\to\mu\mu\mu$).
However, our work contains predictions leading to BR($\tau \to 3 \mu$) below but close to the present experimental limit, $2.1 \times 10^{-8}$, and allowed BR($Z \to \tau \mu$) close to $10^{-7}$ which are not contained in Ref.~\cite{Abada:2014cca}. There are empty spaces in their plots, with no predictions, corresponding to points in the parameter space which are not reached by their scans. It is in some of those regions where we are finding the allowed LFV $Z$ decay rates of order $10^{-7}$ by means of our selected directions in the (3,3)-ISS parameter space.  
Therefore we conclude that, although random scans with the Casas-Ibarra parametrization allow to explore a large region of the parameter space and to study the general features of the model, they are not always optimal to reach specific directions along the parameter space.
In particular, we infer that they seem to be inefficient in generating Yukawa textures leading to scenarios like ours and as a consequence, they lead to lower allowed LFV Z decay rates. 
A similar conclusion was obtained in Ref.~\cite{Arganda:2014dta} in the context of the LFV Higgs decays, where the use of $\mu_X$ parametrization also provided larger allowed rates in these decays than with the Casas-Ibarra parametrization.
Obviously, physics does not depend on the parametrization one chooses. However, the efficiency in reaching specific points/directions/areas in the parameter space does.  
In this sense, our study of the maximum allowed LFV Z decays is complementary to that in Ref.~\cite{Abada:2014cca}.

\section{Conclusions}
\label{conclusions}
In this work we have studied several aspects of cLFV processes in the context of the ISS model. 
Motivated by the strong experimental upper bounds on LFV $\mu$-$e$ transitions, we have discussed a useful geometrical parametrization of the neutrino Yukawa coupling matrix that allows us to easily define ISS scenarios with suppressed ${\rm LFV}_{\mu e}$.
We have studied in full detail the LFV $Z$ decays in these scenarios that are designed to find large rates for processes including a $\tau$ lepton, and we have  investigated those that are allowed by all the present constraints. We have therefore fully explored in parallel also the most relevant constraints within this \modelname ~model.
Important constraints come from experimental upper bounds on the LFV three body lepton decays, since they are strongly correlated to the LFVZD in these scenarios. 
Taking into account all the relevant bounds, we found that heavy ISS neutrinos with masses in the few TeV range can induce maximal rates of BR$(Z\to\tau\mu)\sim 2 \times 10^{-7}$ and BR$(Z\to\tau e)\sim 2 \times 10^{-7}$ in the TM and TE scenarios, respectively, larger than what was found in previous studies. 
These rates are potentially measurable at future linear colliders and FCC-ee. 
Therefore, we have shown that searches for LFVZD at future colliders may be a powerful tool to probe cLFV in low scale seesaw models, in complementarity with low-energy (high-intensity) facilities searching for cLFV processes.
Another appealing feature of our results is that the here presented improved sensitivity to LFVZD rates could come together with the possibility that the heavy neutrinos could be directly produced at LHC.

\section*{Acknowledgments}
We wish to thank warmly Ana M. Teixeira for the very interesting discussions on the subject studied in this work at the starting stages of this research. This work is supported by the European Union through the ITN ELUSIVES H2020-MSCA-ITN-2015//674896 and the RISE INVISIBLESPLUS H2020-MSCA-RISE-2015//690575, by the CICYT through the project FPA2012-31880, by the Spanish Consolider-Ingenio 2010 Programme CPAN (CSD2007-00042) and by the Spanish MINECO's ``Centro de Excelencia Severo Ochoa''  Programme under grant SEV-2012-0249. X. M. is supported through the FPU grant AP-2012-6708.  
F. S. is supported through a grant from the Scuola di Scienze of the Universit\`a di Bologna.

\begin{appendix}
\section{Appendix: LFV lepton decays} 
\label{app:LFVdecays}
For completeness, in this Appendix we provide the needed formulas for the full one-loop computation of the LFV lepton decays in the particle mass basis, both the three body and the radiative decays, which we have implemented in our code.

The expression for the branching ratio BR($\ell_m \to \ell_k \ell_k\ell_k$) is taken from Refs.~\cite{Ilakovac:1994kj,Alonso:2012ji} as well as all the form factors required for its computation.
 
\begin{align}
\text{BR}
(\ell_m \to \ell_k \ell_k \ell_k)=& \frac{\alpha^4_W
}{24576\pi^3}\frac{m^4_{\ell_m}}{m^4_W}\frac{m_{\ell_m}}{\Gamma_{\ell_m}} 
\times \Bigg\{ 2 \left|\frac{1}{2}F^{\ell_m \ell_k \ell_k \ell_k}_{\rm Box}
+F^{\ell_m \ell_k}_Z-2s^2_W\Big(F^{\ell_m \ell_k}_Z-F^{\ell_m \ell_k}_\gamma\Big)\right|^2\nonumber \\
&+ 16 s^2_W \text{Re}\left[ \Big(F^{\ell_m \ell_k}_Z 
+\frac{1}{2}F^{\ell_m \ell_k \ell_k \ell_k}_{\rm Box}\Big) G^{\ell_m \ell_k^{\,\scalebox{.75}{*}}}_\gamma \right]
- 48 s^4_W \text{Re}\left[\Big(F^{\ell_m \ell_k}_Z-F^{\ell_m \ell_k}_\gamma\Big)G^{\ell_m \ell_k^{\,\scalebox{.75}{*}}}_\gamma \right] \nonumber \\ 
&+4 s^4_W \left|F^{\ell_m \ell_k}_Z-F^{\ell_m \ell_k}_\gamma\right|^2 
+32 s^4_W \big|G^{\ell_m \ell_k}_\gamma\big|^2\left[\ln \frac{m^2_{\ell_m}}{m^2_{\ell_k}} -\frac{11}{4}	\right]
\Bigg\}\,.  \label{eq:mueee} 
\end{align}

The BR($\ell_m \to \ell_k \ell_k \ell_k$) contains several form factors, corresponding to the dipole, 
penguin (photon and $Z$) and box diagrams.
The expressions for these form factors are given by~\cite{Ilakovac:1994kj,Alonso:2012ji}:  
\begin{align}
G^{\ell_m \ell_k}_\gamma &= \sum_{i=1}^{9} B_{\ell_k n_i}B^*_{\ell_m
  n_i} G_\gamma(x_i)\,,  \nonumber  \\ 
F^{\ell_m \ell_k}_\gamma &= \sum_{i=1}^{9} B_{\ell_k n_i}B^*_{\ell_m
  n_i} F_\gamma(x_i)\,, \nonumber \\ 
F^{\ell_m \ell_k}_Z &= \sum_{i,j=1}^{9} B_{\ell_k n_i}B^*_{\ell_m n_j}
\left(\delta_{ij} F_Z(x_i) + C_{n_i n_j} G_Z(x_i,x_j) + C^*_{n_i n_j} H_Z(x_i,x_j)   \right)\,, \nonumber \\ 
F^{\ell_m \ell_k \ell_k \ell_k}_{\rm Box}&=  \sum_{i,j=1}^{9}B_{\ell_k n_i} B^*_{\ell_m n_j}\left(B_{\ell_k n_i}B^*_{\ell_k n_j}G_{\rm Box}(x_i,x_j)+2\,B^*_{\ell_k n_i}B_{\ell_k n_j}F_{\rm
  Box}(x_i,x_j)\right)\,. \label{eq:formfact}
\end{align}
where $x_i$ stands for the  dimensionless ratio of masses ($x_i = m^2_{n_i}/m_W^2$).
Moreover, the following loop functions enter in the previous form factors~\cite{Ilakovac:1994kj,Alonso:2012ji}: 
\begin{align}
F_Z(x)&= -\frac{5x}{2(1-x)}-\frac{5x^2}{2(1-x)^2}\ln x \, , \nonumber
\\  
G_Z(x,y)&= -\frac{1}{2(x-y)}\left[	\frac{x^2(1-y)}{1-x}\ln x -
  \frac{y^2(1-x)}{1-y}\ln y	\right]\, ,  \nonumber \\ 
H_Z(x,y)&=  \frac{\sqrt{xy}}{4(x-y)}\left[	\frac{x^2-4x}{1-x}\ln
  x - \frac{y^2-4y}{1-y}\ln y	\right] \, , \nonumber \\ 
F_\gamma(x)&= 	\frac{x(7x^2-x-12)}{12(1-x)^3} -
\frac{x^2(x^2-10x+12)}{6(1-x)^4} \ln x	\, , \nonumber \\ 
G_\gamma(x)&=    -\frac{x(2x^2+5x-1)}{4(1-x)^3} -
\frac{3x^3}{2(1-x)^4} \ln x \,   \nonumber \\	
F_{\rm Box}(x, y) &= \frac{1}{x - y} \bigg\{
\left(1+\frac{x
y}{4}\right)\left[\frac{1}{1-x}+\frac{x^2}{(1-x)^2}\ln
x\right] - 2x
y\left[\frac{1}{1-x}+\frac{x}{(1-x)^2}\ln
x\right] -(x\to y)\bigg\} \, ,  \nonumber \\	
G_{\rm Box}(x, y) &= -\frac{\sqrt{xy}}{x - y}\bigg\{
(4+xy)\left(\frac{1}{1-x}+x \frac{\ln x}{(1-x)^2} \right) 
-2\left(\frac{1}{1-x}+x^2  \frac{\ln x}{(1-x)^2}\right)-(x\to y)\bigg\}\,. 
\label{eq:formfact2}		
\end{align}
In the limit of degenerate neutrino masses ($x=y$), we get the following expressions: 
\begin{align}
G_Z(x,x ) &= {} \left[x (-1 + x - 2 \ln x)/(2 (1-x)) \right]]\, ,  \nonumber\\
H_Z(x,x ) & = {} - \left[ x (4 - 5x + x^2 + (4 - 2x + x^2)\ln
  x)/(4(1 - x)^2) \right] \, , \nonumber\\ 
F_{\rm Box}(x,x )& = \left[ (4-19 x^2+16 x^3-x^4-2 x(-4+4 x+3 x^2) \ln x)/(4(1-x)^3) \right]  , \nonumber\\ 
G_{\rm Box}(x,x)	 &= x \left[\left(6 - 8 x+4 x^2-2 x^3+(4+x^2+x^3) \ln 
 x \right)/(-1 + x)^3 \right] \,\, .\label{limitval2} 
\end{align}

For the LFV radiative decay rates, we use the analytical formulas
appearing in~\cite{Ilakovac:1994kj} and~\cite{Deppisch:2004fa} that have also been implemented in our code:
\begin{equation}
\label{BRradiative}
{\rm BR}(\ell_m\to \ell_k\gamma)=\frac{\alpha^3_W s_W^2}{256\pi^2}\left(\frac{m_{\ell_m}}{m_W}\right)^4\frac{m_{\ell_m}}{\Gamma_{\ell_m}}\big|G_{mk}\big|^2,
\end{equation}
where $\Gamma_{l_m}$ is total decay width of the lepton $l_m$, and
\begin{align}\label{Gmk}
G_{mk}&=\sum_{i=1}^{9}B_{ki} B^*_{mi}\, G_\gamma\left(x_i\right)\,,
\end{align} 
with $G_\gamma(x)$ defined in Eq.~(\ref{eq:formfact2}) and, again, $x_i\equiv m_{n_i}^2/m_W^2$.

\section{Appendix: $\Delta r_k$}
\label{app:Deltark}
In this appendix we give the formulas to calculate the quantity $\Delta r_K$ (see Eq.~(\ref{eq:deltarkandRK})), which parametrizes the deviation from the SM prediction arising from the sterile neutrinos contribution, as a test of lepton flavor universality.
The expression for  $\Delta r_K$ in generic SM extension with sterile neutrinos has been given in~\cite{Abada:2012mc}:
\begin{equation}
\label{eq:deltarK}
\Delta r_K \,= \,\frac{m_\mu^2 (m_K^2 - m_\mu^2)^2}{m_e^2 (m_K^2 - m_e^2)^2}\,
\frac{\operatornamewithlimits{\sum}_{i=1}^{N_\text{max}^{(e)}} 
 |B_{e n_i}|^2\, \left[m_K^2 (m_{n_i}^2+m_{e}^2) - 
 (m_{n_i}^2-m_{e}^2)^2 \right] \lambda^{1/2}(m_{K} ,m_{n_i}, m_{e})}
{\operatornamewithlimits{\sum}_{j=1}^{N_\text{max}^{(\mu)}} 
 |B_{\mu 
 n_j}|^2\,  \left[m_K^2 (m_{n_j}^2+m_{\mu}^2) - 
 (m_{n_j}^2-m_{\mu}^2)^2 \right] \lambda^{1/2}(m_{K} ,m_{n_j}, m_{\mu})}-1 \,,
\end{equation}
where $N_\text{max}^{e, \mu}$ is the heaviest neutrino mass
eigenstate kinematically allowed in association with $e$ or $\mu$ respectively, and the kinematical function $\lambda(m_{K} ,m_{n_i}, m_{\ell})$ reads~\cite{Abada:2012mc}:

\begin{equation}\label{lambdabc}
\lambda(a,b,c) \, = \, (a^2 - b^2 -c^2)^2 -
4\,b^2\,c^2\,.
\end{equation}

\section{Appendix: The $Z$ invisible decay width}
\label{app:Zinv}

The $Z$ invisible decay width in presence of massive Majorana neutrinos, like it is the case of the present ISS model,  reads~\cite{Abada:2013aba}:
\begin{align}
\Gamma(Z\to{\rm inv.})_{\rm ISS}&=
 \sum_{n} \Gamma(Z\to n n)_{\rm ISS}  \nonumber \\ 
 &=\, \operatornamewithlimits{\sum}_{i\leq j=1}^{N_\text{max}} (1 - \frac{1}{2} \delta_{ij})
\frac{\sqrt{2} G_F }{48\, \pi\, m_Z}  \times \lambda^{1/2}(m_{Z} ,m_{n_i}, m_{n_j}) \nonumber \\
&  \times \left[2  |C_{n_i n_j}|^2 \left(2 m_Z^2 - m_{n_i}^2 - m_{n_j}^2 - \frac{(m_{n_i}^2 - m_{n_j}^2)^2}{m_Z^2}\right) - 12 m_{n_i} 
   m_{n_j} \text{Re}\Big[\big(C_{n_i n_j}\big)^2\Big]  \right].
\label{eq:Znunu:sum}
\end{align}
where $N_\text{max}$ is  the heaviest neutrino mass which is kinematically allowed and $\lambda$ is given in Eq.~(\ref{lambdabc}).

\section{Appendix: Oblique parameters: $S, T, U$}
\label{app:STU}
The Majorana neutrino contributions to the $S, T, U$ parameters have been computed in Ref.~\cite{Akhmedov:2013hec}. We apply those formulas to compute the sterile neutrinos contributions to the oblique parameters in the ISS model.\\
The equation for the $T$ parameter reads:
\begin{align}
\label{eq:Tpar}
T_\text{tot}=T_{\rm ISS}+T_{\text{SM}}&=
\frac{-1}{8\pi s^2_W m_W^2}   \Biggl\{
\sum_{\alpha=1}^{3}  m^2_{\ell_\alpha} B_0 (0,m^2_{\ell_\alpha},m^2_{\ell_\alpha})
- 2\,\sum_{i=1}^{9}\sum_{\alpha=1}^{3} \big|B_{\ell_\alpha n_i}\big|^2 \,Q(0,m^2_{n_i},m^2_{\ell_\alpha}) \nonumber\\
&+ \sum_{i,j=1}^{9}\Big(C_{n_i n_j} C_{n_j n_i} Q(0,m^2_{n_i},m^2_{n_j})
+(C_{n_i n_j})^2 m_{n_i} m_{n_j} B_0(0,m^2_{n_i},m^2_{n_j})\Big) \Bigg\} \,,
\end{align}
where the index $\alpha$ refers to the charged leptons and 
\begin{align}
Q(q^2,&m^2_1,m^2_2)\equiv (D-2)B_{00}(q^2,m_1^2,m_2^2)+q^2\bigl[ 
B_1(q^2,m_1^2,m_2^2)+B_{11}(q^2,m_1^2,m_2^2)
\bigr]\,,
\end{align}
with $D\equiv 4-2\epsilon$  ($\epsilon\rightarrow 0$) and  $B_0$, $B_1$, $B_{11}$ and $B_{00}$ are the Passarino-Veltman functions~\cite{Passarino:1978jh} in the {\it LoopTools}~\cite{Hahn:1998yk} notation. 

The SM contribution can be cast as: 
\begin{equation}
\label{eq:TparSM}
 T_{\text{SM}} = -\frac{1}{8\pi s^2_W m_W^2}  \bigg\{3 Q(0,0,0) - 2 \sum_{\alpha=1}^{3} Q(0,0,m^2_{\ell_\alpha}) +\sum_{\alpha=1}^{3}  m^2_{\ell_\alpha} B_0 (0,m^2_{\ell_\alpha},m^2_{\ell_\alpha})\bigg\}\,,
\end{equation}
where it has been used that the active neutrino masses are zero and the leptonic mixing matrix $U$ is unitary in the SM.

The equation for the $S$ parameter is:
\begin{align}
\label{eq:parS}
S_\text{tot}=S_{\rm ISS}+S_{\text{SM}}&
=-\frac{1}{2\pi m_Z^2} \Bigg\{
 \sum_{i,j=1}^{9} C_{n_i n_j} C_{n_j n_i} \Delta Q(m_Z^2,m^2_{n_i},m^2_{n_j}) \nonumber\\
+& \sum_{i,j=1}^{9}(C_{n_i n_j})^2 m_{n_i} m_{n_j} \Big(B_0(0,m^2_{n_i},m^2_{n_j})-B_0(m_Z^2,m^2_{n_i},m^2_{n_j})\Big)\nonumber\\
+&\sum_{\alpha=1}^{3} m^2_{\ell_\alpha} \big(B_0 (0,m^2_{\ell_\alpha},m^2_{\ell_\alpha})-2B_0 (m_Z^2,m^2_{\ell_\alpha},m^2_{\ell_\alpha})\big) +  Q(m_Z^2,m^2_{\ell_\alpha},m^2_{\ell_\alpha})\Bigg\}\,,
\end{align}
where $\Delta Q(q^2,m^2_1,m^2_2)\equiv Q(0,m^2_1,m^2_2)-Q(q^2,m^2_1,m^2_2)$ and

\begin{align}
\label{eq:SparSM}
 S_{\text{SM}} =& -\frac{1}{2\pi m_Z^2}  \bigg\{
 3 \Delta Q(m_Z^2,0,0) \nonumber\\
& + \sum_{\alpha=1}^{3} m^2_{\ell_\alpha} \big(B_0 (0,m^2_{\ell_\alpha},m^2_{\ell_\alpha})-2B_0 (m_Z^2,m^2_{\ell_\alpha},m^2_{\ell_\alpha})\big) +  Q(m_Z^2,m^2_{\ell_\alpha},m^2_{\ell_\alpha})\bigg\}\,.
\end{align}

Finally, the $U$ parameter is given by:
\begin{align}
\label{eq:parU}
U_\text{tot}=U_{\rm ISS}+U_{\text{SM}}
&=\frac{1}{2\pi m_Z^2}\Bigg\{
 \sum_{i,j=1}^{9} C_{n_i n_j} C_{n_j n_i} \Delta Q(m_Z^2,m^2_{n_i},m^2_{n_j}) \nonumber\\
+& \sum_{i,j=1}^{9}(C_{n_i n_j})^2 m_{n_i} m_{n_j} \Big(B_0(0,m^2_{n_i},m^2_{n_j})-B_0(m_Z^2,m^2_{n_i},m^2_{n_j})\Big)\nonumber\\
-&\sum_{i=1}^9 \sum_{\alpha=1}^3 2\,\frac{m_Z^2}{m_W^2} \big|B_{\ell_\alpha n_i}\big|^2 \Delta Q(m_W^2,m^2_{n_i},m^2_{\ell_\alpha})\nonumber\\
+&\sum_{\alpha=1}^{3} m^2_{\ell_\alpha} \big(B_0 (0,m^2_{\ell_\alpha},m^2_{\ell_\alpha})-2B_0 (m_Z^2,m^2_{\ell_\alpha},m^2_{\ell_\alpha})\big) -  Q(m_Z^2,m^2_{\ell_\alpha},m^2_{\ell_\alpha})\Bigg\}\,,
\end{align}
and its SM contribution reads:
\begin{align}
\label{eq:UparSM}
 U_{\text{SM}} = \frac{1}{2\pi m_Z^2} &\bigg\{3 \Delta Q(m_Z^2,0,0) 
 + \sum_{\alpha=1}^{3}\Big( m^2_{\ell_\alpha} \big(B_0 (0,m^2_{\ell_\alpha},m^2_{\ell_\alpha})
 -2B_0 (m_Z^2,m^2_{\ell_\alpha},m^2_{\ell_\alpha})\big) \nonumber\\
 &-  Q(m_Z^2,m^2_{\ell_\alpha},m^2_{\ell_\alpha})
-2\,\frac{m_Z^2}{m_W^2} \Delta Q(m_W^2,0,m^2_{\ell_\alpha}) \Big)
\bigg\}\,.
\end{align}

\section{LFV Z decays}\label{app:LFVZD}

\begin{figure}[t!]
\begin{center}
\includegraphics[width=\textwidth]{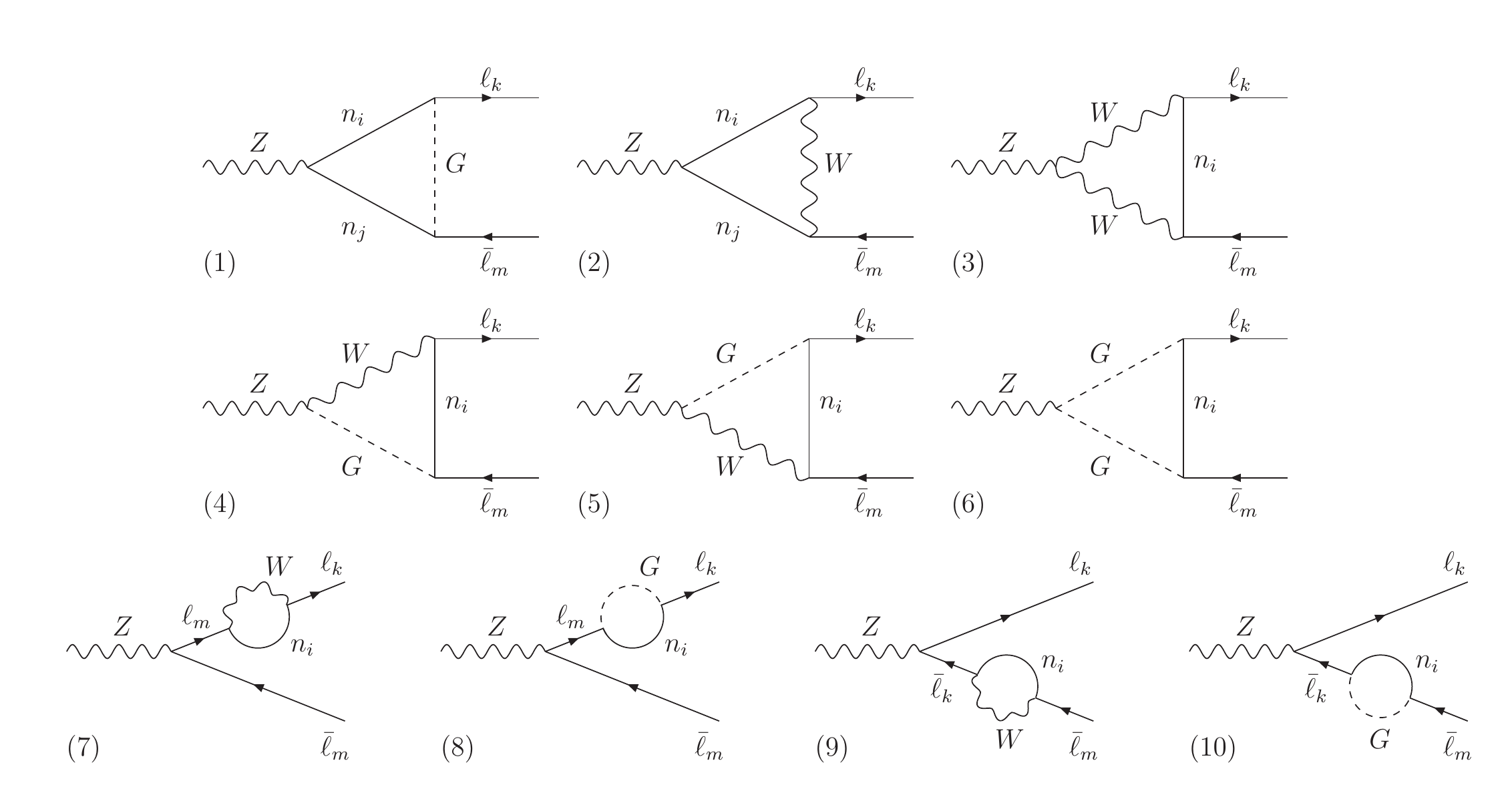}
\caption{One-loop diagrams in the Feynman-t'Hooft gauge for LFV Z decays with massive neutrinos.}\label{Diagrams}
\end{center}
\end{figure}

For completeness, we give here the analytical expressions for LFV Z decay partial widths in the Feynman-t'Hooft gauge, which are obtained by computing the diagrams shown in Figure~ \ref{Diagrams}. We take the results from~\cite{Illana:1999ww,Abada:2014cca} and adapt them to the notation introduced in Sec.~\ref{model} and to the convection of {\it LoopTools}~\cite{Hahn:1998yk} for the loop functions. 
Then, for $k\neq m$, we have

\begin{equation}
\Gamma(Z\to \ell_k \bar\ell_m) = \frac{\alpha_W^3}{192\pi^2 c_W^2}\, m_Z\, \big| \mathcal F_Z \big|^2,
\end{equation}
with
\begin{equation}
\mathcal F_Z = \sum_{a=1}^{10} \mathcal F_{Z}^{(a)}.
\end{equation}
The form factors of the different diagrams are
\begin{equation}
\mathcal F_Z^{(1)} = \frac12\, B_{\ell_k n_i}^{} B^*_{\ell_m n_j}\left\{ -C_{n_i n_j} \,x_i x_j \, m_W^2 C_0 + C_{n_i n_j}^* \sqrt{x_ix_j} \Big[m_Z^2\, C_{12} - 2 C_{00}+\frac12\Big]\right\},
\end{equation}
where $C_{0,12,00}\equiv C_{0,12,00}(0,m_Z^2,0,m_W^2,m_{n_i}^2, m_{n_j}^2)$;
\begin{equation}
\mathcal F_Z^{(2)}  = B_{\ell_k n_i}^{}B^*_{\ell_m n_j}\left\{ -C_{n_i n_j} \Big[m_Z^2 \Big( C_0 + C_1 +C_2 + C_{12} \Big) - 2 C_{00} +1  \Big]+ C_{n_i n_j}^* \sqrt{x_ix_j} \, m_W^2 C_0\right\},
\end{equation}
where $C_{0,1,2,12,00}\equiv C_{0,1,2,12,00}(0,m_Z^2,0,m_W^2,m_{n_i}^2, m_{n_j}^2)$;
\begin{equation}
\mathcal F_Z^{(3)} = 2c_W^2 B_{\ell_k n_i}^{} B^*_{\ell_m n_i}\left\{  m_Z^2 \Big(C_1+C_2+C_{12}\Big)-6 C_{00}+1\right\} ,
\end{equation}
where $C_{1,2,12,00}\equiv C_{1,2,12,00}(0,m_Z^2,0,m_{n_i}^2,m_W^2,m_W^2)$;
\begin{equation}
\mathcal F_Z^{(4)}+\mathcal F_Z^{(5)}  = -2s_W^2\, B_{\ell_k n_i}^{} B^*_{\ell_m n_i}  \, x_i\, m_W^2  C_0 ,
\end{equation}
where $C_{0}\equiv C_{0}(0,m_Z^2,0,m_{n_i}^2,m_W^2,m_W^2)$;
\begin{equation}
\mathcal F_Z^{(6)} = -(1-2s_W^2)\,  B_{\ell_k n_i}^{} B^*_{\ell_m n_i}\,  x_i\, C_{00},
\end{equation}
where  $C_{00}\equiv C_{00}(0,m_Z^2,0,m_{n_i}^2,m_W^2,m_W^2)$;
\begin{equation}
\mathcal F_Z^{(7)}+\mathcal F_Z^{(8)}+\mathcal F_Z^{(9)}+\mathcal F_Z^{(10)}  = \frac12(1-2c_W^2)\, B_{\ell_k n_i}^{} B^*_{\ell_m n_i}\left\{(2+x_i)B_1+1\right\},
\end{equation}
where $B_1\equiv B_1(0,m_{n_i}^2,m_W^2)$.

In all these formulas,  sum over neutrino indices, $i,j=1,... ,9$ has to be understood, $x_i\equiv~m_{n_i}^2/m_W^2$ and  the charged lepton masses have been neglected.

\end{appendix}


\bibliography{bibliography}

\begin{thebibliography}{97}%
\makeatletter
\providecommand \@ifxundefined [1]{%
 \@ifx{#1\undefined}
}%
\providecommand \@ifnum [1]{%
 \ifnum #1\expandafter \@firstoftwo
 \else \expandafter \@secondoftwo
 \fi
}%
\providecommand \@ifx [1]{%
 \ifx #1\expandafter \@firstoftwo
 \else \expandafter \@secondoftwo
 \fi
}%
\providecommand \natexlab [1]{#1}%
\providecommand \enquote  [1]{``#1''}%
\providecommand \bibnamefont  [1]{#1}%
\providecommand \bibfnamefont [1]{#1}%
\providecommand \citenamefont [1]{#1}%
\providecommand \href@noop [0]{\@secondoftwo}%
\providecommand \href [0]{\begingroup \@sanitize@url \@href}%
\providecommand \@href[1]{\@@startlink{#1}\@@href}%
\providecommand \@@href[1]{\endgroup#1\@@endlink}%
\providecommand \@sanitize@url [0]{\catcode `\\12\catcode `\$12\catcode
  `\&12\catcode `\#12\catcode `\^12\catcode `\_12\catcode `\%12\relax}%
\providecommand \@@startlink[1]{}%
\providecommand \@@endlink[0]{}%
\providecommand \url  [0]{\begingroup\@sanitize@url \@url }%
\providecommand \@url [1]{\endgroup\@href {#1}{\urlprefix }}%
\providecommand \urlprefix  [0]{URL }%
\providecommand \Eprint [0]{\href }%
\providecommand \doibase [0]{http://dx.doi.org/}%
\providecommand \selectlanguage [0]{\@gobble}%
\providecommand \bibinfo  [0]{\@secondoftwo}%
\providecommand \bibfield  [0]{\@secondoftwo}%
\providecommand \translation [1]{[#1]}%
\providecommand \BibitemOpen [0]{}%
\providecommand \bibitemStop [0]{}%
\providecommand \bibitemNoStop [0]{.\EOS\space}%
\providecommand \EOS [0]{\spacefactor3000\relax}%
\providecommand \BibitemShut  [1]{\csname bibitem#1\endcsname}%
\let\auto@bib@innerbib\@empty
\bibitem [{\citenamefont {Minkowski}(1977)}]{Minkowski:1977sc}%
  \BibitemOpen
  \bibfield  {author} {\bibinfo {author} {\bibfnamefont {P.}~\bibnamefont
  {Minkowski}},\ }\href {\doibase 10.1016/0370-2693(77)90435-X} {\bibfield
  {journal} {\bibinfo  {journal} {Phys. Lett.}\ }\textbf {\bibinfo {volume}
  {B67}},\ \bibinfo {pages} {421} (\bibinfo {year} {1977})}\BibitemShut
  {NoStop}%
\bibitem [{\citenamefont {Gell-Mann}\ \emph {et~al.}(1979)\citenamefont
  {Gell-Mann}, \citenamefont {Ramond},\ and\ \citenamefont
  {Slansky}}]{GellMann}%
  \BibitemOpen
  \bibfield  {author} {\bibinfo {author} {\bibfnamefont {M.}~\bibnamefont
  {Gell-Mann}}, \bibinfo {author} {\bibfnamefont {P.}~\bibnamefont {Ramond}}, \
  and\ \bibinfo {author} {\bibfnamefont {R.}~\bibnamefont {Slansky}},\ }in\
  \href@noop {} {\emph {\bibinfo {booktitle} {Supergravity proceedings}}},\
  \bibinfo {editor} {edited by\ \bibinfo {editor} {\bibfnamefont {P.~V.}\
  \bibnamefont {Nieuwenhuizen}}\ and\ \bibinfo {editor} {\bibfnamefont {D.~Z.}\
  \bibnamefont {Freedman}}}\ (\bibinfo {year} {1979})\BibitemShut {NoStop}%
\bibitem [{\citenamefont {Yanagida}(1979)}]{Yanagida}%
  \BibitemOpen
  \bibfield  {author} {\bibinfo {author} {\bibfnamefont {T.}~\bibnamefont
  {Yanagida}},\ }in\ \href@noop {} {\emph {\bibinfo {booktitle} {Proceedings of
  the Workshop on the Baryon Number of the Universe and Unified Theories}}},\
  \bibinfo {editor} {edited by\ \bibinfo {editor} {\bibfnamefont
  {O.}~\bibnamefont {Sawada}}\ and\ \bibinfo {editor} {\bibfnamefont
  {A.}~\bibnamefont {Sugamoto}}}\ (\bibinfo {year} {1979})\BibitemShut
  {NoStop}%
\bibitem [{\citenamefont {Mohapatra}\ and\ \citenamefont
  {Senjanovic}(1980)}]{Mohapatra:1979ia}%
  \BibitemOpen
  \bibfield  {author} {\bibinfo {author} {\bibfnamefont {R.~N.}\ \bibnamefont
  {Mohapatra}}\ and\ \bibinfo {author} {\bibfnamefont {G.}~\bibnamefont
  {Senjanovic}},\ }\href {\doibase 10.1103/PhysRevLett.44.912} {\bibfield
  {journal} {\bibinfo  {journal} {Phys. Rev. Lett.}\ }\textbf {\bibinfo
  {volume} {44}},\ \bibinfo {pages} {912} (\bibinfo {year} {1980})}\BibitemShut
  {NoStop}%
\bibitem [{\citenamefont {Schechter}\ and\ \citenamefont
  {Valle}(1980)}]{Schechter:1980gr}%
  \BibitemOpen
  \bibfield  {author} {\bibinfo {author} {\bibfnamefont {J.}~\bibnamefont
  {Schechter}}\ and\ \bibinfo {author} {\bibfnamefont {J.~W.~F.}\ \bibnamefont
  {Valle}},\ }\href {\doibase 10.1103/PhysRevD.22.2227} {\bibfield  {journal}
  {\bibinfo  {journal} {Phys. Rev.}\ }\textbf {\bibinfo {volume} {D22}},\
  \bibinfo {pages} {2227} (\bibinfo {year} {1980})}\BibitemShut {NoStop}%
\bibitem [{\citenamefont {Ilakovac}\ and\ \citenamefont
  {Pilaftsis}(1995)}]{Ilakovac:1994kj}%
  \BibitemOpen
  \bibfield  {author} {\bibinfo {author} {\bibfnamefont {A.}~\bibnamefont
  {Ilakovac}}\ and\ \bibinfo {author} {\bibfnamefont {A.}~\bibnamefont
  {Pilaftsis}},\ }\href {\doibase 10.1016/0550-3213(94)00567-X} {\bibfield
  {journal} {\bibinfo  {journal} {Nucl. Phys.}\ }\textbf {\bibinfo {volume}
  {B437}},\ \bibinfo {pages} {491} (\bibinfo {year} {1995})},\ \Eprint
  {http://arxiv.org/abs/hep-ph/9403398} {arXiv:hep-ph/9403398 [hep-ph]}
  \BibitemShut {NoStop}%
\bibitem [{\citenamefont {Arganda}\ \emph {et~al.}(2005)\citenamefont
  {Arganda}, \citenamefont {Curiel}, \citenamefont {Herrero},\ and\
  \citenamefont {Temes}}]{Arganda:2004bz}%
  \BibitemOpen
  \bibfield  {author} {\bibinfo {author} {\bibfnamefont {E.}~\bibnamefont
  {Arganda}}, \bibinfo {author} {\bibfnamefont {A.~M.}\ \bibnamefont {Curiel}},
  \bibinfo {author} {\bibfnamefont {M.~J.}\ \bibnamefont {Herrero}}, \ and\
  \bibinfo {author} {\bibfnamefont {D.}~\bibnamefont {Temes}},\ }\href
  {\doibase 10.1103/PhysRevD.71.035011} {\bibfield  {journal} {\bibinfo
  {journal} {Phys. Rev.}\ }\textbf {\bibinfo {volume} {D71}},\ \bibinfo {pages}
  {035011} (\bibinfo {year} {2005})},\ \Eprint
  {http://arxiv.org/abs/hep-ph/0407302} {arXiv:hep-ph/0407302 [hep-ph]}
  \BibitemShut {NoStop}%
\bibitem [{\citenamefont {{For a review on Majorana/Dirac neutrinos see, for
  instance, E. Akhmedov}}()}]{Akhmedov:2014kxa}%
  \BibitemOpen
  \bibfield  {author} {\bibinfo {author} {\bibnamefont {{For a review on
  Majorana/Dirac neutrinos see, for instance, E. Akhmedov}}},\ }\href
  {http://inspirehep.net/record/1333680/files/arXiv:1412.3320.pdf} {\ }\Eprint
  {http://arxiv.org/abs/1412.3320} {arXiv:1412.3320 [hep-ph]} \BibitemShut
  {NoStop}%
\bibitem [{\citenamefont {Aad}\ \emph {et~al.}(2014)\citenamefont {Aad} \emph
  {et~al.}}]{Aad:2014bca}%
  \BibitemOpen
  \bibfield  {author} {\bibinfo {author} {\bibfnamefont {G.}~\bibnamefont
  {Aad}} \emph {et~al.} (\bibinfo {collaboration} {ATLAS}),\ }\href {\doibase
  10.1103/PhysRevD.90.072010} {\bibfield  {journal} {\bibinfo  {journal} {Phys.
  Rev.}\ }\textbf {\bibinfo {volume} {D90}},\ \bibinfo {pages} {072010}
  (\bibinfo {year} {2014})},\ \Eprint {http://arxiv.org/abs/1408.5774}
  {arXiv:1408.5774 [hep-ex]} \BibitemShut {NoStop}%
\bibitem [{\citenamefont {Wilson}(1998)}]{Wilson:I}%
  \BibitemOpen
  \bibfield  {author} {\bibinfo {author} {\bibfnamefont {G.}~\bibnamefont
  {Wilson}},\ }in\ \href@noop {} {\emph {\bibinfo {booktitle} {DESY-ECFA LC
  Workshops, Frascati}}}\ (\bibinfo {year} {November 1998})\BibitemShut
  {NoStop}%
\bibitem [{\citenamefont {Wilson}(1999)}]{Wilson:II}%
  \BibitemOpen
  \bibfield  {author} {\bibinfo {author} {\bibfnamefont {G.}~\bibnamefont
  {Wilson}},\ }in\ \href@noop {} {\emph {\bibinfo {booktitle} {DESY-ECFA LC
  Workshops, Oxford}}}\ (\bibinfo {year} {March 1999})\BibitemShut {NoStop}%
\bibitem [{\citenamefont {Blondel}\ \emph {et~al.}(2016)\citenamefont
  {Blondel}, \citenamefont {Graverini}, \citenamefont {Serra},\ and\
  \citenamefont {Shaposhnikov}}]{Blondel:2014bra}%
  \BibitemOpen
  \bibfield  {author} {\bibinfo {author} {\bibfnamefont {A.}~\bibnamefont
  {Blondel}}, \bibinfo {author} {\bibfnamefont {E.}~\bibnamefont {Graverini}},
  \bibinfo {author} {\bibfnamefont {N.}~\bibnamefont {Serra}}, \ and\ \bibinfo
  {author} {\bibfnamefont {M.}~\bibnamefont {Shaposhnikov}} (\bibinfo
  {collaboration} {FCC-ee study Team}),\ }in\ \href {\doibase
  10.1016/j.nuclphysbps.2015.09.304} {\emph {\bibinfo {booktitle}
  {{Proceedings, 37th International Conference on High Energy Physics (ICHEP
  2014)}}}}\ (\bibinfo {year} {2016})\ \Eprint {http://arxiv.org/abs/1411.5230}
  {arXiv:1411.5230 [hep-ex]} \BibitemShut {NoStop}%
\bibitem [{\citenamefont {Mann}\ and\ \citenamefont
  {Riemann}(1984)}]{Mann:1983dv}%
  \BibitemOpen
  \bibfield  {author} {\bibinfo {author} {\bibfnamefont {G.}~\bibnamefont
  {Mann}}\ and\ \bibinfo {author} {\bibfnamefont {T.}~\bibnamefont {Riemann}},\
  }\href@noop {} {\bibfield  {journal} {\bibinfo  {journal} {Annalen Phys.}\
  }\textbf {\bibinfo {volume} {40}},\ \bibinfo {pages} {334} (\bibinfo {year}
  {1984})}\BibitemShut {NoStop}%
\bibitem [{\citenamefont {Bernabeu}\ \emph {et~al.}(1987)\citenamefont
  {Bernabeu}, \citenamefont {Santamaria}, \citenamefont {Vidal}, \citenamefont
  {Mendez},\ and\ \citenamefont {Valle}}]{Bernabeu:1987gr}%
  \BibitemOpen
  \bibfield  {author} {\bibinfo {author} {\bibfnamefont {J.}~\bibnamefont
  {Bernabeu}}, \bibinfo {author} {\bibfnamefont {A.}~\bibnamefont
  {Santamaria}}, \bibinfo {author} {\bibfnamefont {J.}~\bibnamefont {Vidal}},
  \bibinfo {author} {\bibfnamefont {A.}~\bibnamefont {Mendez}}, \ and\ \bibinfo
  {author} {\bibfnamefont {J.~W.~F.}\ \bibnamefont {Valle}},\ }\href {\doibase
  10.1016/0370-2693(87)91100-2} {\bibfield  {journal} {\bibinfo  {journal}
  {Phys. Lett.}\ }\textbf {\bibinfo {volume} {B187}},\ \bibinfo {pages} {303}
  (\bibinfo {year} {1987})}\BibitemShut {NoStop}%
\bibitem [{\citenamefont {Dittmar}\ \emph {et~al.}(1990)\citenamefont
  {Dittmar}, \citenamefont {Santamaria}, \citenamefont {Gonzalez-Garcia},\ and\
  \citenamefont {Valle}}]{Dittmar:1989yg}%
  \BibitemOpen
  \bibfield  {author} {\bibinfo {author} {\bibfnamefont {M.}~\bibnamefont
  {Dittmar}}, \bibinfo {author} {\bibfnamefont {A.}~\bibnamefont {Santamaria}},
  \bibinfo {author} {\bibfnamefont {M.~C.}\ \bibnamefont {Gonzalez-Garcia}}, \
  and\ \bibinfo {author} {\bibfnamefont {J.~W.~F.}\ \bibnamefont {Valle}},\
  }\href {\doibase 10.1016/0550-3213(90)90028-C} {\bibfield  {journal}
  {\bibinfo  {journal} {Nucl. Phys.}\ }\textbf {\bibinfo {volume} {B332}},\
  \bibinfo {pages} {1} (\bibinfo {year} {1990})}\BibitemShut {NoStop}%
\bibitem [{\citenamefont {Korner}\ \emph {et~al.}(1993)\citenamefont {Korner},
  \citenamefont {Pilaftsis},\ and\ \citenamefont {Schilcher}}]{Korner:1992an}%
  \BibitemOpen
  \bibfield  {author} {\bibinfo {author} {\bibfnamefont {J.~G.}\ \bibnamefont
  {Korner}}, \bibinfo {author} {\bibfnamefont {A.}~\bibnamefont {Pilaftsis}}, \
  and\ \bibinfo {author} {\bibfnamefont {K.}~\bibnamefont {Schilcher}},\ }\href
  {\doibase 10.1016/0370-2693(93)91350-V} {\bibfield  {journal} {\bibinfo
  {journal} {Phys. Lett.}\ }\textbf {\bibinfo {volume} {B300}},\ \bibinfo
  {pages} {381} (\bibinfo {year} {1993})},\ \Eprint
  {http://arxiv.org/abs/hep-ph/9301290} {arXiv:hep-ph/9301290 [hep-ph]}
  \BibitemShut {NoStop}%
\bibitem [{\citenamefont {Ilakovac}(2000)}]{Ilakovac:1999md}%
  \BibitemOpen
  \bibfield  {author} {\bibinfo {author} {\bibfnamefont {A.}~\bibnamefont
  {Ilakovac}},\ }\href {\doibase 10.1103/PhysRevD.62.036010} {\bibfield
  {journal} {\bibinfo  {journal} {Phys. Rev.}\ }\textbf {\bibinfo {volume}
  {D62}},\ \bibinfo {pages} {036010} (\bibinfo {year} {2000})},\ \Eprint
  {http://arxiv.org/abs/hep-ph/9910213} {arXiv:hep-ph/9910213 [hep-ph]}
  \BibitemShut {NoStop}%
\bibitem [{\citenamefont {Illana}\ \emph {et~al.}(1999)\citenamefont {Illana},
  \citenamefont {Jack},\ and\ \citenamefont {Riemann}}]{Illana:1999ww}%
  \BibitemOpen
  \bibfield  {author} {\bibinfo {author} {\bibfnamefont {J.~I.}\ \bibnamefont
  {Illana}}, \bibinfo {author} {\bibfnamefont {M.}~\bibnamefont {Jack}}, \ and\
  \bibinfo {author} {\bibfnamefont {T.}~\bibnamefont {Riemann}}\ }(\bibinfo
  {year} {1999})\ \Eprint {http://arxiv.org/abs/hep-ph/0001273}
  {arXiv:hep-ph/0001273 [hep-ph]} \BibitemShut {NoStop}%
\bibitem [{\citenamefont {Illana}\ and\ \citenamefont
  {Riemann}(2001)}]{Illana:2000ic}%
  \BibitemOpen
  \bibfield  {author} {\bibinfo {author} {\bibfnamefont {J.~I.}\ \bibnamefont
  {Illana}}\ and\ \bibinfo {author} {\bibfnamefont {T.}~\bibnamefont
  {Riemann}},\ }\href {\doibase 10.1103/PhysRevD.63.053004} {\bibfield
  {journal} {\bibinfo  {journal} {Phys. Rev.}\ }\textbf {\bibinfo {volume}
  {D63}},\ \bibinfo {pages} {053004} (\bibinfo {year} {2001})},\ \Eprint
  {http://arxiv.org/abs/hep-ph/0010193} {arXiv:hep-ph/0010193 [hep-ph]}
  \BibitemShut {NoStop}%
\bibitem [{\citenamefont {Perez}\ \emph {et~al.}(2004)\citenamefont {Perez},
  \citenamefont {Tavares-Velasco},\ and\ \citenamefont
  {Toscano}}]{Perez:2003ad}%
  \BibitemOpen
  \bibfield  {author} {\bibinfo {author} {\bibfnamefont {M.~A.}\ \bibnamefont
  {Perez}}, \bibinfo {author} {\bibfnamefont {G.}~\bibnamefont
  {Tavares-Velasco}}, \ and\ \bibinfo {author} {\bibfnamefont {J.~J.}\
  \bibnamefont {Toscano}},\ }\href {\doibase 10.1142/S0217751X04017100}
  {\bibfield  {journal} {\bibinfo  {journal} {Int. J. Mod. Phys.}\ }\textbf
  {\bibinfo {volume} {A19}},\ \bibinfo {pages} {159} (\bibinfo {year}
  {2004})},\ \Eprint {http://arxiv.org/abs/hep-ph/0305227}
  {arXiv:hep-ph/0305227 [hep-ph]} \BibitemShut {NoStop}%
\bibitem [{\citenamefont {Flores-Tlalpa}\ \emph {et~al.}(2002)\citenamefont
  {Flores-Tlalpa}, \citenamefont {Hernandez}, \citenamefont {Tavares-Velasco},\
  and\ \citenamefont {Toscano}}]{FloresTlalpa:2001sp}%
  \BibitemOpen
  \bibfield  {author} {\bibinfo {author} {\bibfnamefont {A.}~\bibnamefont
  {Flores-Tlalpa}}, \bibinfo {author} {\bibfnamefont {J.~M.}\ \bibnamefont
  {Hernandez}}, \bibinfo {author} {\bibfnamefont {G.}~\bibnamefont
  {Tavares-Velasco}}, \ and\ \bibinfo {author} {\bibfnamefont {J.~J.}\
  \bibnamefont {Toscano}},\ }\href {\doibase 10.1103/PhysRevD.65.073010}
  {\bibfield  {journal} {\bibinfo  {journal} {Phys. Rev.}\ }\textbf {\bibinfo
  {volume} {D65}},\ \bibinfo {pages} {073010} (\bibinfo {year} {2002})},\
  \Eprint {http://arxiv.org/abs/hep-ph/0112065} {arXiv:hep-ph/0112065 [hep-ph]}
  \BibitemShut {NoStop}%
\bibitem [{\citenamefont {Delepine}\ and\ \citenamefont
  {Vissani}(2001)}]{Delepine:2001di}%
  \BibitemOpen
  \bibfield  {author} {\bibinfo {author} {\bibfnamefont {D.}~\bibnamefont
  {Delepine}}\ and\ \bibinfo {author} {\bibfnamefont {F.}~\bibnamefont
  {Vissani}},\ }\href {\doibase 10.1016/S0370-2693(01)01254-0} {\bibfield
  {journal} {\bibinfo  {journal} {Phys. Lett.}\ }\textbf {\bibinfo {volume}
  {B522}},\ \bibinfo {pages} {95} (\bibinfo {year} {2001})},\ \Eprint
  {http://arxiv.org/abs/hep-ph/0106287} {arXiv:hep-ph/0106287 [hep-ph]}
  \BibitemShut {NoStop}%
\bibitem [{\citenamefont {Davidson}\ \emph {et~al.}(2012)\citenamefont
  {Davidson}, \citenamefont {Lacroix},\ and\ \citenamefont
  {Verdier}}]{Davidson:2012wn}%
  \BibitemOpen
  \bibfield  {author} {\bibinfo {author} {\bibfnamefont {S.}~\bibnamefont
  {Davidson}}, \bibinfo {author} {\bibfnamefont {S.}~\bibnamefont {Lacroix}}, \
  and\ \bibinfo {author} {\bibfnamefont {P.}~\bibnamefont {Verdier}},\ }\href
  {\doibase 10.1007/JHEP09(2012)092} {\bibfield  {journal} {\bibinfo  {journal}
  {JHEP}\ }\textbf {\bibinfo {volume} {09}},\ \bibinfo {pages} {092} (\bibinfo
  {year} {2012})},\ \Eprint {http://arxiv.org/abs/1207.4894} {arXiv:1207.4894
  [hep-ph]} \BibitemShut {NoStop}%
\bibitem [{\citenamefont {Baldini}\ \emph {et~al.}(2016)\citenamefont {Baldini}
  \emph {et~al.}}]{TheMEG:2016wtm}%
  \BibitemOpen
  \bibfield  {author} {\bibinfo {author} {\bibfnamefont {A.~M.}\ \bibnamefont
  {Baldini}} \emph {et~al.} (\bibinfo {collaboration} {MEG}),\ }\href {\doibase
  10.1140/epjc/s10052-016-4271-x} {\bibfield  {journal} {\bibinfo  {journal}
  {Eur. Phys. J.}\ }\textbf {\bibinfo {volume} {C76}},\ \bibinfo {pages} {434}
  (\bibinfo {year} {2016})},\ \Eprint {http://arxiv.org/abs/1605.05081}
  {arXiv:1605.05081 [hep-ex]} \BibitemShut {NoStop}%
\bibitem [{\citenamefont {Baldini}\ \emph {et~al.}(2013)\citenamefont {Baldini}
  \emph {et~al.}}]{Baldini:2013ke}%
  \BibitemOpen
  \bibfield  {author} {\bibinfo {author} {\bibfnamefont {A.~M.}\ \bibnamefont
  {Baldini}} \emph {et~al.},\ }\href@noop {} {\  (\bibinfo {year} {2013})},\
  \Eprint {http://arxiv.org/abs/1301.7225} {arXiv:1301.7225 [physics.ins-det]}
  \BibitemShut {NoStop}%
\bibitem [{\citenamefont {Aubert}\ \emph {et~al.}(2010)\citenamefont {Aubert}
  \emph {et~al.}}]{Aubert:2009ag}%
  \BibitemOpen
  \bibfield  {author} {\bibinfo {author} {\bibfnamefont {B.}~\bibnamefont
  {Aubert}} \emph {et~al.} (\bibinfo {collaboration} {BaBar}),\ }\href
  {\doibase 10.1103/PhysRevLett.104.021802} {\bibfield  {journal} {\bibinfo
  {journal} {Phys. Rev. Lett.}\ }\textbf {\bibinfo {volume} {104}},\ \bibinfo
  {pages} {021802} (\bibinfo {year} {2010})},\ \Eprint
  {http://arxiv.org/abs/0908.2381} {arXiv:0908.2381 [hep-ex]} \BibitemShut
  {NoStop}%
\bibitem [{\citenamefont {Aushev}\ \emph {et~al.}(2010)\citenamefont {Aushev}
  \emph {et~al.}}]{Aushev:2010bq}%
  \BibitemOpen
  \bibfield  {author} {\bibinfo {author} {\bibfnamefont {T.}~\bibnamefont
  {Aushev}} \emph {et~al.},\ }\href@noop {} {\  (\bibinfo {year} {2010})},\
  \Eprint {http://arxiv.org/abs/1002.5012} {arXiv:1002.5012 [hep-ex]}
  \BibitemShut {NoStop}%
\bibitem [{\citenamefont {Bellgardt}\ \emph {et~al.}(1988)\citenamefont
  {Bellgardt} \emph {et~al.}}]{Bellgardt:1987du}%
  \BibitemOpen
  \bibfield  {author} {\bibinfo {author} {\bibfnamefont {U.}~\bibnamefont
  {Bellgardt}} \emph {et~al.} (\bibinfo {collaboration} {SINDRUM}),\ }\href
  {\doibase 10.1016/0550-3213(88)90462-2} {\bibfield  {journal} {\bibinfo
  {journal} {Nucl. Phys.}\ }\textbf {\bibinfo {volume} {B299}},\ \bibinfo
  {pages} {1} (\bibinfo {year} {1988})}\BibitemShut {NoStop}%
\bibitem [{\citenamefont {Blondel}\ \emph {et~al.}(2013)\citenamefont {Blondel}
  \emph {et~al.}}]{Blondel:2013ia}%
  \BibitemOpen
  \bibfield  {author} {\bibinfo {author} {\bibfnamefont {A.}~\bibnamefont
  {Blondel}} \emph {et~al.},\ }\href@noop {} {\  (\bibinfo {year} {2013})},\
  \Eprint {http://arxiv.org/abs/1301.6113} {arXiv:1301.6113 [physics.ins-det]}
  \BibitemShut {NoStop}%
\bibitem [{\citenamefont {Hayasaka}\ \emph {et~al.}(2010)\citenamefont
  {Hayasaka} \emph {et~al.}}]{Hayasaka:2010np}%
  \BibitemOpen
  \bibfield  {author} {\bibinfo {author} {\bibfnamefont {K.}~\bibnamefont
  {Hayasaka}} \emph {et~al.},\ }\href {\doibase 10.1016/j.physletb.2010.03.037}
  {\bibfield  {journal} {\bibinfo  {journal} {Phys. Lett.}\ }\textbf {\bibinfo
  {volume} {B687}},\ \bibinfo {pages} {139} (\bibinfo {year} {2010})},\ \Eprint
  {http://arxiv.org/abs/1001.3221} {arXiv:1001.3221 [hep-ex]} \BibitemShut
  {NoStop}%
\bibitem [{\citenamefont {Hayasaka}(2010)}]{Hayasaka:2010et}%
  \BibitemOpen
  \bibfield  {author} {\bibinfo {author} {\bibfnamefont {K.}~\bibnamefont
  {Hayasaka}},\ }in\ \href
  {https://inspirehep.net/record/873396/files/arXiv:1010.3746.pdf} {\emph
  {\bibinfo {booktitle} {{22nd Rencontres de Blois on Particle Physics and
  Cosmology Blois, Loire Valley, France, July 15-20, 2010}}}}\ (\bibinfo {year}
  {2010})\ \Eprint {http://arxiv.org/abs/1010.3746} {arXiv:1010.3746 [hep-ex]}
  \BibitemShut {NoStop}%
\bibitem [{\citenamefont {Bertl}\ \emph {et~al.}(2006)\citenamefont {Bertl}
  \emph {et~al.}}]{Bertl:2006up}%
  \BibitemOpen
  \bibfield  {author} {\bibinfo {author} {\bibfnamefont {W.~H.}\ \bibnamefont
  {Bertl}} \emph {et~al.} (\bibinfo {collaboration} {SINDRUM II}),\ }\href
  {\doibase 10.1140/epjc/s2006-02582-x} {\bibfield  {journal} {\bibinfo
  {journal} {Eur. Phys. J.}\ }\textbf {\bibinfo {volume} {C47}},\ \bibinfo
  {pages} {337} (\bibinfo {year} {2006})}\BibitemShut {NoStop}%
\bibitem [{\citenamefont {Dohmen}\ \emph {et~al.}(1993)\citenamefont {Dohmen}
  \emph {et~al.}}]{Dohmen:1993mp}%
  \BibitemOpen
  \bibfield  {author} {\bibinfo {author} {\bibfnamefont {C.}~\bibnamefont
  {Dohmen}} \emph {et~al.} (\bibinfo {collaboration} {SINDRUM II}),\ }\href
  {\doibase 10.1016/0370-2693(93)91383-X} {\bibfield  {journal} {\bibinfo
  {journal} {Phys. Lett.}\ }\textbf {\bibinfo {volume} {B317}},\ \bibinfo
  {pages} {631} (\bibinfo {year} {1993})}\BibitemShut {NoStop}%
\bibitem [{\citenamefont {Alekou}\ \emph {et~al.}(2013)\citenamefont {Alekou}
  \emph {et~al.}}]{Alekou:2013eta}%
  \BibitemOpen
  \bibfield  {author} {\bibinfo {author} {\bibfnamefont {A.}~\bibnamefont
  {Alekou}} \emph {et~al.},\ }in\ \href
  {https://inspirehep.net/record/1256506/files/arXiv:1310.0804.pdf} {\emph
  {\bibinfo {booktitle} {{Community Summer Study 2013: Snowmass on the
  Mississippi (CSS2013) Minneapolis, MN, USA, July 29-August 6, 2013}}}}\
  (\bibinfo {year} {2013})\ \Eprint {http://arxiv.org/abs/1310.0804}
  {arXiv:1310.0804 [physics.acc-ph]} \BibitemShut {NoStop}%
\bibitem [{\citenamefont {Kuno}(2013)}]{Kuno:2013mha}%
  \BibitemOpen
  \bibfield  {author} {\bibinfo {author} {\bibfnamefont {Y.}~\bibnamefont
  {Kuno}} (\bibinfo {collaboration} {COMET}),\ }\href {\doibase
  10.1093/ptep/pts089} {\bibfield  {journal} {\bibinfo  {journal} {PTEP}\
  }\textbf {\bibinfo {volume} {2013}},\ \bibinfo {pages} {022C01} (\bibinfo
  {year} {2013})}\BibitemShut {NoStop}%
\bibitem [{\citenamefont {Carey}\ \emph {et~al.}(2008)\citenamefont {Carey}
  \emph {et~al.}}]{Carey:2008zz}%
  \BibitemOpen
  \bibfield  {author} {\bibinfo {author} {\bibfnamefont {R.~M.}\ \bibnamefont
  {Carey}} \emph {et~al.} (\bibinfo {collaboration} {Mu2e}),\ }\href@noop {} {}
  (\bibinfo {year} {2008}),\ \bibinfo {note} {{FERMILAB-PROPOSAL-0973
  (2008)}}\BibitemShut {NoStop}%
\bibitem [{\citenamefont {Akers}\ \emph {et~al.}(1995)\citenamefont {Akers}
  \emph {et~al.}}]{Akers:1995gz}%
  \BibitemOpen
  \bibfield  {author} {\bibinfo {author} {\bibfnamefont {R.}~\bibnamefont
  {Akers}} \emph {et~al.} (\bibinfo {collaboration} {OPAL}),\ }\href {\doibase
  10.1007/BF01553981} {\bibfield  {journal} {\bibinfo  {journal} {Z. Phys.}\
  }\textbf {\bibinfo {volume} {C67}},\ \bibinfo {pages} {555} (\bibinfo {year}
  {1995})}\BibitemShut {NoStop}%
\bibitem [{\citenamefont {Abreu}\ \emph {et~al.}(1997)\citenamefont {Abreu}
  \emph {et~al.}}]{Abreu:1996mj}%
  \BibitemOpen
  \bibfield  {author} {\bibinfo {author} {\bibfnamefont {P.}~\bibnamefont
  {Abreu}} \emph {et~al.} (\bibinfo {collaboration} {DELPHI}),\ }\href
  {\doibase 10.1007/s002880050313} {\bibfield  {journal} {\bibinfo  {journal}
  {Z. Phys.}\ }\textbf {\bibinfo {volume} {C73}},\ \bibinfo {pages} {243}
  (\bibinfo {year} {1997})}\BibitemShut {NoStop}%
\bibitem [{\citenamefont {Aad}\ \emph {et~al.}(2016)\citenamefont {Aad} \emph
  {et~al.}}]{Aad:2016blu}%
  \BibitemOpen
  \bibfield  {author} {\bibinfo {author} {\bibfnamefont {G.}~\bibnamefont
  {Aad}} \emph {et~al.} (\bibinfo {collaboration} {ATLAS}),\ }\href@noop {} {\
  (\bibinfo {year} {2016})},\ \Eprint {http://arxiv.org/abs/1604.07730}
  {arXiv:1604.07730 [hep-ex]} \BibitemShut {NoStop}%
\bibitem [{\citenamefont {{CMS Collaboration}}(2015)}]{CMS:2015udp}%
  \BibitemOpen
  \bibfield  {author} {\bibinfo {author} {\bibnamefont {{CMS Collaboration}}}
  (\bibinfo {collaboration} {CMS}),\ }\href@noop {} {} (\bibinfo {year}
  {2015}),\ \bibinfo {note} {{CMS-PAS-HIG-14-040} (2015)}\BibitemShut {NoStop}%
\bibitem [{\citenamefont {Khachatryan}\ \emph {et~al.}(2015)\citenamefont
  {Khachatryan} \emph {et~al.}}]{Khachatryan:2015kon}%
  \BibitemOpen
  \bibfield  {author} {\bibinfo {author} {\bibfnamefont {V.}~\bibnamefont
  {Khachatryan}} \emph {et~al.} (\bibinfo {collaboration} {CMS}),\ }\href
  {\doibase 10.1016/j.physletb.2015.07.053} {\bibfield  {journal} {\bibinfo
  {journal} {Phys. Lett.}\ }\textbf {\bibinfo {volume} {B749}},\ \bibinfo
  {pages} {337} (\bibinfo {year} {2015})},\ \Eprint
  {http://arxiv.org/abs/1502.07400} {arXiv:1502.07400 [hep-ex]} \BibitemShut
  {NoStop}%
\bibitem [{\citenamefont {Mohapatra}(1986)}]{Mohapatra:1986aw}%
  \BibitemOpen
  \bibfield  {author} {\bibinfo {author} {\bibfnamefont {R.~N.}\ \bibnamefont
  {Mohapatra}},\ }\href {\doibase 10.1103/PhysRevLett.56.561} {\bibfield
  {journal} {\bibinfo  {journal} {Phys. Rev. Lett.}\ }\textbf {\bibinfo
  {volume} {56}},\ \bibinfo {pages} {561} (\bibinfo {year} {1986})}\BibitemShut
  {NoStop}%
\bibitem [{\citenamefont {Mohapatra}\ and\ \citenamefont
  {Valle}(1986)}]{Mohapatra:1986bd}%
  \BibitemOpen
  \bibfield  {author} {\bibinfo {author} {\bibfnamefont {R.~N.}\ \bibnamefont
  {Mohapatra}}\ and\ \bibinfo {author} {\bibfnamefont {J.~W.~F.}\ \bibnamefont
  {Valle}},\ }\bibfield  {booktitle} {\emph {\bibinfo {booktitle}
  {{Proceedings, 23RD International Conference on High Energy Physics, JULY
  16-23, 1986, Berkeley, CA}}},\ }\href {\doibase 10.1103/PhysRevD.34.1642}
  {\bibfield  {journal} {\bibinfo  {journal} {Phys. Rev.}\ }\textbf {\bibinfo
  {volume} {D34}},\ \bibinfo {pages} {1642} (\bibinfo {year}
  {1986})}\BibitemShut {NoStop}%
\bibitem [{\citenamefont {Gonzalez-Garcia}\ and\ \citenamefont
  {Valle}(1992)}]{GonzalezGarcia:1991be}%
  \BibitemOpen
  \bibfield  {author} {\bibinfo {author} {\bibfnamefont {M.~C.}\ \bibnamefont
  {Gonzalez-Garcia}}\ and\ \bibinfo {author} {\bibfnamefont {J.~W.~F.}\
  \bibnamefont {Valle}},\ }\href {\doibase 10.1142/S0217732392000434}
  {\bibfield  {journal} {\bibinfo  {journal} {Mod. Phys. Lett.}\ }\textbf
  {\bibinfo {volume} {A7}},\ \bibinfo {pages} {477} (\bibinfo {year}
  {1992})}\BibitemShut {NoStop}%
\bibitem [{\citenamefont {Abada}\ \emph {et~al.}(2013)\citenamefont {Abada},
  \citenamefont {Das}, \citenamefont {Teixeira}, \citenamefont {Vicente},\ and\
  \citenamefont {Weiland}}]{Abada:2012mc}%
  \BibitemOpen
  \bibfield  {author} {\bibinfo {author} {\bibfnamefont {A.}~\bibnamefont
  {Abada}}, \bibinfo {author} {\bibfnamefont {D.}~\bibnamefont {Das}}, \bibinfo
  {author} {\bibfnamefont {A.~M.}\ \bibnamefont {Teixeira}}, \bibinfo {author}
  {\bibfnamefont {A.}~\bibnamefont {Vicente}}, \ and\ \bibinfo {author}
  {\bibfnamefont {C.}~\bibnamefont {Weiland}},\ }\href {\doibase
  10.1007/JHEP02(2013)048} {\bibfield  {journal} {\bibinfo  {journal} {JHEP}\
  }\textbf {\bibinfo {volume} {02}},\ \bibinfo {pages} {048} (\bibinfo {year}
  {2013})},\ \Eprint {http://arxiv.org/abs/1211.3052} {arXiv:1211.3052
  [hep-ph]} \BibitemShut {NoStop}%
\bibitem [{\citenamefont {Abada}\ \emph
  {et~al.}(2014{\natexlab{a}})\citenamefont {Abada}, \citenamefont {Teixeira},
  \citenamefont {Vicente},\ and\ \citenamefont {Weiland}}]{Abada:2013aba}%
  \BibitemOpen
  \bibfield  {author} {\bibinfo {author} {\bibfnamefont {A.}~\bibnamefont
  {Abada}}, \bibinfo {author} {\bibfnamefont {A.~M.}\ \bibnamefont {Teixeira}},
  \bibinfo {author} {\bibfnamefont {A.}~\bibnamefont {Vicente}}, \ and\
  \bibinfo {author} {\bibfnamefont {C.}~\bibnamefont {Weiland}},\ }\href
  {\doibase 10.1007/JHEP02(2014)091} {\bibfield  {journal} {\bibinfo  {journal}
  {JHEP}\ }\textbf {\bibinfo {volume} {02}},\ \bibinfo {pages} {091} (\bibinfo
  {year} {2014}{\natexlab{a}})},\ \Eprint {http://arxiv.org/abs/1311.2830}
  {arXiv:1311.2830 [hep-ph]} \BibitemShut {NoStop}%
\bibitem [{\citenamefont {Abada}\ and\ \citenamefont
  {Toma}(2016{\natexlab{a}})}]{Abada:2015trh}%
  \BibitemOpen
  \bibfield  {author} {\bibinfo {author} {\bibfnamefont {A.}~\bibnamefont
  {Abada}}\ and\ \bibinfo {author} {\bibfnamefont {T.}~\bibnamefont {Toma}},\
  }\href {\doibase 10.1007/JHEP02(2016)174} {\bibfield  {journal} {\bibinfo
  {journal} {JHEP}\ }\textbf {\bibinfo {volume} {02}},\ \bibinfo {pages} {174}
  (\bibinfo {year} {2016}{\natexlab{a}})},\ \Eprint
  {http://arxiv.org/abs/1511.03265} {arXiv:1511.03265 [hep-ph]} \BibitemShut
  {NoStop}%
\bibitem [{\citenamefont {Abada}\ and\ \citenamefont
  {Toma}(2016{\natexlab{b}})}]{Abada:2016awd}%
  \BibitemOpen
  \bibfield  {author} {\bibinfo {author} {\bibfnamefont {A.}~\bibnamefont
  {Abada}}\ and\ \bibinfo {author} {\bibfnamefont {T.}~\bibnamefont {Toma}},\
  }\href@noop {} {\  (\bibinfo {year} {2016}{\natexlab{b}})},\ \Eprint
  {http://arxiv.org/abs/1605.07643} {arXiv:1605.07643 [hep-ph]} \BibitemShut
  {NoStop}%
\bibitem [{\citenamefont {Abada}\ \emph
  {et~al.}(2014{\natexlab{b}})\citenamefont {Abada}, \citenamefont
  {De~Romeri},\ and\ \citenamefont {Teixeira}}]{Abada:2014nwa}%
  \BibitemOpen
  \bibfield  {author} {\bibinfo {author} {\bibfnamefont {A.}~\bibnamefont
  {Abada}}, \bibinfo {author} {\bibfnamefont {V.}~\bibnamefont {De~Romeri}}, \
  and\ \bibinfo {author} {\bibfnamefont {A.~M.}\ \bibnamefont {Teixeira}},\
  }\href {\doibase 10.1007/JHEP09(2014)074} {\bibfield  {journal} {\bibinfo
  {journal} {JHEP}\ }\textbf {\bibinfo {volume} {09}},\ \bibinfo {pages} {074}
  (\bibinfo {year} {2014}{\natexlab{b}})},\ \Eprint
  {http://arxiv.org/abs/1406.6978} {arXiv:1406.6978 [hep-ph]} \BibitemShut
  {NoStop}%
\bibitem [{\citenamefont {Chen}\ and\ \citenamefont {Dev}(2012)}]{Chen:2011hc}%
  \BibitemOpen
  \bibfield  {author} {\bibinfo {author} {\bibfnamefont {C.-Y.}\ \bibnamefont
  {Chen}}\ and\ \bibinfo {author} {\bibfnamefont {P.~S.~B.}\ \bibnamefont
  {Dev}},\ }\href {\doibase 10.1103/PhysRevD.85.093018} {\bibfield  {journal}
  {\bibinfo  {journal} {Phys. Rev.}\ }\textbf {\bibinfo {volume} {D85}},\
  \bibinfo {pages} {093018} (\bibinfo {year} {2012})},\ \Eprint
  {http://arxiv.org/abs/1112.6419} {arXiv:1112.6419 [hep-ph]} \BibitemShut
  {NoStop}%
\bibitem [{\citenamefont {Bhupal~Dev}\ \emph {et~al.}(2012)\citenamefont
  {Bhupal~Dev}, \citenamefont {Franceschini},\ and\ \citenamefont
  {Mohapatra}}]{BhupalDev:2012zg}%
  \BibitemOpen
  \bibfield  {author} {\bibinfo {author} {\bibfnamefont {P.~S.}\ \bibnamefont
  {Bhupal~Dev}}, \bibinfo {author} {\bibfnamefont {R.}~\bibnamefont
  {Franceschini}}, \ and\ \bibinfo {author} {\bibfnamefont {R.~N.}\
  \bibnamefont {Mohapatra}},\ }\href {\doibase 10.1103/PhysRevD.86.093010}
  {\bibfield  {journal} {\bibinfo  {journal} {Phys. Rev.}\ }\textbf {\bibinfo
  {volume} {D86}},\ \bibinfo {pages} {093010} (\bibinfo {year} {2012})},\
  \Eprint {http://arxiv.org/abs/1207.2756} {arXiv:1207.2756 [hep-ph]}
  \BibitemShut {NoStop}%
\bibitem [{\citenamefont {Das}\ and\ \citenamefont {Okada}(2013)}]{Das:2012ze}%
  \BibitemOpen
  \bibfield  {author} {\bibinfo {author} {\bibfnamefont {A.}~\bibnamefont
  {Das}}\ and\ \bibinfo {author} {\bibfnamefont {N.}~\bibnamefont {Okada}},\
  }\href {\doibase 10.1103/PhysRevD.88.113001} {\bibfield  {journal} {\bibinfo
  {journal} {Phys. Rev.}\ }\textbf {\bibinfo {volume} {D88}},\ \bibinfo {pages}
  {113001} (\bibinfo {year} {2013})},\ \Eprint {http://arxiv.org/abs/1207.3734}
  {arXiv:1207.3734 [hep-ph]} \BibitemShut {NoStop}%
\bibitem [{\citenamefont {Das}\ \emph {et~al.}(2014)\citenamefont {Das},
  \citenamefont {Bhupal~Dev},\ and\ \citenamefont {Okada}}]{Das:2014jxa}%
  \BibitemOpen
  \bibfield  {author} {\bibinfo {author} {\bibfnamefont {A.}~\bibnamefont
  {Das}}, \bibinfo {author} {\bibfnamefont {P.~S.}\ \bibnamefont {Bhupal~Dev}},
  \ and\ \bibinfo {author} {\bibfnamefont {N.}~\bibnamefont {Okada}},\ }\href
  {\doibase 10.1016/j.physletb.2014.06.058} {\bibfield  {journal} {\bibinfo
  {journal} {Phys. Lett.}\ }\textbf {\bibinfo {volume} {B735}},\ \bibinfo
  {pages} {364} (\bibinfo {year} {2014})},\ \Eprint
  {http://arxiv.org/abs/1405.0177} {arXiv:1405.0177 [hep-ph]} \BibitemShut
  {NoStop}%
\bibitem [{\citenamefont {Arganda}\ \emph
  {et~al.}(2016{\natexlab{a}})\citenamefont {Arganda}, \citenamefont {Herrero},
  \citenamefont {Marcano},\ and\ \citenamefont {Weiland}}]{Arganda:2015ija}%
  \BibitemOpen
  \bibfield  {author} {\bibinfo {author} {\bibfnamefont {E.}~\bibnamefont
  {Arganda}}, \bibinfo {author} {\bibfnamefont {M.~J.}\ \bibnamefont
  {Herrero}}, \bibinfo {author} {\bibfnamefont {X.}~\bibnamefont {Marcano}}, \
  and\ \bibinfo {author} {\bibfnamefont {C.}~\bibnamefont {Weiland}},\ }\href
  {\doibase 10.1016/j.physletb.2015.11.013} {\bibfield  {journal} {\bibinfo
  {journal} {Phys. Lett.}\ }\textbf {\bibinfo {volume} {B752}},\ \bibinfo
  {pages} {46} (\bibinfo {year} {2016}{\natexlab{a}})},\ \Eprint
  {http://arxiv.org/abs/1508.05074} {arXiv:1508.05074 [hep-ph]} \BibitemShut
  {NoStop}%
\bibitem [{\citenamefont {Das}\ and\ \citenamefont
  {Okada}(2016)}]{Das:2015toa}%
  \BibitemOpen
  \bibfield  {author} {\bibinfo {author} {\bibfnamefont {A.}~\bibnamefont
  {Das}}\ and\ \bibinfo {author} {\bibfnamefont {N.}~\bibnamefont {Okada}},\
  }\href {\doibase 10.1103/PhysRevD.93.033003} {\bibfield  {journal} {\bibinfo
  {journal} {Phys. Rev.}\ }\textbf {\bibinfo {volume} {D93}},\ \bibinfo {pages}
  {033003} (\bibinfo {year} {2016})},\ \Eprint
  {http://arxiv.org/abs/1510.04790} {arXiv:1510.04790 [hep-ph]} \BibitemShut
  {NoStop}%
\bibitem [{\citenamefont {Das}\ \emph {et~al.}(2016)\citenamefont {Das},
  \citenamefont {Konar},\ and\ \citenamefont {Majhi}}]{Das:2016hof}%
  \BibitemOpen
  \bibfield  {author} {\bibinfo {author} {\bibfnamefont {A.}~\bibnamefont
  {Das}}, \bibinfo {author} {\bibfnamefont {P.}~\bibnamefont {Konar}}, \ and\
  \bibinfo {author} {\bibfnamefont {S.}~\bibnamefont {Majhi}},\ }\href
  {\doibase 10.1007/JHEP06(2016)019} {\bibfield  {journal} {\bibinfo  {journal}
  {JHEP}\ }\textbf {\bibinfo {volume} {06}},\ \bibinfo {pages} {019} (\bibinfo
  {year} {2016})},\ \Eprint {http://arxiv.org/abs/1604.00608} {arXiv:1604.00608
  [hep-ph]} \BibitemShut {NoStop}%
\bibitem [{\citenamefont {Abada}\ \emph
  {et~al.}(2014{\natexlab{c}})\citenamefont {Abada}, \citenamefont {Arcadi},\
  and\ \citenamefont {Lucente}}]{Abada:2014zra}%
  \BibitemOpen
  \bibfield  {author} {\bibinfo {author} {\bibfnamefont {A.}~\bibnamefont
  {Abada}}, \bibinfo {author} {\bibfnamefont {G.}~\bibnamefont {Arcadi}}, \
  and\ \bibinfo {author} {\bibfnamefont {M.}~\bibnamefont {Lucente}},\ }\href
  {\doibase 10.1088/1475-7516/2014/10/001} {\bibfield  {journal} {\bibinfo
  {journal} {JCAP}\ }\textbf {\bibinfo {volume} {1410}},\ \bibinfo {pages}
  {001} (\bibinfo {year} {2014}{\natexlab{c}})},\ \Eprint
  {http://arxiv.org/abs/1406.6556} {arXiv:1406.6556 [hep-ph]} \BibitemShut
  {NoStop}%
\bibitem [{\citenamefont {Arganda}\ \emph {et~al.}(2015)\citenamefont
  {Arganda}, \citenamefont {Herrero}, \citenamefont {Marcano},\ and\
  \citenamefont {Weiland}}]{Arganda:2014dta}%
  \BibitemOpen
  \bibfield  {author} {\bibinfo {author} {\bibfnamefont {E.}~\bibnamefont
  {Arganda}}, \bibinfo {author} {\bibfnamefont {M.~J.}\ \bibnamefont
  {Herrero}}, \bibinfo {author} {\bibfnamefont {X.}~\bibnamefont {Marcano}}, \
  and\ \bibinfo {author} {\bibfnamefont {C.}~\bibnamefont {Weiland}},\ }\href
  {\doibase 10.1103/PhysRevD.91.015001} {\bibfield  {journal} {\bibinfo
  {journal} {Phys. Rev.}\ }\textbf {\bibinfo {volume} {D91}},\ \bibinfo {pages}
  {015001} (\bibinfo {year} {2015})},\ \Eprint {http://arxiv.org/abs/1405.4300}
  {arXiv:1405.4300 [hep-ph]} \BibitemShut {NoStop}%
\bibitem [{\citenamefont {Arganda}\ \emph
  {et~al.}(2016{\natexlab{b}})\citenamefont {Arganda}, \citenamefont {Herrero},
  \citenamefont {Marcano},\ and\ \citenamefont {Weiland}}]{Arganda:2015naa}%
  \BibitemOpen
  \bibfield  {author} {\bibinfo {author} {\bibfnamefont {E.}~\bibnamefont
  {Arganda}}, \bibinfo {author} {\bibfnamefont {M.~J.}\ \bibnamefont
  {Herrero}}, \bibinfo {author} {\bibfnamefont {X.}~\bibnamefont {Marcano}}, \
  and\ \bibinfo {author} {\bibfnamefont {C.}~\bibnamefont {Weiland}},\ }\href
  {\doibase 10.1103/PhysRevD.93.055010} {\bibfield  {journal} {\bibinfo
  {journal} {Phys. Rev.}\ }\textbf {\bibinfo {volume} {D93}},\ \bibinfo {pages}
  {055010} (\bibinfo {year} {2016}{\natexlab{b}})},\ \Eprint
  {http://arxiv.org/abs/1508.04623} {arXiv:1508.04623 [hep-ph]} \BibitemShut
  {NoStop}%
\bibitem [{\citenamefont {Abada}\ \emph {et~al.}(2012)\citenamefont {Abada},
  \citenamefont {Das}, \citenamefont {Vicente},\ and\ \citenamefont
  {Weiland}}]{Abada:2012cq}%
  \BibitemOpen
  \bibfield  {author} {\bibinfo {author} {\bibfnamefont {A.}~\bibnamefont
  {Abada}}, \bibinfo {author} {\bibfnamefont {D.}~\bibnamefont {Das}}, \bibinfo
  {author} {\bibfnamefont {A.}~\bibnamefont {Vicente}}, \ and\ \bibinfo
  {author} {\bibfnamefont {C.}~\bibnamefont {Weiland}},\ }\href {\doibase
  10.1007/JHEP09(2012)015} {\bibfield  {journal} {\bibinfo  {journal} {JHEP}\
  }\textbf {\bibinfo {volume} {09}},\ \bibinfo {pages} {015} (\bibinfo {year}
  {2012})},\ \Eprint {http://arxiv.org/abs/1206.6497} {arXiv:1206.6497
  [hep-ph]} \BibitemShut {NoStop}%
\bibitem [{\citenamefont {Abada}\ \emph
  {et~al.}(2014{\natexlab{d}})\citenamefont {Abada}, \citenamefont {Krauss},
  \citenamefont {Porod}, \citenamefont {Staub}, \citenamefont {Vicente},\ and\
  \citenamefont {Weiland}}]{Abada:2014kba}%
  \BibitemOpen
  \bibfield  {author} {\bibinfo {author} {\bibfnamefont {A.}~\bibnamefont
  {Abada}}, \bibinfo {author} {\bibfnamefont {M.~E.}\ \bibnamefont {Krauss}},
  \bibinfo {author} {\bibfnamefont {W.}~\bibnamefont {Porod}}, \bibinfo
  {author} {\bibfnamefont {F.}~\bibnamefont {Staub}}, \bibinfo {author}
  {\bibfnamefont {A.}~\bibnamefont {Vicente}}, \ and\ \bibinfo {author}
  {\bibfnamefont {C.}~\bibnamefont {Weiland}},\ }\href {\doibase
  10.1007/JHEP11(2014)048} {\bibfield  {journal} {\bibinfo  {journal} {JHEP}\
  }\textbf {\bibinfo {volume} {11}},\ \bibinfo {pages} {048} (\bibinfo {year}
  {2014}{\natexlab{d}})},\ \Eprint {http://arxiv.org/abs/1408.0138}
  {arXiv:1408.0138 [hep-ph]} \BibitemShut {NoStop}%
\bibitem [{\citenamefont {Abada}\ \emph {et~al.}(2016)\citenamefont {Abada},
  \citenamefont {De~Romeri},\ and\ \citenamefont {Teixeira}}]{Abada:2015oba}%
  \BibitemOpen
  \bibfield  {author} {\bibinfo {author} {\bibfnamefont {A.}~\bibnamefont
  {Abada}}, \bibinfo {author} {\bibfnamefont {V.}~\bibnamefont {De~Romeri}}, \
  and\ \bibinfo {author} {\bibfnamefont {A.~M.}\ \bibnamefont {Teixeira}},\
  }\href {\doibase 10.1007/JHEP02(2016)083} {\bibfield  {journal} {\bibinfo
  {journal} {JHEP}\ }\textbf {\bibinfo {volume} {02}},\ \bibinfo {pages} {083}
  (\bibinfo {year} {2016})},\ \Eprint {http://arxiv.org/abs/1510.06657}
  {arXiv:1510.06657 [hep-ph]} \BibitemShut {NoStop}%
\bibitem [{\citenamefont {Abada}\ \emph
  {et~al.}(2015{\natexlab{a}})\citenamefont {Abada}, \citenamefont {De~Romeri},
  \citenamefont {Monteil}, \citenamefont {Orloff},\ and\ \citenamefont
  {Teixeira}}]{Abada:2014cca}%
  \BibitemOpen
  \bibfield  {author} {\bibinfo {author} {\bibfnamefont {A.}~\bibnamefont
  {Abada}}, \bibinfo {author} {\bibfnamefont {V.}~\bibnamefont {De~Romeri}},
  \bibinfo {author} {\bibfnamefont {S.}~\bibnamefont {Monteil}}, \bibinfo
  {author} {\bibfnamefont {J.}~\bibnamefont {Orloff}}, \ and\ \bibinfo {author}
  {\bibfnamefont {A.~M.}\ \bibnamefont {Teixeira}},\ }\href {\doibase
  10.1007/JHEP04(2015)051} {\bibfield  {journal} {\bibinfo  {journal} {JHEP}\
  }\textbf {\bibinfo {volume} {04}},\ \bibinfo {pages} {051} (\bibinfo {year}
  {2015}{\natexlab{a}})},\ \Eprint {http://arxiv.org/abs/1412.6322}
  {arXiv:1412.6322 [hep-ph]} \BibitemShut {NoStop}%
\bibitem [{\citenamefont {Abada}\ \emph
  {et~al.}(2015{\natexlab{b}})\citenamefont {Abada}, \citenamefont {Be{\v
  c}irevi{\'c}}, \citenamefont {Lucente},\ and\ \citenamefont
  {Sumensari}}]{Abada:2015zea}%
  \BibitemOpen
  \bibfield  {author} {\bibinfo {author} {\bibfnamefont {A.}~\bibnamefont
  {Abada}}, \bibinfo {author} {\bibfnamefont {D.}~\bibnamefont {Be{\v
  c}irevi{\'c}}}, \bibinfo {author} {\bibfnamefont {M.}~\bibnamefont
  {Lucente}}, \ and\ \bibinfo {author} {\bibfnamefont {O.}~\bibnamefont
  {Sumensari}},\ }\href {\doibase 10.1103/PhysRevD.91.113013} {\bibfield
  {journal} {\bibinfo  {journal} {Phys. Rev.}\ }\textbf {\bibinfo {volume}
  {D91}},\ \bibinfo {pages} {113013} (\bibinfo {year} {2015}{\natexlab{b}})},\
  \Eprint {http://arxiv.org/abs/1503.04159} {arXiv:1503.04159 [hep-ph]}
  \BibitemShut {NoStop}%
\bibitem [{\citenamefont {Pontecorvo}(1957)}]{Pontecorvo:1957cp}%
  \BibitemOpen
  \bibfield  {author} {\bibinfo {author} {\bibfnamefont {B.}~\bibnamefont
  {Pontecorvo}},\ }\href@noop {} {\bibfield  {journal} {\bibinfo  {journal}
  {Sov. Phys. JETP}\ }\textbf {\bibinfo {volume} {6}},\ \bibinfo {pages} {429}
  (\bibinfo {year} {1957})},\ \bibinfo {note} {[Zh. Eksp. Teor.
  Fiz.33,549(1957)]}\BibitemShut {NoStop}%
\bibitem [{\citenamefont {Maki}\ \emph {et~al.}(1962)\citenamefont {Maki},
  \citenamefont {Nakagawa},\ and\ \citenamefont {Sakata}}]{Maki:1962mu}%
  \BibitemOpen
  \bibfield  {author} {\bibinfo {author} {\bibfnamefont {Z.}~\bibnamefont
  {Maki}}, \bibinfo {author} {\bibfnamefont {M.}~\bibnamefont {Nakagawa}}, \
  and\ \bibinfo {author} {\bibfnamefont {S.}~\bibnamefont {Sakata}},\ }\href
  {\doibase 10.1143/PTP.28.870} {\bibfield  {journal} {\bibinfo  {journal}
  {Prog. Theor. Phys.}\ }\textbf {\bibinfo {volume} {28}},\ \bibinfo {pages}
  {870} (\bibinfo {year} {1962})}\BibitemShut {NoStop}%
\bibitem [{\citenamefont {Gonzalez-Garcia}\ \emph {et~al.}(2014)\citenamefont
  {Gonzalez-Garcia}, \citenamefont {Maltoni},\ and\ \citenamefont
  {Schwetz}}]{Gonzalez-Garcia:2014bfa}%
  \BibitemOpen
  \bibfield  {author} {\bibinfo {author} {\bibfnamefont {M.~C.}\ \bibnamefont
  {Gonzalez-Garcia}}, \bibinfo {author} {\bibfnamefont {M.}~\bibnamefont
  {Maltoni}}, \ and\ \bibinfo {author} {\bibfnamefont {T.}~\bibnamefont
  {Schwetz}},\ }\href {\doibase 10.1007/JHEP11(2014)052} {\bibfield  {journal}
  {\bibinfo  {journal} {JHEP}\ }\textbf {\bibinfo {volume} {11}},\ \bibinfo
  {pages} {052} (\bibinfo {year} {2014})},\ \Eprint
  {http://arxiv.org/abs/1409.5439} {arXiv:1409.5439 [hep-ph]} \BibitemShut
  {NoStop}%
\bibitem [{\citenamefont {Antusch}\ \emph {et~al.}(2006)\citenamefont
  {Antusch}, \citenamefont {Biggio}, \citenamefont {Fernandez-Martinez},
  \citenamefont {Gavela},\ and\ \citenamefont {Lopez-Pavon}}]{Antusch:2006vwa}%
  \BibitemOpen
  \bibfield  {author} {\bibinfo {author} {\bibfnamefont {S.}~\bibnamefont
  {Antusch}}, \bibinfo {author} {\bibfnamefont {C.}~\bibnamefont {Biggio}},
  \bibinfo {author} {\bibfnamefont {E.}~\bibnamefont {Fernandez-Martinez}},
  \bibinfo {author} {\bibfnamefont {M.~B.}\ \bibnamefont {Gavela}}, \ and\
  \bibinfo {author} {\bibfnamefont {J.}~\bibnamefont {Lopez-Pavon}},\ }\href
  {\doibase 10.1088/1126-6708/2006/10/084} {\bibfield  {journal} {\bibinfo
  {journal} {JHEP}\ }\textbf {\bibinfo {volume} {10}},\ \bibinfo {pages} {084}
  (\bibinfo {year} {2006})},\ \Eprint {http://arxiv.org/abs/hep-ph/0607020}
  {arXiv:hep-ph/0607020 [hep-ph]} \BibitemShut {NoStop}%
\bibitem [{\citenamefont {Fernandez-Martinez}\ \emph
  {et~al.}(2007)\citenamefont {Fernandez-Martinez}, \citenamefont {Gavela},
  \citenamefont {Lopez-Pavon},\ and\ \citenamefont
  {Yasuda}}]{FernandezMartinez:2007ms}%
  \BibitemOpen
  \bibfield  {author} {\bibinfo {author} {\bibfnamefont {E.}~\bibnamefont
  {Fernandez-Martinez}}, \bibinfo {author} {\bibfnamefont {M.~B.}\ \bibnamefont
  {Gavela}}, \bibinfo {author} {\bibfnamefont {J.}~\bibnamefont {Lopez-Pavon}},
  \ and\ \bibinfo {author} {\bibfnamefont {O.}~\bibnamefont {Yasuda}},\ }\href
  {\doibase 10.1016/j.physletb.2007.03.069} {\bibfield  {journal} {\bibinfo
  {journal} {Phys. Lett.}\ }\textbf {\bibinfo {volume} {B649}},\ \bibinfo
  {pages} {427} (\bibinfo {year} {2007})},\ \Eprint
  {http://arxiv.org/abs/hep-ph/0703098} {arXiv:hep-ph/0703098 [hep-ph]}
  \BibitemShut {NoStop}%
\bibitem [{\citenamefont {del Aguila}\ \emph {et~al.}(2008)\citenamefont {del
  Aguila}, \citenamefont {de~Blas},\ and\ \citenamefont
  {Perez-Victoria}}]{delAguila:2008pw}%
  \BibitemOpen
  \bibfield  {author} {\bibinfo {author} {\bibfnamefont {F.}~\bibnamefont {del
  Aguila}}, \bibinfo {author} {\bibfnamefont {J.}~\bibnamefont {de~Blas}}, \
  and\ \bibinfo {author} {\bibfnamefont {M.}~\bibnamefont {Perez-Victoria}},\
  }\href {\doibase 10.1103/PhysRevD.78.013010} {\bibfield  {journal} {\bibinfo
  {journal} {Phys. Rev.}\ }\textbf {\bibinfo {volume} {D78}},\ \bibinfo {pages}
  {013010} (\bibinfo {year} {2008})},\ \Eprint {http://arxiv.org/abs/0803.4008}
  {arXiv:0803.4008 [hep-ph]} \BibitemShut {NoStop}%
\bibitem [{\citenamefont {Antusch}\ \emph {et~al.}(2009)\citenamefont
  {Antusch}, \citenamefont {Baumann},\ and\ \citenamefont
  {Fernandez-Martinez}}]{Antusch:2008tz}%
  \BibitemOpen
  \bibfield  {author} {\bibinfo {author} {\bibfnamefont {S.}~\bibnamefont
  {Antusch}}, \bibinfo {author} {\bibfnamefont {J.~P.}\ \bibnamefont
  {Baumann}}, \ and\ \bibinfo {author} {\bibfnamefont {E.}~\bibnamefont
  {Fernandez-Martinez}},\ }\href {\doibase 10.1016/j.nuclphysb.2008.11.018}
  {\bibfield  {journal} {\bibinfo  {journal} {Nucl. Phys.}\ }\textbf {\bibinfo
  {volume} {B810}},\ \bibinfo {pages} {369} (\bibinfo {year} {2009})},\ \Eprint
  {http://arxiv.org/abs/0807.1003} {arXiv:0807.1003 [hep-ph]} \BibitemShut
  {NoStop}%
\bibitem [{\citenamefont {Antusch}\ and\ \citenamefont
  {Fischer}(2014)}]{Antusch:2014woa}%
  \BibitemOpen
  \bibfield  {author} {\bibinfo {author} {\bibfnamefont {S.}~\bibnamefont
  {Antusch}}\ and\ \bibinfo {author} {\bibfnamefont {O.}~\bibnamefont
  {Fischer}},\ }\href {\doibase 10.1007/JHEP10(2014)094} {\bibfield  {journal}
  {\bibinfo  {journal} {JHEP}\ }\textbf {\bibinfo {volume} {10}},\ \bibinfo
  {pages} {094} (\bibinfo {year} {2014})},\ \Eprint
  {http://arxiv.org/abs/1407.6607} {arXiv:1407.6607 [hep-ph]} \BibitemShut
  {NoStop}%
\bibitem [{\citenamefont {Fernandez-Martinez}\ \emph
  {et~al.}(2016)\citenamefont {Fernandez-Martinez}, \citenamefont
  {Hernandez-Garcia},\ and\ \citenamefont
  {Lopez-Pavon}}]{Fernandez-Martinez:2016lgt}%
  \BibitemOpen
  \bibfield  {author} {\bibinfo {author} {\bibfnamefont {E.}~\bibnamefont
  {Fernandez-Martinez}}, \bibinfo {author} {\bibfnamefont {J.}~\bibnamefont
  {Hernandez-Garcia}}, \ and\ \bibinfo {author} {\bibfnamefont
  {J.}~\bibnamefont {Lopez-Pavon}},\ }\href {\doibase 10.1007/JHEP08(2016)033}
  {\bibfield  {journal} {\bibinfo  {journal} {JHEP}\ }\textbf {\bibinfo
  {volume} {08}},\ \bibinfo {pages} {033} (\bibinfo {year} {2016})},\ \Eprint
  {http://arxiv.org/abs/1605.08774} {arXiv:1605.08774 [hep-ph]} \BibitemShut
  {NoStop}%
\bibitem [{\citenamefont {Alonso}\ \emph {et~al.}(2013)\citenamefont {Alonso},
  \citenamefont {Dhen}, \citenamefont {Gavela},\ and\ \citenamefont
  {Hambye}}]{Alonso:2012ji}%
  \BibitemOpen
  \bibfield  {author} {\bibinfo {author} {\bibfnamefont {R.}~\bibnamefont
  {Alonso}}, \bibinfo {author} {\bibfnamefont {M.}~\bibnamefont {Dhen}},
  \bibinfo {author} {\bibfnamefont {M.~B.}\ \bibnamefont {Gavela}}, \ and\
  \bibinfo {author} {\bibfnamefont {T.}~\bibnamefont {Hambye}},\ }\href
  {\doibase 10.1007/JHEP01(2013)118} {\bibfield  {journal} {\bibinfo  {journal}
  {JHEP}\ }\textbf {\bibinfo {volume} {01}},\ \bibinfo {pages} {118} (\bibinfo
  {year} {2013})},\ \Eprint {http://arxiv.org/abs/1209.2679} {arXiv:1209.2679
  [hep-ph]} \BibitemShut {NoStop}%
\bibitem [{\citenamefont {Cirigliano}\ and\ \citenamefont
  {Rosell}(2007)}]{Cirigliano:2007xi}%
  \BibitemOpen
  \bibfield  {author} {\bibinfo {author} {\bibfnamefont {V.}~\bibnamefont
  {Cirigliano}}\ and\ \bibinfo {author} {\bibfnamefont {I.}~\bibnamefont
  {Rosell}},\ }\href {\doibase 10.1103/PhysRevLett.99.231801} {\bibfield
  {journal} {\bibinfo  {journal} {Phys. Rev. Lett.}\ }\textbf {\bibinfo
  {volume} {99}},\ \bibinfo {pages} {231801} (\bibinfo {year} {2007})},\
  \Eprint {http://arxiv.org/abs/0707.3439} {arXiv:0707.3439 [hep-ph]}
  \BibitemShut {NoStop}%
\bibitem [{\citenamefont {Finkemeier}(1996)}]{Finkemeier:1995gi}%
  \BibitemOpen
  \bibfield  {author} {\bibinfo {author} {\bibfnamefont {M.}~\bibnamefont
  {Finkemeier}},\ }\bibfield  {booktitle} {\emph {\bibinfo {booktitle} {{2nd
  Workshop on Physics and Detectors for DAPHNE (DAPHNE 95) Frascati, Italy,
  April 4-7, 1995}}},\ }\href {\doibase 10.1016/0370-2693(96)01030-1}
  {\bibfield  {journal} {\bibinfo  {journal} {Phys. Lett.}\ }\textbf {\bibinfo
  {volume} {B387}},\ \bibinfo {pages} {391} (\bibinfo {year} {1996})},\ \Eprint
  {http://arxiv.org/abs/hep-ph/9505434} {arXiv:hep-ph/9505434 [hep-ph]}
  \BibitemShut {NoStop}%
\bibitem [{\citenamefont {Goudzovski}(2011)}]{Goudzovski:2011tc}%
  \BibitemOpen
  \bibfield  {author} {\bibinfo {author} {\bibfnamefont {E.}~\bibnamefont
  {Goudzovski}} (\bibinfo {collaboration} {NA48/2, NA62}),\ }\bibfield
  {booktitle} {\emph {\bibinfo {booktitle} {{Proceedings, 21st International
  Europhysics Conference on High energy physics (EPS-HEP 2011)}}},\ }\href@noop
  {} {\bibfield  {journal} {\bibinfo  {journal} {PoS}\ }\textbf {\bibinfo
  {volume} {EPS-HEP2011}},\ \bibinfo {pages} {181} (\bibinfo {year} {2011})},\
  \Eprint {http://arxiv.org/abs/1111.2818} {arXiv:1111.2818 [hep-ex]}
  \BibitemShut {NoStop}%
\bibitem [{\citenamefont {Lazzeroni}\ \emph {et~al.}(2013)\citenamefont
  {Lazzeroni} \emph {et~al.}}]{Lazzeroni:2012cx}%
  \BibitemOpen
  \bibfield  {author} {\bibinfo {author} {\bibfnamefont {C.}~\bibnamefont
  {Lazzeroni}} \emph {et~al.} (\bibinfo {collaboration} {NA62}),\ }\href
  {\doibase 10.1016/j.physletb.2013.01.037} {\bibfield  {journal} {\bibinfo
  {journal} {Phys. Lett.}\ }\textbf {\bibinfo {volume} {B719}},\ \bibinfo
  {pages} {326} (\bibinfo {year} {2013})},\ \Eprint
  {http://arxiv.org/abs/1212.4012} {arXiv:1212.4012 [hep-ex]} \BibitemShut
  {NoStop}%
\bibitem [{\citenamefont {Olive}\ \emph {et~al.}(2014)\citenamefont {Olive}
  \emph {et~al.}}]{Agashe:2014kda}%
  \BibitemOpen
  \bibfield  {author} {\bibinfo {author} {\bibfnamefont {K.~A.}\ \bibnamefont
  {Olive}} \emph {et~al.} (\bibinfo {collaboration} {Particle Data Group}),\
  }\href {\doibase 10.1088/1674-1137/38/9/090001} {\bibfield  {journal}
  {\bibinfo  {journal} {Chin. Phys.}\ }\textbf {\bibinfo {volume} {C38}},\
  \bibinfo {pages} {090001} (\bibinfo {year} {2014})}\BibitemShut {NoStop}%
\bibitem [{\citenamefont {Fernandez-Martinez}\ \emph
  {et~al.}(2015)\citenamefont {Fernandez-Martinez}, \citenamefont
  {Hernandez-Garcia}, \citenamefont {Lopez-Pavon},\ and\ \citenamefont
  {Lucente}}]{Fernandez-Martinez:2015hxa}%
  \BibitemOpen
  \bibfield  {author} {\bibinfo {author} {\bibfnamefont {E.}~\bibnamefont
  {Fernandez-Martinez}}, \bibinfo {author} {\bibfnamefont {J.}~\bibnamefont
  {Hernandez-Garcia}}, \bibinfo {author} {\bibfnamefont {J.}~\bibnamefont
  {Lopez-Pavon}}, \ and\ \bibinfo {author} {\bibfnamefont {M.}~\bibnamefont
  {Lucente}},\ }\href {\doibase 10.1007/JHEP10(2015)130} {\bibfield  {journal}
  {\bibinfo  {journal} {JHEP}\ }\textbf {\bibinfo {volume} {10}},\ \bibinfo
  {pages} {130} (\bibinfo {year} {2015})},\ \Eprint
  {http://arxiv.org/abs/1508.03051} {arXiv:1508.03051 [hep-ph]} \BibitemShut
  {NoStop}%
\bibitem [{\citenamefont {Benes}\ \emph {et~al.}(2005)\citenamefont {Benes},
  \citenamefont {Faessler}, \citenamefont {Simkovic},\ and\ \citenamefont
  {Kovalenko}}]{Benes:2005hn}%
  \BibitemOpen
  \bibfield  {author} {\bibinfo {author} {\bibfnamefont {P.}~\bibnamefont
  {Benes}}, \bibinfo {author} {\bibfnamefont {A.}~\bibnamefont {Faessler}},
  \bibinfo {author} {\bibfnamefont {F.}~\bibnamefont {Simkovic}}, \ and\
  \bibinfo {author} {\bibfnamefont {S.}~\bibnamefont {Kovalenko}},\ }\href
  {\doibase 10.1103/PhysRevD.71.077901} {\bibfield  {journal} {\bibinfo
  {journal} {Phys. Rev.}\ }\textbf {\bibinfo {volume} {D71}},\ \bibinfo {pages}
  {077901} (\bibinfo {year} {2005})},\ \Eprint
  {http://arxiv.org/abs/hep-ph/0501295} {arXiv:hep-ph/0501295 [hep-ph]}
  \BibitemShut {NoStop}%
\bibitem [{\citenamefont {Blennow}\ \emph {et~al.}(2010)\citenamefont
  {Blennow}, \citenamefont {Fernandez-Martinez}, \citenamefont {Lopez-Pavon},\
  and\ \citenamefont {Menendez}}]{Blennow:2010th}%
  \BibitemOpen
  \bibfield  {author} {\bibinfo {author} {\bibfnamefont {M.}~\bibnamefont
  {Blennow}}, \bibinfo {author} {\bibfnamefont {E.}~\bibnamefont
  {Fernandez-Martinez}}, \bibinfo {author} {\bibfnamefont {J.}~\bibnamefont
  {Lopez-Pavon}}, \ and\ \bibinfo {author} {\bibfnamefont {J.}~\bibnamefont
  {Menendez}},\ }\href {\doibase 10.1007/JHEP07(2010)096} {\bibfield  {journal}
  {\bibinfo  {journal} {JHEP}\ }\textbf {\bibinfo {volume} {07}},\ \bibinfo
  {pages} {096} (\bibinfo {year} {2010})},\ \Eprint
  {http://arxiv.org/abs/1005.3240} {arXiv:1005.3240 [hep-ph]} \BibitemShut
  {NoStop}%
\bibitem [{\citenamefont {Abada}\ and\ \citenamefont
  {Lucente}(2014)}]{Abada:2014vea}%
  \BibitemOpen
  \bibfield  {author} {\bibinfo {author} {\bibfnamefont {A.}~\bibnamefont
  {Abada}}\ and\ \bibinfo {author} {\bibfnamefont {M.}~\bibnamefont
  {Lucente}},\ }\href {\doibase 10.1016/j.nuclphysb.2014.06.003} {\bibfield
  {journal} {\bibinfo  {journal} {Nucl. Phys.}\ }\textbf {\bibinfo {volume}
  {B885}},\ \bibinfo {pages} {651} (\bibinfo {year} {2014})},\ \Eprint
  {http://arxiv.org/abs/1401.1507} {arXiv:1401.1507 [hep-ph]} \BibitemShut
  {NoStop}%
\bibitem [{\citenamefont {Agostini}\ \emph {et~al.}(2013)\citenamefont
  {Agostini} \emph {et~al.}}]{Agostini:2013mzu}%
  \BibitemOpen
  \bibfield  {author} {\bibinfo {author} {\bibfnamefont {M.}~\bibnamefont
  {Agostini}} \emph {et~al.} (\bibinfo {collaboration} {GERDA}),\ }\href
  {\doibase 10.1103/PhysRevLett.111.122503} {\bibfield  {journal} {\bibinfo
  {journal} {Phys. Rev. Lett.}\ }\textbf {\bibinfo {volume} {111}},\ \bibinfo
  {pages} {122503} (\bibinfo {year} {2013})},\ \Eprint
  {http://arxiv.org/abs/1307.4720} {arXiv:1307.4720 [nucl-ex]} \BibitemShut
  {NoStop}%
\bibitem [{\citenamefont {Auger}\ \emph {et~al.}(2012)\citenamefont {Auger}
  \emph {et~al.}}]{Auger:2012ar}%
  \BibitemOpen
  \bibfield  {author} {\bibinfo {author} {\bibfnamefont {M.}~\bibnamefont
  {Auger}} \emph {et~al.} (\bibinfo {collaboration} {EXO-200}),\ }\href
  {\doibase 10.1103/PhysRevLett.109.032505} {\bibfield  {journal} {\bibinfo
  {journal} {Phys. Rev. Lett.}\ }\textbf {\bibinfo {volume} {109}},\ \bibinfo
  {pages} {032505} (\bibinfo {year} {2012})},\ \Eprint
  {http://arxiv.org/abs/1205.5608} {arXiv:1205.5608 [hep-ex]} \BibitemShut
  {NoStop}%
\bibitem [{\citenamefont {Albert}\ \emph {et~al.}(2014)\citenamefont {Albert}
  \emph {et~al.}}]{Albert:2014awa}%
  \BibitemOpen
  \bibfield  {author} {\bibinfo {author} {\bibfnamefont {J.~B.}\ \bibnamefont
  {Albert}} \emph {et~al.} (\bibinfo {collaboration} {EXO-200}),\ }\href
  {\doibase 10.1038/nature13432} {\bibfield  {journal} {\bibinfo  {journal}
  {Nature}\ }\textbf {\bibinfo {volume} {510}},\ \bibinfo {pages} {229}
  (\bibinfo {year} {2014})},\ \Eprint {http://arxiv.org/abs/1402.6956}
  {arXiv:1402.6956 [nucl-ex]} \BibitemShut {NoStop}%
\bibitem [{\citenamefont {Gando}\ \emph {et~al.}(2013)\citenamefont {Gando}
  \emph {et~al.}}]{Gando:2012zm}%
  \BibitemOpen
  \bibfield  {author} {\bibinfo {author} {\bibfnamefont {A.}~\bibnamefont
  {Gando}} \emph {et~al.} (\bibinfo {collaboration} {KamLAND-Zen}),\ }\href
  {\doibase 10.1103/PhysRevLett.110.062502} {\bibfield  {journal} {\bibinfo
  {journal} {Phys. Rev. Lett.}\ }\textbf {\bibinfo {volume} {110}},\ \bibinfo
  {pages} {062502} (\bibinfo {year} {2013})},\ \Eprint
  {http://arxiv.org/abs/1211.3863} {arXiv:1211.3863 [hep-ex]} \BibitemShut
  {NoStop}%
\bibitem [{\citenamefont {Peskin}\ and\ \citenamefont
  {Takeuchi}(1992)}]{Peskin:1991sw}%
  \BibitemOpen
  \bibfield  {author} {\bibinfo {author} {\bibfnamefont {M.~E.}\ \bibnamefont
  {Peskin}}\ and\ \bibinfo {author} {\bibfnamefont {T.}~\bibnamefont
  {Takeuchi}},\ }\href {\doibase 10.1103/PhysRevD.46.381} {\bibfield  {journal}
  {\bibinfo  {journal} {Phys. Rev.}\ }\textbf {\bibinfo {volume} {D46}},\
  \bibinfo {pages} {381} (\bibinfo {year} {1992})}\BibitemShut {NoStop}%
\bibitem [{\citenamefont {Akhmedov}\ \emph {et~al.}(2013)\citenamefont
  {Akhmedov}, \citenamefont {Kartavtsev}, \citenamefont {Lindner},
  \citenamefont {Michaels},\ and\ \citenamefont {Smirnov}}]{Akhmedov:2013hec}%
  \BibitemOpen
  \bibfield  {author} {\bibinfo {author} {\bibfnamefont {E.}~\bibnamefont
  {Akhmedov}}, \bibinfo {author} {\bibfnamefont {A.}~\bibnamefont
  {Kartavtsev}}, \bibinfo {author} {\bibfnamefont {M.}~\bibnamefont {Lindner}},
  \bibinfo {author} {\bibfnamefont {L.}~\bibnamefont {Michaels}}, \ and\
  \bibinfo {author} {\bibfnamefont {J.}~\bibnamefont {Smirnov}},\ }\href
  {\doibase 10.1007/JHEP05(2013)081} {\bibfield  {journal} {\bibinfo  {journal}
  {JHEP}\ }\textbf {\bibinfo {volume} {05}},\ \bibinfo {pages} {081} (\bibinfo
  {year} {2013})},\ \Eprint {http://arxiv.org/abs/1302.1872} {arXiv:1302.1872
  [hep-ph]} \BibitemShut {NoStop}%
\bibitem [{\citenamefont {Atre}\ \emph {et~al.}(2009)\citenamefont {Atre},
  \citenamefont {Han}, \citenamefont {Pascoli},\ and\ \citenamefont
  {Zhang}}]{Atre:2009rg}%
  \BibitemOpen
  \bibfield  {author} {\bibinfo {author} {\bibfnamefont {A.}~\bibnamefont
  {Atre}}, \bibinfo {author} {\bibfnamefont {T.}~\bibnamefont {Han}}, \bibinfo
  {author} {\bibfnamefont {S.}~\bibnamefont {Pascoli}}, \ and\ \bibinfo
  {author} {\bibfnamefont {B.}~\bibnamefont {Zhang}},\ }\href {\doibase
  10.1088/1126-6708/2009/05/030} {\bibfield  {journal} {\bibinfo  {journal}
  {JHEP}\ }\textbf {\bibinfo {volume} {05}},\ \bibinfo {pages} {030} (\bibinfo
  {year} {2009})},\ \Eprint {http://arxiv.org/abs/0901.3589} {arXiv:0901.3589
  [hep-ph]} \BibitemShut {NoStop}%
\bibitem [{\citenamefont {Forero}\ \emph {et~al.}(2012)\citenamefont {Forero},
  \citenamefont {Tortola},\ and\ \citenamefont {Valle}}]{Tortola:2012te}%
  \BibitemOpen
  \bibfield  {author} {\bibinfo {author} {\bibfnamefont {D.~V.}\ \bibnamefont
  {Forero}}, \bibinfo {author} {\bibfnamefont {M.}~\bibnamefont {Tortola}}, \
  and\ \bibinfo {author} {\bibfnamefont {J.~W.~F.}\ \bibnamefont {Valle}},\
  }\href {\doibase 10.1103/PhysRevD.86.073012} {\bibfield  {journal} {\bibinfo
  {journal} {Phys. Rev.}\ }\textbf {\bibinfo {volume} {D86}},\ \bibinfo {pages}
  {073012} (\bibinfo {year} {2012})},\ \Eprint {http://arxiv.org/abs/1205.4018}
  {arXiv:1205.4018 [hep-ph]} \BibitemShut {NoStop}%
\bibitem [{\citenamefont {Fogli}\ \emph {et~al.}(2012)\citenamefont {Fogli},
  \citenamefont {Lisi}, \citenamefont {Marrone}, \citenamefont {Montanino},
  \citenamefont {Palazzo},\ and\ \citenamefont {Rotunno}}]{Fogli:2012ua}%
  \BibitemOpen
  \bibfield  {author} {\bibinfo {author} {\bibfnamefont {G.~L.}\ \bibnamefont
  {Fogli}}, \bibinfo {author} {\bibfnamefont {E.}~\bibnamefont {Lisi}},
  \bibinfo {author} {\bibfnamefont {A.}~\bibnamefont {Marrone}}, \bibinfo
  {author} {\bibfnamefont {D.}~\bibnamefont {Montanino}}, \bibinfo {author}
  {\bibfnamefont {A.}~\bibnamefont {Palazzo}}, \ and\ \bibinfo {author}
  {\bibfnamefont {A.~M.}\ \bibnamefont {Rotunno}},\ }\href {\doibase
  10.1103/PhysRevD.86.013012} {\bibfield  {journal} {\bibinfo  {journal} {Phys.
  Rev.}\ }\textbf {\bibinfo {volume} {D86}},\ \bibinfo {pages} {013012}
  (\bibinfo {year} {2012})},\ \Eprint {http://arxiv.org/abs/1205.5254}
  {arXiv:1205.5254 [hep-ph]} \BibitemShut {NoStop}%
\bibitem [{\citenamefont {Forero}\ \emph {et~al.}(2014)\citenamefont {Forero},
  \citenamefont {Tortola},\ and\ \citenamefont {Valle}}]{Forero:2014bxa}%
  \BibitemOpen
  \bibfield  {author} {\bibinfo {author} {\bibfnamefont {D.~V.}\ \bibnamefont
  {Forero}}, \bibinfo {author} {\bibfnamefont {M.}~\bibnamefont {Tortola}}, \
  and\ \bibinfo {author} {\bibfnamefont {J.~W.~F.}\ \bibnamefont {Valle}},\
  }\href {\doibase 10.1103/PhysRevD.90.093006} {\bibfield  {journal} {\bibinfo
  {journal} {Phys. Rev.}\ }\textbf {\bibinfo {volume} {D90}},\ \bibinfo {pages}
  {093006} (\bibinfo {year} {2014})},\ \Eprint {http://arxiv.org/abs/1405.7540}
  {arXiv:1405.7540 [hep-ph]} \BibitemShut {NoStop}%
\bibitem [{\citenamefont {Adhikari}\ \emph {et~al.}()\citenamefont {Adhikari}
  \emph {et~al.}}]{Adhikari:2016bei}%
  \BibitemOpen
  \bibfield  {author} {\bibinfo {author} {\bibfnamefont {R.}~\bibnamefont
  {Adhikari}} \emph {et~al.},\ }\href@noop {} {\ }\Eprint
  {http://arxiv.org/abs/1602.04816} {arXiv:1602.04816 [hep-ph]} \BibitemShut
  {NoStop}%
\bibitem [{\citenamefont {Hahn}\ and\ \citenamefont
  {Perez-Victoria}(1999)}]{Hahn:1998yk}%
  \BibitemOpen
  \bibfield  {author} {\bibinfo {author} {\bibfnamefont {T.}~\bibnamefont
  {Hahn}}\ and\ \bibinfo {author} {\bibfnamefont {M.}~\bibnamefont
  {Perez-Victoria}},\ }\href {\doibase 10.1016/S0010-4655(98)00173-8}
  {\bibfield  {journal} {\bibinfo  {journal} {Comput. Phys. Commun.}\ }\textbf
  {\bibinfo {volume} {118}},\ \bibinfo {pages} {153} (\bibinfo {year}
  {1999})},\ \Eprint {http://arxiv.org/abs/hep-ph/9807565}
  {arXiv:hep-ph/9807565 [hep-ph]} \BibitemShut {NoStop}%
\bibitem [{\citenamefont {Deppisch}\ and\ \citenamefont
  {Valle}(2005)}]{Deppisch:2004fa}%
  \BibitemOpen
  \bibfield  {author} {\bibinfo {author} {\bibfnamefont {F.}~\bibnamefont
  {Deppisch}}\ and\ \bibinfo {author} {\bibfnamefont {J.~W.~F.}\ \bibnamefont
  {Valle}},\ }\href {\doibase 10.1103/PhysRevD.72.036001} {\bibfield  {journal}
  {\bibinfo  {journal} {Phys. Rev.}\ }\textbf {\bibinfo {volume} {D72}},\
  \bibinfo {pages} {036001} (\bibinfo {year} {2005})},\ \Eprint
  {http://arxiv.org/abs/hep-ph/0406040} {arXiv:hep-ph/0406040 [hep-ph]}
  \BibitemShut {NoStop}%
\bibitem [{\citenamefont {Passarino}\ and\ \citenamefont
  {Veltman}(1979)}]{Passarino:1978jh}%
  \BibitemOpen
  \bibfield  {author} {\bibinfo {author} {\bibfnamefont {G.}~\bibnamefont
  {Passarino}}\ and\ \bibinfo {author} {\bibfnamefont {M.~J.~G.}\ \bibnamefont
  {Veltman}},\ }\href {\doibase 10.1016/0550-3213(79)90234-7} {\bibfield
  {journal} {\bibinfo  {journal} {Nucl. Phys.}\ }\textbf {\bibinfo {volume}
  {B160}},\ \bibinfo {pages} {151} (\bibinfo {year} {1979})}\BibitemShut
  {NoStop}%
\end{thebibliography}%

\end{document}